\documentclass[iop,12pt]{emulateapj}
\usepackage[normalem]{ulem}
\usepackage{epsfig}
\usepackage{natbib}
\usepackage{amsfonts}
\usepackage{amsmath}
\usepackage{multirow}
\usepackage{enumerate}
\usepackage[plainpages=false, colorlinks=true, anchorcolor=blue, linkcolor=blue, citecolor=blue, bookmarks=false]{hyperref}
\usepackage[usenames,dvipsnames]{color}
\newcommand{\z}{\phantom{0}}

\def\Dwa{$\,$\uppercase\expandafter{\romannumeral5}$\,$}

\def\sless{\lower2pt\hbox{$\buildrel {\scriptstyle <}
   \over {\scriptstyle\sim}$}}

\def\sgreat{\lower2pt\hbox{$\buildrel {\scriptstyle >}
   \over {\scriptstyle\sim}$}}
\def\sharpnull#1{}

\newcommand{\code}[1]{\texttt{#1}}

\begin{document}
\slugcomment{Accepted to ApJ. June 4, 2015}

\title{Neutrino-driven Turbulent Convection and Standing Accretion
  Shock Instability in \\ Three-Dimensional Core-Collapse Supernovae}

\author{Ernazar Abdikamalov\altaffilmark{1,2},
  Christian D. Ott\altaffilmark{1,*},
  David Radice\altaffilmark{1},
  Luke F. Roberts\altaffilmark{1,+},
  Roland Haas\altaffilmark{1,3},\\
  Christian Reisswig\altaffilmark{1,+},
  Philipp M\"osta\altaffilmark{1}, 
  Hannah Klion\altaffilmark{1},
  and Erik Schnetter\altaffilmark{4,5,6}}
  \altaffiltext{1}{TAPIR, Walter Burke Institute for Theoretical Physics, 
  Mailcode 350-17,
  California Institute of Technology, Pasadena, CA 91125, USA, 
  cott@tapir.caltech.edu}
\altaffiltext{*}{Alfred P. Sloan Research Fellow}
\altaffiltext{+}{NASA Einstein Fellow}
\altaffiltext{2}{Department of Physics, School of Science and Technology, Nazarbayev University, Astana 010000, Kazakhstan}
\altaffiltext{3}{Max-Planck-Institut f\"ur Gravitationsphysik, Albert-Einstein-Institut, 14476 Golm, Germany}
\altaffiltext{4}{Perimeter Institute for Theoretical Physics, Waterloo, ON, Canada}
\altaffiltext{5}{Department of Physics, University of Guelph, Guelph, ON, Canada}
\altaffiltext{6}{Center for Computation \& Technology, Louisiana State
  University, Baton Rouge, LA, USA}

\begin{abstract}
We conduct a series of numerical experiments into the nature of
three-dimensional (3D) hydrodynamics in the postbounce stalled-shock
phase of core-collapse supernovae using 3D general-relativistic
hydrodynamic simulations of a $27$-$M_\odot$ progenitor star with a
neutrino leakage/heating scheme. We vary the strength of neutrino
heating and find three cases of 3D dynamics: (1) neutrino-driven
convection, (2) initially neutrino-driven convection and subsequent
development of the standing accretion shock instability (SASI), (3)
SASI dominated evolution. This confirms previous 3D results of Hanke
et al.~(2013), ApJ 770:66 and Couch \& Connor (2014), ApJ 785:123. We
carry out simulations with resolutions differing by up to a factor of
$\sim$4 and demonstrate that low resolution is
artificially favorable for explosion in the 3D convection-dominated
case, since it decreases the efficiency of energy transport to small
scales. Low resolution results in higher radial convective fluxes of
energy and enthalpy, more fully buoyant mass, and stronger neutrino
heating. In the SASI-dominated case, lower resolution damps SASI
oscillations.  In the convection-dominated case, a quasi-stationary
angular kinetic energy spectrum $E(\ell)$ develops in the heating
layer. Like other 3D studies, we find $E(\ell) \propto \ell^{-1}$ in
the ``inertial range,'' while theory and local simulations argue for
$E(\ell) \propto \ell^{-5/3}$.  We argue that current 3D
simulations do not resolve the inertial range of turbulence and are
affected by numerical viscosity up to the energy containing scale,
creating a ``bottleneck'' that prevents an efficient turbulent
cascade.
\end{abstract}


\section{Introduction}

Multi-dimensional dynamics is, quite literally, at the heart of
core-collapse supernovae from massive stars. Decades of theoretical
and computational studies have shown that the hydrodynamic shock
formed at core bounce always stalls and fails to be revived by
neutrino energy deposition in simulations that assume spherical
symmetry (1D;
\citealt{bethe:90,liebendoerfer:05,thompson:03,ramppjanka:00,sumiyoshi:05}).
The advent of detailed axisymmetric (2D) simulations led to the
realization that neutrino-driven convection
\citep{herant:95,bhf:95,janka:96} and the advective-acoustic standing
accretion shock instability (SASI;
\citealt{blondin:03,foglizzo:07,scheck:08}) may both play an important
facilitating role in the neutrino mechanism for core-collapse
supernova explosions. The nonradial dynamics associated with these
instabilities can increase the time material spends in the layer near
the stalled shock where net neutrino energy absorption occurs (the
``gain layer''). This, in turn, increases the neutrino heating
efficiency and creates conditions favorable for launching an explosion
(e.g., \citealt{murphy:08}). Rising convective plumes and large
high-entropy bubbles created by SASI-induced secondary shocks can
exert mechanical force on the shock and push it out
\citep{bhf:95,dolence:13,couch:13b,fernandez:14}. As recently pointed
out by \cite{murphy:13} and \cite{couch:15a}, turbulent flow, which is
both unavoidable and ubiquitous in the gain layer, provides an
effective pressure that adds to the pressure budget behind the shock
and thus further helps the multi-D neutrino mechanism.

The set of recent detailed ab initio 2D neutrino
radiation-hydrodynamics simulations yields successful explosions in
multiple cases and codes (e.g.,
\citealt{marek:09,mueller:12a,mueller:12b,bruenn:13}), but failures in
some others { (e.g., \citealt{ott:08,dolence:15}, who used
  different approximations for radiation transport and microphysics)}.
One must not rest on the partial 2D success of the neutrino
mechanism. Nature is 3D, so are core-collapse supernovae, and so is
the multi-D dynamics in their postbounce cores.  3D work was pioneered
by the smooth-particle hydrodynamics simulations of
\cite{fryerwarren:02}, but grid-based 3D simulations had to await the
broad availability of petascale computing resources and have become
possible only recently. Most current 3D simulations do not yet reach
the level of their 2D counterparts in implemented and captured
physics, and in numerical resolution. Yet they are beginning to yield
results that elucidate the 3D hydrodynamics of core-collapse
supernovae and differences between 2D and 3D (e.g.,
\citealt{hanke:12,burrows:12,murphy:13,dolence:13, couch:13b,ott:13a,
  couch:13d,handy:14,couch:14a,couch:15a,takiwaki:14a}).

\cite{hanke:13} and \cite{tamborra:14} carried out the only 3D studies
to-date with accurate energy-dependent neutrino transport, which they
implement not in 3D, but along many 1D rays. The angular resolution of
these simulations is $\sim$2$^\circ$ for both hydrodynamics and
neutrinos.  Current 3D Cartesian adaptive-mesh-refinement (AMR)
simulations with a more approximate neutrino treatment reach much
finer effective angular resolutions of $0.4^\circ-0.8^\circ$ in the
gain layer (e.g., \citealt{couch:14a,ott:13a,dolence:13}).

While there is still much tension between the detailed results (and
their interpretation) of current 3D simulations obtained with
different approximations and codes, there is consensus that
the development of large-scale, high-entropy regions (by neutrino
heating or SASI) and, generally, kinetic energy at large scales is
required for a neutrino-driven explosion to succeed
\citep{burrows:12,hanke:12,hanke:13,murphy:13,
  dolence:13,ott:13a,couch:13d,couch:14a,couch:15a}.

In this work, we systematically study the qualitative and quantitative
dependence of 3D postbounce hydrodynamics on the strength of neutrino
heating and on numerical resolution. For this, we employ our 3D fully
general-relativistic core-collapse supernova simulation code
\texttt{Zelmani} introduced in \cite{ott:12a} and \cite{ott:13a}. This
code includes a three-species neutrino leakage scheme, which allows us
to control the local efficiency of neutrino heating. We carry out
simulations of the postbounce evolution of the $27$-$M_\odot$
progenitor model of \cite{whw:02}, which has been {
  considered by multiple recent studies}.  Its structure results in a
high postbounce accretion rate, which leads to a small radius of the
stalled shock, favoring the development of SASI
\citep{mueller:12b,ott:13a,couch:14a,hanke:13}.

We are particularly interested in (\emph{i}) the prominence of 3D
neutrino-driven convection and 3D SASI, their interplay, and their
dependence on neutrino heating; (\emph{ii}) the resolution dependence
of postbounce hydrodynamics, neutrino heating, and the development of
an explosion; and (\emph{iii}) the nature of turbulence under
neutrino-driven convection dominated conditions and its dependence on
resolution.

We find three general regimes of postbounce 3D hydrodynamics: (1)
neutrino-driven convection and onset of explosion (for strong neutrino
heating; e.g., \citealt{dolence:13,ott:13a,couch:14a}), (2) initially
neutrino-driven convection that subsides and is replaced by strong
SASI with spiral modes and no explosion (for moderate neutrino
heating; consistent with \citealt{hanke:13} and \citealt{couch:14a}),
and (3) complete absence of neutrino-driven convection, SASI-dominated
dynamics with spiral modes and no explosion (for weak neutrino
heating).  The results of our resolution study show that low numerical
resolution artificially damps SASI oscillations in the SASI-dominated
case. In the neutrino-driven convection dominated case, we show that
low resolution leads to artificially favorable conditions for
explosion. The lower the resolution, the less efficient the cascade of
turbulent kinetic energy to small scales (as previously noted by
\cite{hanke:12} on the basis of their simpler ``light-bulb''
simulations).  Low resolution simulations have higher radial
convective kinetic energy and enthalpy fluxes, more buoyant mass in
the gain layer, higher neutrino heating rates, larger average shock
radii, and transition to explosion earlier than more finely resolved
simulations. Analyzing the angular spectra $E(\ell)$ of turbulence in
our simulations, we find a scaling $E(\ell) \propto \ell^{-1}$
(cf.~\citealt{dolence:13,couch:14a}) at spherical harmonic mode
$\ell$ that should belong to
the inertial range of turbulence. By comparison with the literature on
local mildly compressible turbulence, we argue that our and other
global 3D simulations similar to ours do not resolve the inertial
range of neutrino-driven turbulent convection. Instead, numerical
viscosity creates a bottleneck that hinders the efficient cascade of
turbulent kinetic energy to small scales. Energy is thus kept at large
scales, which may, incorrectly and artificially, promote explosion.

We begin in Section~\ref{sec:methods} with a discussion of our
numerical approach and lay out our simulation plan in
Section~\ref{sec:id}. In Section~\ref{sec:heating}, we present results
from our simulations in the strong, moderate, and weak neutrino
heating regimes and provide detailed analyses of neutrino-driven
convection, SASI, and turbulence in these simulations. In
Section~\ref{sec:res}, we present and discuss the results of our
extensive resolution study. In Section~\ref{sec:conclusions}, we put
our results into the broader context of the current discussion of the
multi-D neutrino mechanism of core-collapse supernovae and conclude.

\section{Methods}
\label{sec:methods}

\begin{figure}[t]
\centering
\includegraphics[width=1.0\columnwidth]{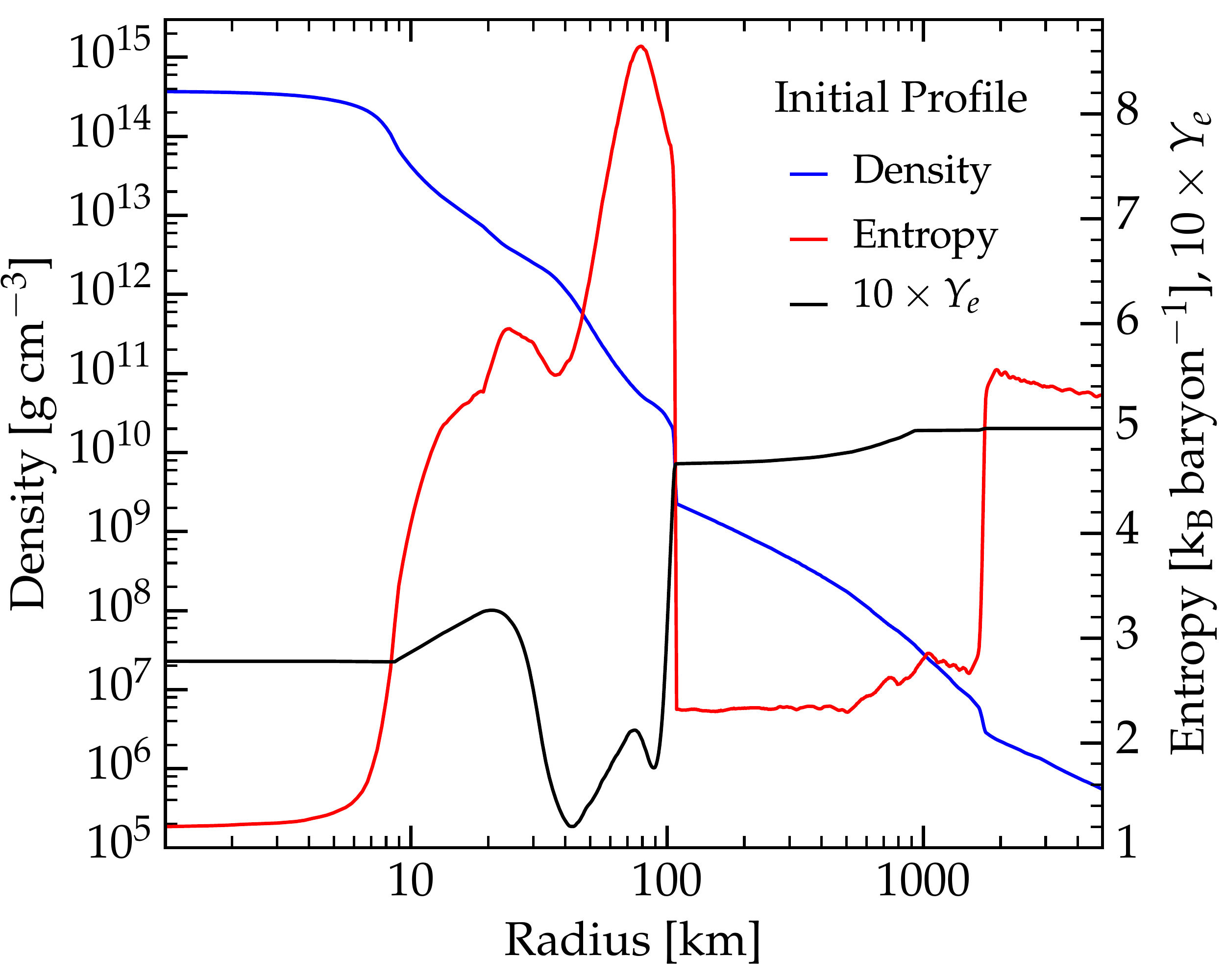}
\caption{Density (left ordinate), specific entropy and electron
  fraction $Y_e$ (both right ordinate) profiles from the \code{GR1D}
  simulation at the time of mapping to 3D, $20\,\mathrm{ms}$ after
  bounce.}
\label{fig:initial_profile}
\vspace{0.5ex}
\end{figure}

{ We simulate core collapse and postbounce evolution of the
  nonrotating $27$-$M_\odot$ solar-metallicity model of
  \cite{whw:02}. We follow collapse, bounce, and the first
  $20\,\mathrm{ms}$ in spherical symmetry using {\tt GR1D}
  \citep{oconnor:10} with neutrino leakage and a heating factor
  $f_\mathrm{heat} = 1.05$ (see below for a definition of
  $f_\mathrm{heat}$). At $20\,\mathrm{ms}$ after bounce, the shock has
  almost stalled.}  {Figure~\ref{fig:initial_profile} shows
  the spherically-symmetric density, specific entropy, and electron
  fraction $Y_e$ profiles at $20\,\mathrm{ms}$ after bounce}. We then map this
configuration to our 3D grid and continue the evolution in full 3D.
We choose this 1D--3D approach to save computer time during the
spherical collapse phase and to avoid having the shock cross the
boundaries of the two innermost mesh refinement levels of the 3D grid,
which could generate significant numerical error \citep{ott:13a}. By
mapping at $\sim$$20\,\mathrm{ms}$, we miss the earliest part of
prompt postbounce convection due to the negative entropy left behind
by the weakening shock.  Since we are not interested in studying this
prompt convection, we believe that our approach is appropriate for the
simulations at hand. At the time of mapping, the shock has reached
$\sim$$110\,\mathrm{km}$.

The subsequent 3D evolution is performed with the \code{Zelmani}
core-collapse simulation package \citep{ott:12a,ott:13a}. It is based
on the \code{Cactus Computational Toolkit} \citep{goodale:03} and it
uses modules of the open-source \code{Einstein
  Toolkit}\footnote{\url{http://www.einsteintoolkit.org}}
(\citealt{et:12,moesta:14a}).  We employ a cubed-sphere multiblock
adaptive-mesh-refinement (AMR) system that consists of a set of
overlapping curvilinear grid blocks adapted to the overall spherical
geometry of the problem \citep{pollney:11,reisswig:13a}. {
  The qualitative grid setup is very similar to the one described in
  \cite{ott:13a} and we refer the reader to their Figure~1 that
  visualizes the overall structure of our grid}. The inner $\sim
532\,\mathrm{km}$ (along one of the coordinate axes), which contain
the protoneutron star and the entire shocked region including the
shock, are covered by a cubic Cartesian mesh. This Cartesian region
contains four additional co-centric cubic refinement
levels. Initially, these levels have radial extents (along the
coordinate axes) of $(286, 161, 43, 21 )\,\mathrm{km}$.  Throughout
the 3D simulation, the shock is contained on the third finest level
whose outer boundary automatically adapts to the shock's position.  In
our baseline resolution, the grid on the finest AMR level has a linear
cell width of $0.354\,\mathrm{km}$. The third-finest level containing
the entire postshock region and the shock has a linear cell width of
$\sim 1.416\,\mathrm{km}$. This corresponds to an effective angular
resolution of $0.81^\circ$ at $100\,\mathrm{km}$ and $0.54^\circ$ at
$150\,\mathrm{km}$.

The outer regions are covered by a shell of six angular grid blocks
that stretch to $15,000\,\mathrm{km}$. The angular blocks are arranged
such that the two angular coordinate directions at each lateral edge
of each block always coincide with those from neighboring patches
\citep{reisswig:13a}.  The angular resolution in those patches is
$\sim$$3^\circ$, which is sufficient since matter in those regions
remains spherically symmetric. The radial resolution at the inner
boundary of the angular patches is chosen to be the same as that of
the coarsest AMR level, which, for the baseline resolution, is a
linear cell width of $5.67\,\mathrm{km}$. The resolution decreases
gradually with radius, reaching $189\,\mathrm{km}$ at the outer
boundary. An important advantange of this multi-block system is that
it does not suffer from any coordinate pathologies unlike standard
spherical-polar and cylindrical grids.

\begin{deluxetable*}{ccccccccc}
\tablecolumns{5} \tablewidth{0pc} \tablecaption{Key Simulation
  Parameters and Results.}  \tablehead{
  Model&$f_\mathrm{heat}$&$dx_\mathrm{shock}$&$d\theta,d\phi$&$t_\mathrm{end}$&$R_\mathrm{shock,max}$&$R_\mathrm{shock,avg}$&$R_\mathrm{shock,min}$&Numerical\\ &
  &(km) &@100\,km &(ms) & @$t_\mathrm{end}$ &@$t_\mathrm{end}$ &
  @$t_\mathrm{end}$ &Reynolds\\ & & &(degrees) & & (km) & (km) & (km)
  & Number } \startdata $s27\mathrm{\,U\,L\,R}f_\mathrm{heat}1.05$
&1.05 & 3.784 & 2.16 & 160 & 295 & 321 & 224 & 53.25
\\ $s27\mathrm{\,L\,R}f_\mathrm{heat}1.05$ &1.05 & 1.892 & 1.08 & 138
& 248 & 202 & 171 & 62.06 \\ $s27\mathrm{MR}f_\mathrm{heat}1.05$ &1.05
& 1.416 & 0.81 & 131 & 233 & 192 & 167 & 68.14
\\ $s27\mathrm{\,\,I\,\,R}f_\mathrm{heat}1.05$&1.05 & 1.240 & 0.71 &
142 & 229 & 190 & 156 & 70.03 \\ $s27\mathrm{HR}f_\mathrm{heat}1.05$
&1.05 & 1.064 & 0.61 & 142 & 215 & 182 & 158 & 72.21 \\[0.5em]
\hline\\[-0.4em] \z$s27\mathrm{MR}f_\mathrm{heat}0.95\z$ &0.95 & 1.416
& 0.81 & 262 &\z79 &\z70 &\z62 & ---\\[0.5em] \hline\\[-0.4em]
$s27\mathrm{\,L\,R}f_\mathrm{heat}0.8\z$ &0.8\z& 1.892 & 1.08 & 215
&\z82 &\z72 &\z63 & ---\\ $s27\mathrm{MR}f_\mathrm{heat}0.8\z$ &0.8\z&
1.416 & 0.81 & 255 &\z85 &\z71 &\z52 & --- \enddata
\tablecomments{$f_\mathrm{heat}$ is the scaling factor that controls
  the neutrino heating rate (cf.~Equation~\ref{eq:heating}),
  $dx_\mathrm{shock}$ is the linear cell width on the AMR level that
  contains the shock, $d\theta,d\phi$ @ $100\,\mathrm{km}$ is the
  effective angular resolution at a distance of $100\,\mathrm{km}$
  from the origin, $t_\mathrm{end}$ is the time after core bounce when
  the simulation is terminated, and $R_\mathrm{shock,min}$,
  $R_\mathrm{shock,avg}$, and $R_\mathrm{shock,max}$ are the minimum,
  average, and maximum shock radii at the end of our simulations,
  respectively. The procedure for calculating the numerical Reynolds
  number is discussed in Appendix~\ref{sec:reynolds_number}. We quote
  its approximate value at $90\,\mathrm{ms}$ after bounce for models
  whose postbounce hydrodynamics is dominated by neutrino-driven
  convection.}
\label{tab:results}
\end{deluxetable*}

We solve the 3D general-relativistic hydrodynamics equations in a
flux-conservative form \citep{banyuls:97} using the finite-volume
general-relativistic hydrodynamics code \code{GRHydro} \citep{et:12}.
It is an improved version of the legacy code \code{Whisky}
\citep{baiotti:05}, which itself is largely based on the
\code{GR-Astro/MAHC} code \citep{font:00}.  We use a customized
version of the piecewise-parabolic method (PPM; \citealt{colella:84})
for the reconstruction of physical states at cell boundaries. The
propagation of a quasi-spherical shock on a Cartesian grid creates
numerical perturbations that could seed convection at a possibly
unphysically high level (\citealt{ott:13a}; but see, e.g.,
\citealt{couch:13d}). To minimize numerical perturbations, we use the
original PPM scheme \citep{colella:84} on the AMR level that contains
the shock. We employ the more aggressive, lower-dissipation enhanced
PPM scheme \citep{mccorquodale:11,reisswig:13a} on finer levels, since
it outperforms the original PPM scheme in capturing the steep
gradients at the edge of the protoneutron star, and, importantly,
maintains the smooth physical density maximum at the center of the
protoneutron star. The intercell fluxes are calculated via solving
approximate Riemann problems with the HLLE solver \citep{HLLE:88}.

We evolve the $3+1$ Einstein equations with the BSSN formulation of
numerical relativity \citep{baumgarte:99,shibata:95}. We
use a $1+\log$ slicing \citep{alcubierre:00} and a modified
$\Gamma$-driver \citep{alcubierre:03a} to evolve the lapse function
$\alpha$ and the shift vector $\beta^i$, respectively. The BSSN
equations and the gauge conditions are evolved using the {\tt CTGamma}
code \citep{pollney:11,reisswig:13a}.

The hydrodynamics and Einstein equations are evolved in time in a
coupled manner using the Method of Lines
\citep{Hyman-1976-Courant-MOL-report}. The latter uses a multi-rate
Runge-Kutta scheme, which is second-order in time for hydrodynamics
and fourth-order in time for spacetime evolution
\citep{reisswig:13a}. We use a Courant-Friedrichs-Levy factor of $0.4$
in all of our simulations and the timestep taken on each refinement
level is governed by the light travel time along a linear
computational cell width.

We employ the tabulated finite-temperature nuclear EOS of
\cite{lseos:91} with $K=220\,\mathrm{MeV}$, generated
by~\cite{oconnor:10}\footnote{Available for download
  at~\url{http://www.stellarcollapse.org}.}.  During collapse, we use
the parameterized $Y_e(\rho)$ deleptonization scheme of
\cite{liebendoerfer:05fakenu} with the same parameters as
\cite{ott:13a}, while in the postbounce phase, we use a three-species
($\nu_e$, $\bar{\nu}_e$, $\nu_x = \{\nu_\mu, \bar{\nu}_\mu, \nu_\tau,
\bar{\nu}_\tau\}$) neutrino leakage/heating scheme that approximates
deleptonization, cooling, and heating in the gain region
\citep{oconnor:10,ott:12a,ott:13a,couch:14a}. { The scheme
  first computes the energy-averaged neutrino optical depths along
  radial rays.  Then, local estimates of energy and lepton loss rates
  are computed.}  The 3D implementation of this scheme in {\tt
  Zelmani} is discussed in detail in \cite{ott:12a,ott:13a}. In
contrast to these previous works, we do not include neutrino pressure
contributions in this study, since the implementations of the neutrino
pressure terms are slightly different in {\tt GR1D} and {\tt Zelmani}
and tests show that this leads to spurious oscillations of the
protoneutron star upon mapping, { which, in turn, due to grid
perturbations, artificially drives unphysically strong prompt convection upon
  mapping.} Neglecting the neutrino pressure,
which contributes $\sim$$10-20\%$ of the pressure in a narrow density
regime from $\sim$$10^{12.5}-10^{14}\,\mathrm{g\,cm}^{-3}$
\citep{kaplan:14}, results in a slightly more compact protoneutron
star, but should not otherwise affect our results.

We approximate the neutrino heating rate $Q^{\mathrm{heat}}_{\nu_i}$
in the gain region by
\begin{equation}
Q^{\mathrm{heat}}_{\nu_i} = f_\mathrm{heat} \frac{L_{\nu_i}(r)}{4\pi r^2}
S_{\nu} \langle\epsilon^2_{\nu_i}\rangle\, {\rho\over m_n}
  X_i \left\langle {1 \over
  F_{\nu_i}} \right\rangle e^{-2\tau_{\nu_i}} \,\,.\label{eq:heating}
\end{equation}
Here $L_{\nu_i}$ is the neutrino luminosity emerging from below as
predicted by auxiliary leakage calculations along radial rays,
$S_{\nu} = 0.25 (1 + 3\alpha^2) \sigma_0 (m_e c^2)^{-2}$, $\sigma_0 =
1.76\times 10^{-44}\, \mathrm{cm}^2$, $\alpha = 1.23$, $m_e$ is the
electron mass and $c$ is the speed of light, $\rho$ is the rest-mass
density, $m_n$ is the neutron mass, $X_i$ is the neutron (proton) mass
fraction for electron neutrinos (antineutrinos), $\langle
\epsilon^2_{\nu_i} \rangle$ is the mean-squared energy of $\nu_i$
neutrinos, $\left\langle F_{\nu_i}^{-1} \right\rangle$ is the mean
inverse flux factor. $f_\mathrm{heat}$ is a free parameter, which we
refer to as the \emph{heating factor}.  We estimate $\langle
\epsilon^2_{\nu_i} \rangle$ based on the temperature at the
neutrinosphere (see \citealt{oconnor:10}) and we parameterize
$\left\langle F_{\nu_i}^{-1} \right\rangle$ as a function of optical
depth $\tau_{\nu_i}$ based on the angle-dependent radiation fields of
the neutrino transport calculations of \cite{ott:08} and set
$\left\langle F_{\nu_i}^{-1} \right\rangle = 4.275
\tau_{\nu_i}+1.15$. { Note that in this parameterization,
  the flux factor levels off at 1.15 at low optical depth in the outer
  postshock region. We choose the latter value instead of 1, because
  the radiation field becomes fully forward peaked only outside the
  shock \citep{ott:08} and because the linear interpolation in
  $\tau_{\nu_i}$ drops off too quickly compared to full
  radiation-hydrodynamics simulations.  Hence the higher floor
  value to compensate \citep{oconnor:10}}. Finally, the factor
  $e^{-2\tau_{\nu_i}}$ in Equation~\ref{eq:heating} is used to
  strongly suppress heating at $\tau_{\nu_i} > 1$. Further details are
  given in \cite{oconnor:10}, \cite{ott:12a}, and \cite{ott:13a}.

\section{Simulated Models}
\label{sec:id}

We carry out a set of eight full 3D simulations, varying heating
factors and numerical resolution as discussed below and summarized in
Table~\ref{tab:results}.

We consider strong, moderate, and weak neutrino heating by dialing in
heating scale factors $f_\mathrm{heat} = \{1.05, 0.95, 0.8\}$,
expressed in the following model names:
$s27\mathrm{MR}f_\mathrm{heat}1.05$ (strong heating),
$s27\mathrm{MR}f_\mathrm{heat}0.95$ (moderate heating), and
$s27\mathrm{MR}f_\mathrm{heat}0.8$ (weak heating). All of these models
have medium numerical resolution, as denoted by ``MR'' in their model
names. This is our baseline resolution discussed in \S\ref{sec:methods}.

To test for dependence on numerical resolution in the scenario of
strong neutrino heating, we take model
$s27\mathrm{MR}f_\mathrm{heat}1.05$ as the reference model and re-run
it with four additional resolutions. We characterize these additional
simulations by their linear computational cell width on the refinement
level that covers the postshock gain layer and contains the shock.
Together with our baseline MR (``medium resolution'') simulation, we
have: ULR (ultra-low resolution, $dx_\mathrm{shock} =
3.784\,\mathrm{km}$), LR (low resolution, $dx_\mathrm{shock} =
1.892\,\mathrm{km}$), MR (medium resolution, $dx_\mathrm{shock} =
1.416\,\mathrm{km}$), IR (intermediate resolution, $dx_\mathrm{shock}
= 1.240\,\mathrm{km}$), HR (high resolution, $dx_\mathrm{shock} =
1.064\,\mathrm{km}$).  Note that for the ULR simulation, we have
simply taken the LR AMR grid setup and moved the outer boundary of the
refinement level covering the shock in the LR simulation down into the
cooling layer. In this way, the ULR simulation has the same resolution
as the LR simulation in the protoneutron star, but a factor of two
lower resolution in the postshock gain layer. All other simulations have
systematically changed resolution on all refinement levels.

For testing resolution dependence in the case of weak neutrino
heating, we use model $s27\mathrm{MR}f_\mathrm{heat}0.8$ as the
reference model and add one more simulation, model
$s27\mathrm{LR}f_\mathrm{heat}0.8$, with $\sim 30\,\%$ lower
resolution than baseline.

\section{Results: Dependence on Neutrino Heating}
\label{sec:heating}

\subsection{Overview}
\label{sec:overview}

\begin{figure}[t]
\centering
\includegraphics[width=1.0\columnwidth]{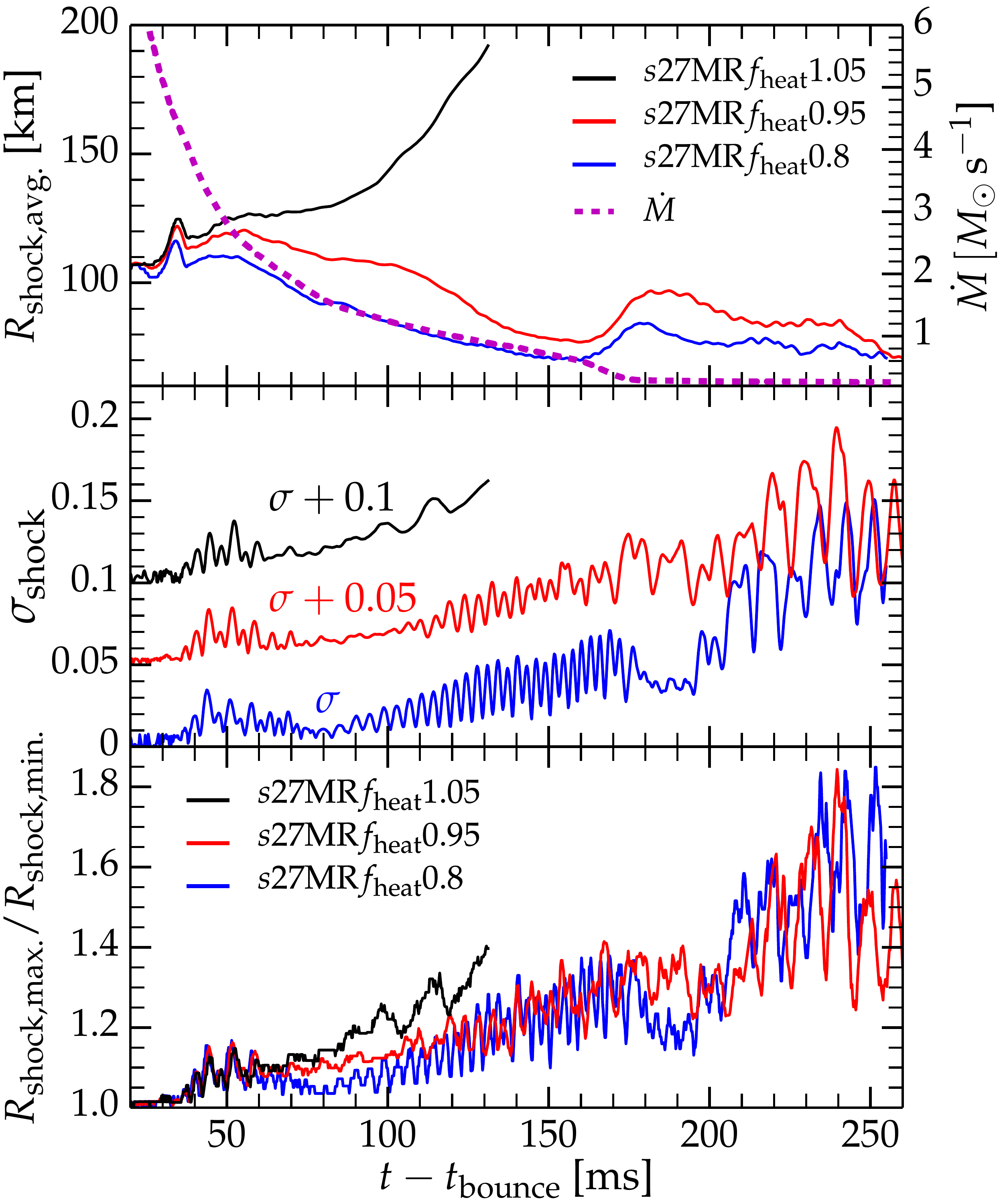}
\caption{{\bf Top panel}: Average shock radius evolution for models
  with strong ($s27\mathrm{MR}f_\mathrm{heat}1.05$), medium
  ($s27\mathrm{MR}f_\mathrm{heat}0.95$), and weak
  ($s27\mathrm{MR}f_\mathrm{heat}0.8$) neutrino heating. Model
  $s27\mathrm{MR}f_\mathrm{heat}1.05$, due to its strong neutrino
  heating, shows the onset of an explosion already
  $\sim$$100\,\mathrm{ms}$ after bounce. The models with moderate and
  weak neutrino heating fail to show signs of explosion, but exhibit a
  transient shock expansion when the accretion rate ($\dot{M}$,
  dashed magenta line) drops at the time the silicon interface accretes
  through the stalled shock. {\bf Center panel}: Normalized root mean
  square deviation $\sigma_\mathrm{shock}$ of the shock radius from
  its angle averaged value. {\bf Bottom panel}: Ratio of the maximum
  shock radius to the minimum shock radius. The with moderate and weak
  neutrino-heating exhibit strong periodic oscillations in the shock
  radius ratio and in $\sigma_\mathrm{shock}$. These variations are
  the tell-tale signs of SASI activity in these models.}
\label{fig:shockradfheat}
\vspace{0.5ex}
\end{figure}

The top panel of Figure~\ref{fig:shockradfheat} shows the time evolution
of the angle-averaged shock radius $R_\mathrm{shock,avg}$ in our three
baseline-resolution simulations $s27\mathrm{MR}f_\mathrm{heat}1.05$,
$s27\mathrm{MR}f_\mathrm{heat}0.95$, and
$s27\mathrm{MR}f_\mathrm{heat}0.8$ with strong, medium, and weak
neutrino heating, respectively. We show only the part of the evolution
tracked in 3D. At early times ($t-t_\mathrm{bounce} \lesssim
50-60\,\mathrm{ms}$) the shock undergoes some transient oscillations as
it relaxes on the 3D grid, which is reflected in
$R_\mathrm{shock,avg}$ of all models. 

The average shock radius in model $s27\mathrm{MR}f_\mathrm{heat}1.05$
grows secularly from $\sim 105\,\mathrm{km}$ to $125\,\mathrm{km}$ in
the first $90 \, \mathrm{ms}$ of postbounce evolution. Subsequently,
the shock expansion accelerates and $R_\mathrm{shock,avg}$ reaches
$\sim 195\,\mathrm{km}$ by $\sim 130\,\mathrm{ms}$ after bounce, which
is when we stop following this model's evolution. The maximum shock
radius at this time is $\sim 220\,\mathrm{km}$. The expansion has
become dynamical and is most likely transitioning to explosion. In
contrast, models $s27\mathrm{MR}f_\mathrm{heat}0.95$ and
$s27\mathrm{MR}f_\mathrm{heat}0.8$ do not show any sign of explosion
within the simulated time. The average shock radius in these models
decreases gradually until $\sim 160\,\mathrm{ms}$ after bounce,
reaching $\sim 97\,\mathrm{km}$ and $\sim 75\,\mathrm{km}$,
respectively. At this point, the silicon shell reaches the shock
front, leading to a sudden decrease of the accretion rate (cf.~the
accretion rate shown in the top panel of
Figure~\ref{fig:shockradfheat}) and thus of the ram pressure
experienced by the shock. This leads to a transient expansion of the
shock by $\sim 10\,\mathrm{km}$ within $\sim 15\,\mathrm{ms}$, after
which it starts retreating again in both models and continues to do so
until the end of our simulations. Due to the weaker heating in model
$s27\mathrm{MR}f_\mathrm{heat}0.8$, $R_\mathrm{shock,avg}$ remains
always smaller than in model $s27\mathrm{MR}f_\mathrm{heat}0.95$, but
has the same qualitative evolution.

\begin{figure}[t]
\centering
\includegraphics[width=1\columnwidth]{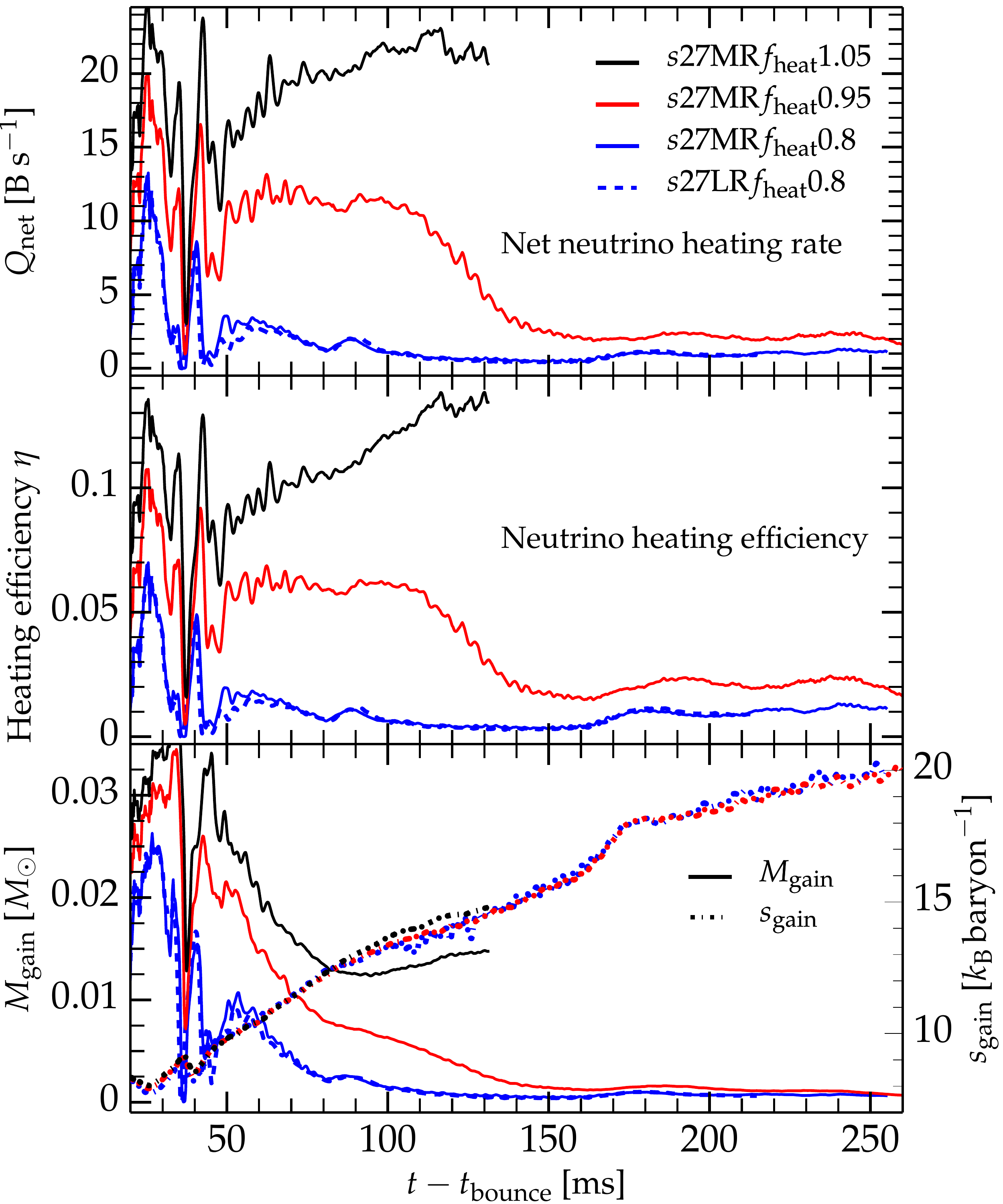}
\caption{Integral quantities characterizing the strength of neutrino
  heating in models $s27\mathrm{MR}f_\mathrm{heat}1.05$,
  $s27\mathrm{MR}f_\mathrm{heat}0.95$, and
  $s27\mathrm{MR}f_\mathrm{heat}0.8$ with three different
  $f_\mathrm{heat}$. { We also show results for the
    low-resolution model $s27\mathrm{LR}f_\mathrm{heat}0.8$}. {\bf Top
    panel}: { integral} net neutrino heating rate
  $Q^\mathrm{net}$ (heating minus cooling) in
  $\mathrm{B}\,\mathrm{s}^{-1}$, where $1\,\mathrm{B}ethe =
  10^{51}\,\mathrm{erg}$.  {\bf Center panel}: heating efficiency
  $\eta$ defined as $Q_\mathrm{net}$ divided by the sum of the $\nu_e$
  and $\bar{\nu}_e$ luminosities emerging from below the inner
  boundary of the gain region. {\bf Bottom panel}: Mass
  $M_\mathrm{gain}$ (left ordinate) and average specific entropy
  $s_\mathrm{gain}$ (right ordinate; { not shown for model
  $s27\mathrm{LR}f_\mathrm{heat}0.8$}) in the gain
  region. $Q_\mathrm{net}$, $\eta$, $M_\mathrm{gain}$ all increase
  with increasing local heating factor $f_\mathrm{heat}$.}
\label{fig:heat}
\vspace{0.5ex}
\end{figure}

The shock radius evolution shown in Figure~\ref{fig:shockradfheat} can
be directly linked to the strength of neutrino heating. We quantify
the latter by a set of metrics shown in Figure~\ref{fig:heat}: the
{ integral} net neutrino heating rate $Q_\mathrm{net}$, the
heating efficiency ($\eta = Q_\mathrm{net}\,(L_{\nu_e} +
L_{\bar{\nu}_e})^{-1}$, where $L_{\nu_e}$ and $L_{\bar{\nu}_e}$ are
the electron neutrino and anti-electron neutrino luminosities incident
from below the inner boundary of the gain region)\footnote{We note in
  passing that the heating efficiencies shown in our previous
  \cite{ott:13a} study were incorrectly underestimated by about a
  factor of $1.7$, because we normalized by the total neutrino
  luminosity and not just by $L_{\nu_e} + L_{\bar{\nu}_e}$.}  and the
mass $M_\mathrm{gain}$ in the gain region. The oscillations in these
quantities at early times are a combined artifact of the
leakage/heating scheme, { which is unreliable in highly
  dynamical situations}, and of the shock settling on the 3D
grid. { Note that the outgoing luminosities are only mildly
  affected and the main effect comes from variations in the mean
  neutrino energies, cf.\ \cite{ott:13a}. Similar features are present
  in the leakage simulations of \cite{couch:14a}}.  As expected, the
larger $f_\mathrm{heat}$ (see Equation~\ref{eq:heating}), the larger
{ the integral} net heating, heating efficiency, and the
mass in the gain region. Note, however, how strong this relationship
is: An increase of $f_\mathrm{heat}$ from $0.95$ to $1.05$
($\sim$10.5\%) results in approximately twice as high
$Q_\mathrm{net}$, $\eta$, and $M_\mathrm{gain}$ around
$50-100\,\mathrm{ms}$ after bounce.  This is a consequence of the fact
that more intense neutrino heating extends the region of net
absorption to smaller radii. It also increases the thermal pressure
and the vigor of turblence (and thus the effective turbulent ram
pressure; \citealt{couch:15a}) throughout this region. This, in turn,
pushes the shock out, further increasing the volume of the gain region
and leading to more net neutrino energy absorption. This nonlinear
feedback shows just how extremely sensitive core-collapse supernovae
near the critical line between explosion and no explosion are to the
details of neutrino transport and neutrino--matter coupling.

The general trends in neutrino heating with $f_\mathrm{heat}$
described in the above hold throughout the postbounce phase. However,
as the shock radii in models $s27\mathrm{MR}f_\mathrm{heat}0.95$ and
$s27\mathrm{MR}f_\mathrm{heat}0.8$ recede and their gain regions
shrink, their values of their neutrino heating variables shown in
Figure~\ref{fig:heat} approach each other. The sudden reduction
of the ram pressure at the silicon interface, which has a significant
effect on the shock radius (Figure~\ref{fig:shockradfheat}), is 
barely noticable in the neutrino heating.

We plot the average mass-weighted specific entropy in the gain region
($s_\mathrm{gain}$) on the right ordinate of the bottom panel of
Figure~\ref{fig:heat}. In agreement with what was found in previous work
(e.g., \citealt{hanke:12,dolence:13,couch:13b,ott:13a}),
$s_\mathrm{gain}$ is largely independent of the shock radius and the
strength of neutrino heating in the postbounce preexplosion phase
simulated here. We attribute this to two competing effects that affect
the averaged quantity $s_\mathrm{gain}$: While  strong neutrino
heating (larger $f_\mathrm{heat}$ in our simulations) leads to locally
higher specific entropy in the region of strongest heating, this is
compensated by the overall larger volume (and mass) of the gain
region, which includes material of lower specific entropy that
contributes to the average. 

After considering the above range of indicative angle-averaged and/or
volume-averaged quantities, it is now useful to study deviations from averaged
dynamics. The center and bottom panel of Figure~\ref{fig:shockradfheat} depict
the normalized root mean square angular deviation
{
of the shock radius from its
mean ($R_\mathrm{shock,avg.}$) , $\sigma_\mathrm{shock}$ defined as
\begin{equation}
 \sigma_\mathrm{shock} =
 \frac{1}{R_\mathrm{shock,avg.}}
    \sqrt{\frac{1}{4\pi}\int_{4\pi} \left[
      R_\mathrm{shock} (\theta,\phi)-R_\mathrm{shock,avg}\right]^2
      d\Omega}\,\,,
\end{equation}}
and the ratio of maximum and mimimum shock radius
$R_\mathrm{shock,max.}/R_\mathrm{shock,min.}$, respectively.  Both
diagnostics yield qualitatively similar results, but the latter is more
sensitive to small local
variations. { In the initial settling phase on the 3D grid,
  all three simulations shown in Figure~\ref{fig:shockradfheat}
  exhibit nearly identical shock deviations from sphericity.  These
  are due to moderate-amplitude cubed ($\ell = 4$--symmetric) shock
  oscillations as the models relax from spherical geometry to our 3D
  Cartesian grid.} Subsequently, the shock deviations begin to differ
between models. Model $s27\mathrm{MR}f_\mathrm{heat}1.05$ (strong
neutrino heating) shows more or less steadily increasing asphericity
as its shock gradually expands and develops large-scale deviations
{ in the maximum shock radius $R_\mathrm{shock,max.}$ from
  $R_\mathrm{shock,avg.}$ driven by expanding localized high-entropy
  bubbles. However, the overall asymmetry, expressed by
  $\sigma_\mathrm{shock}$ in Figure~\ref{fig:shockradfheat} is
  relatively small, due to the strong neutrino heating that
leads to a rather global shock expansion.}

The shock asphericity in models $s27\mathrm{MR}f_\mathrm{heat}0.95$
and $s27\mathrm{MR}f_\mathrm{heat}0.8$ exhibits strong oscillations
with clear (if temporally varying) periodicity -- a tell-tale sign of
active SASI. In model $s27\mathrm{MR}f_\mathrm{heat}0.95$ (moderate
neutrino heating), the oscillations set in around $\sim
105\,\mathrm{ms}$ after bounce while in model
$s27\mathrm{MR}f_\mathrm{heat}0.8$ (weak neutrino heating), they are
already present at $\sim 80\,\mathrm{ms}$. In both models, the period
of the oscillations changes when the silicon interface reaches the
shock front around $160\,\mathrm{ms}$ after bounce. We will analyze
SASI in these models in more detail in \S\ref{sec:sasi}.

\begin{figure}[t]
\centering
\includegraphics[width=1\columnwidth]{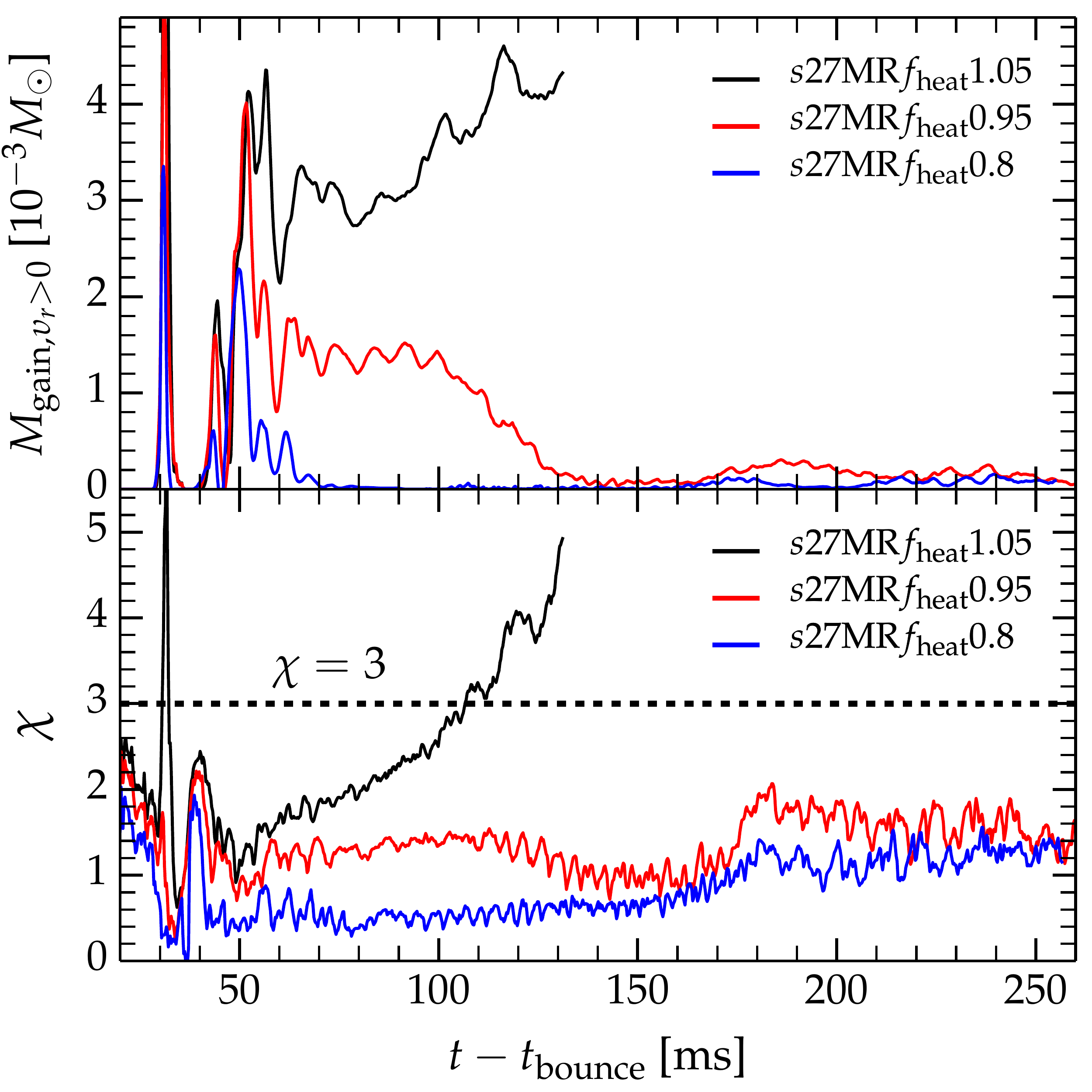}
\caption{{\bf Top panel}: The mass $M_{\mathrm{gain},\upsilon>0}$ in
  the gain region with positive radial velocity (``buoyant mass'') in
  models $s27\mathrm{MR}f_\mathrm{heat}1.05$ (strong neutrino
  heating), $s27\mathrm{MR}f_\mathrm{heat}0.95$ (moderate neutrino
  heating), and $s27\mathrm{MR}f_\mathrm{heat}0.8$ (weak neutrino
  heating). {\bf Bottom panel}: The Foglizzo $\chi$ parameter
  (cf.\ Equation~\ref{eq:chi}) as a function of postbounce time for
  the three models. The horizontal line at $\chi = 3$ marks the
  point where convection is expected to develop in the gain region.
  We calculate $\chi$ on the basis of
  angle-averaged, but not time-averaged quantities.}
\label{fig:chi}
\vspace{0.5ex}
\end{figure}

Whenever our simulations experience strong neutrino heating,
i.e.\ $\eta \gtrsim 0.05$ and $Q_\mathrm{heat} \gtrsim
10^{52}\,\mathrm{erg\,s^{-1}}$, we find neutrino-driven convection in
the postshock region. This is quantified by the top panel of
Figure~\ref{fig:chi}, which shows the buoyant mass in the gain region,
$M_{\mathrm{gain},\upsilon_r>0}$, which we define as the mass of
material with \emph{positive} radial
velocity. $M_{\mathrm{gain},\upsilon_r>0}$ correlates strongly with
$\eta$ and $Q_\mathrm{net}$. Phases of strong neutrino heating
(cf.~Figure~\ref{fig:heat}) correspond to strong neutrino-driven
convection. Model $s27\mathrm{MR}f_\mathrm{heat}1.05$ undergoes
convection throughout its postbounce evolution, while model
$s27\mathrm{MR}f_\mathrm{heat}0.95$ exhibits strong neutrino-driven
convection only until $\sim 110\,\mathrm{ms}$ after bounce. Convective
activity is clearly visible in the 2D ($x$-$z$ plane) entropy
colormaps of models $s27\mathrm{MR}f_\mathrm{heat}1.05$ and
$s27\mathrm{MR}f_\mathrm{heat}0.95$ at various postbounce times in
Figure~\ref{fig:entropy_slices}. We will further analyze
neutrino-driven convection in our models in \S\ref{sec:convection}.

\subsection{SASI}
\label{sec:sasi}

There are three defining characteristics of SASI: (1) 
low-$(\ell,m)$ oscillations of the shock front \cite[e.g.,][]{iwakami:08},
(2) exponential growth of the (spherical-harmonics) mode amplitudes in
the linear phase \cite[e.g.,][]{blondin:03}, and (3) saturation of the
amplitudes once they reach the nonlinear phase
\cite[e.g.,][]{guilet:10}. In order to identify these features in our
simulations, we decompose the shock front
$R_\mathrm{shock}(\theta,\phi)$ into spherical harmonics:
\begin{equation}
\label{eq:alm}
  a_{\ell m} = \frac{(-1)^{|m|}}{\sqrt{4\pi(2\ell+1)}} \int_{4\pi} R_\mathrm{shock}
  (\theta,\phi) Y^m_\ell (\theta,\phi) d\Omega \ ,
\end{equation}
where $Y^m_\ell$ are the standard real spherical harmonics (e.g.,
\citealt{boyd:01}). We employ the normalization convention used in
\cite{burrows:12}, in which $a_{00}$ corresponds to the average shock
radius, while $a_{11}$, $a_{10}$, and $a_{1-1}$ correspond to the
average $x$, $z$, and $y$ Cartesian coordinates of the shock front,
respectively.

\begin{figure*}[t]
\centering
\includegraphics[width=1\textwidth]{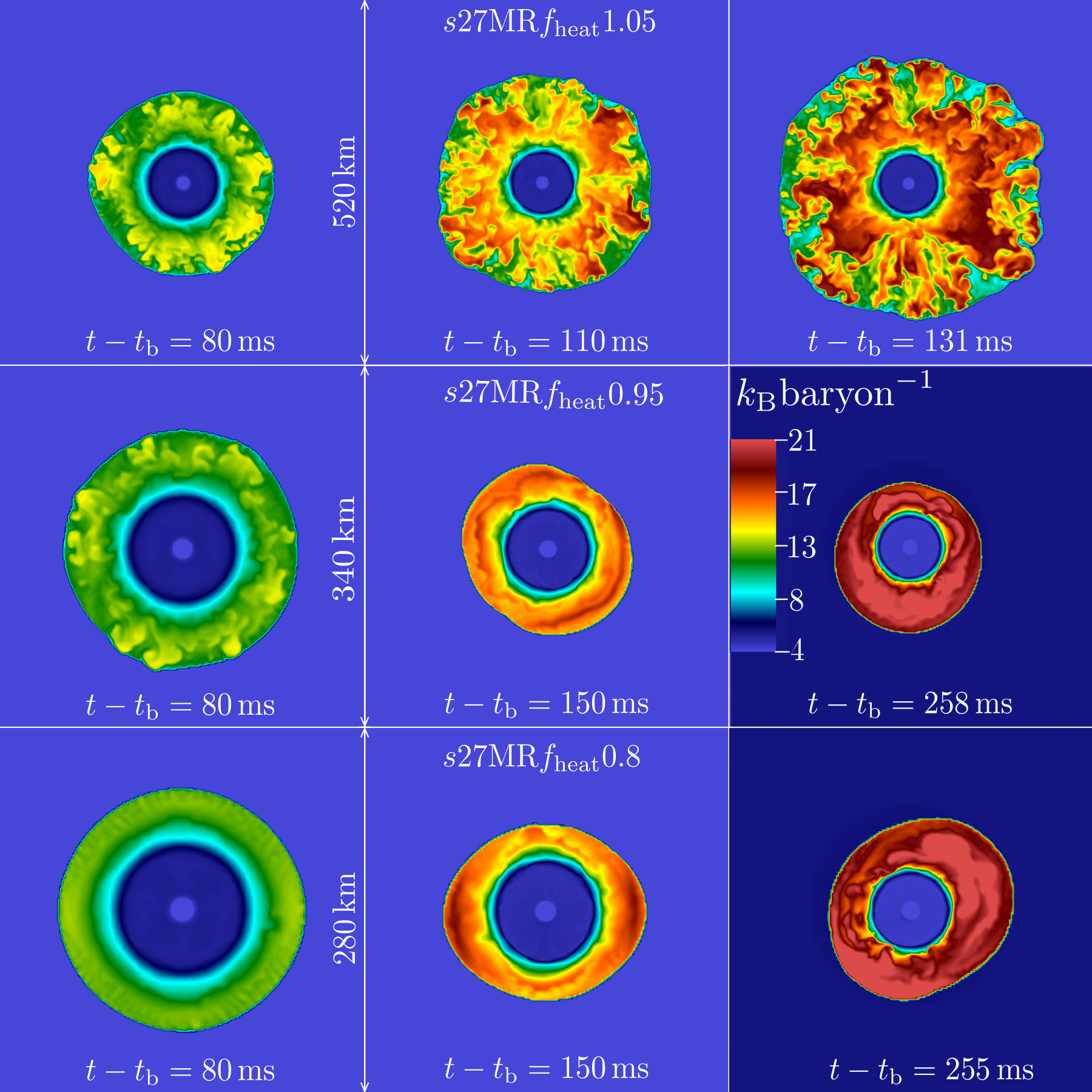}
\caption{Colormaps of specific entropy in the $x$-$z$ plane in models
  $s27\mathrm{MR}f_\mathrm{heat}1.05$ (strong neutrino heating; top
  row), $s27\mathrm{MR}f_\mathrm{heat}0.95$ (moderate neutrino
  heating; center row), and $s27\mathrm{MR}f_\mathrm{heat}0.8$ (weak
  neutrino heating; bottom row) at a range of postbounce times. Note
  that the scale of the region shown is different for each
  model. Model $s27\mathrm{MR}f_\mathrm{heat}1.05$ is dominated by
  neutrino-driven convection.  Model
  $s27\mathrm{MR}f_\mathrm{heat}0.95$ shows neutrino-driven convection
  at early times, but subsequently shows signs of coherent shock
  dynamics typical for SASI. Model
  $s27\mathrm{MR}f_\mathrm{heat}0.8$ never develops significant
  neutrino-driven convection and becomes dominated by SASI.}
\label{fig:entropy_slices}
\end{figure*}

\begin{figure*}[t]
\centering
\includegraphics[width=0.47\textwidth]{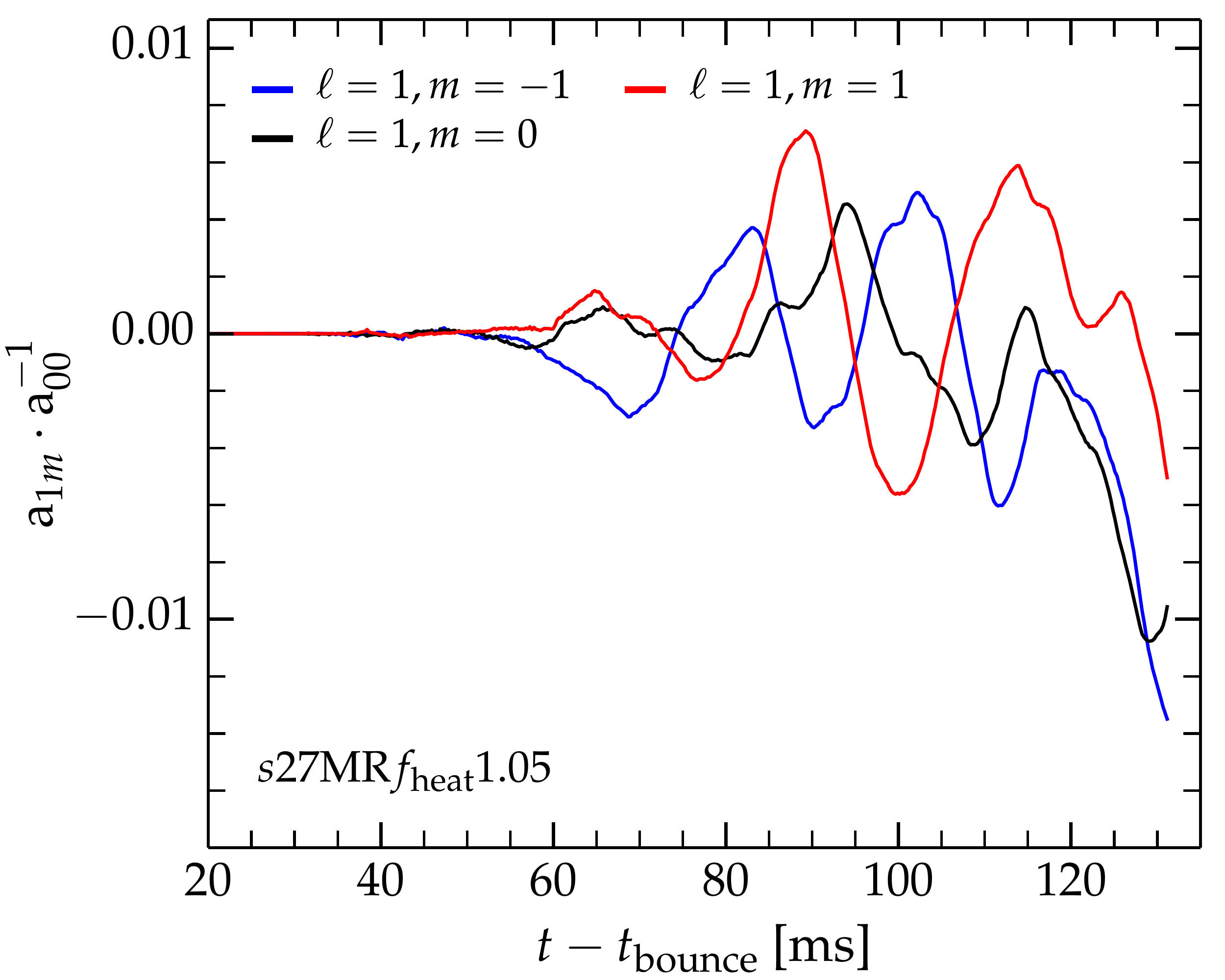}
\includegraphics[width=0.47\textwidth]{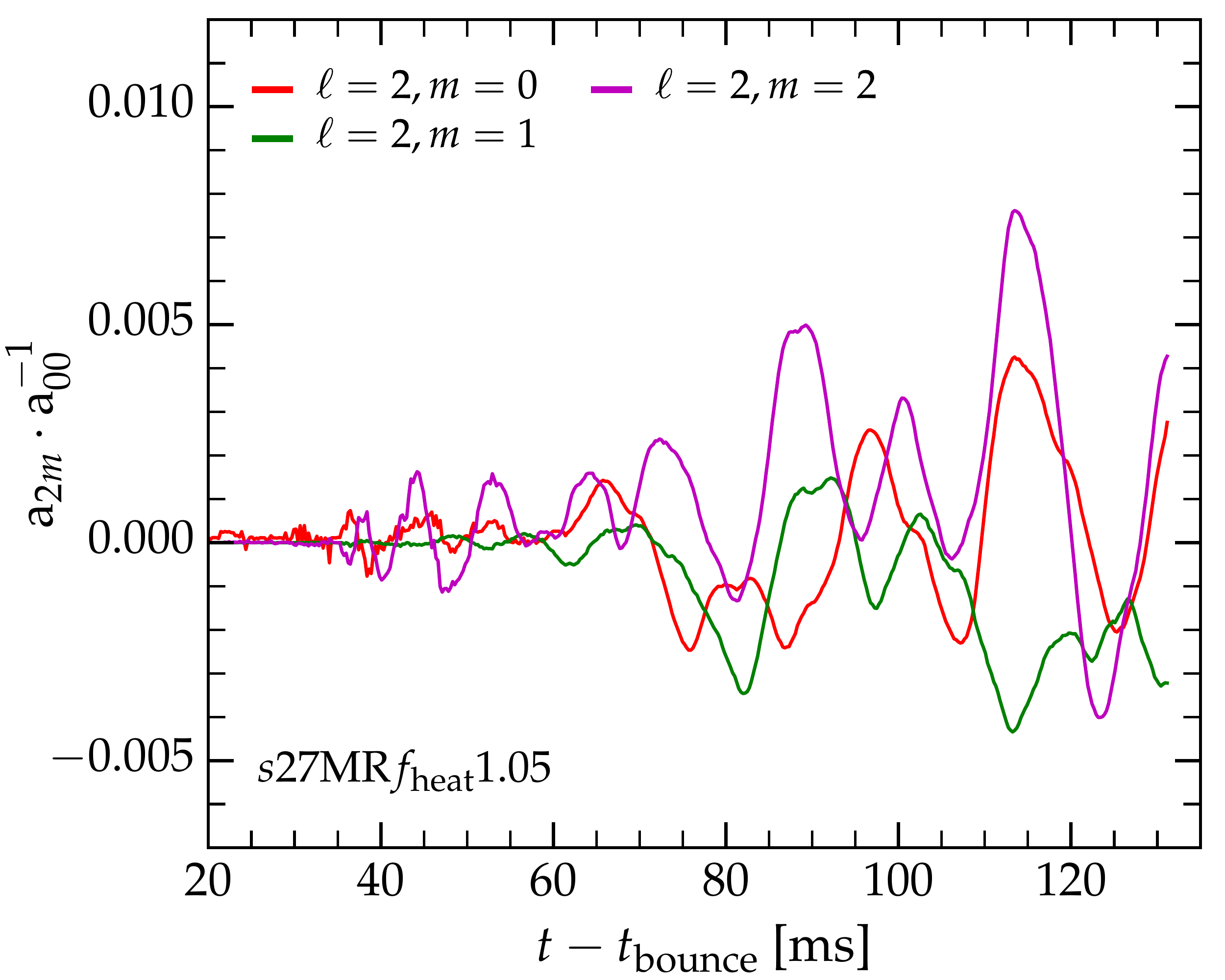}
\includegraphics[width=0.47\textwidth]{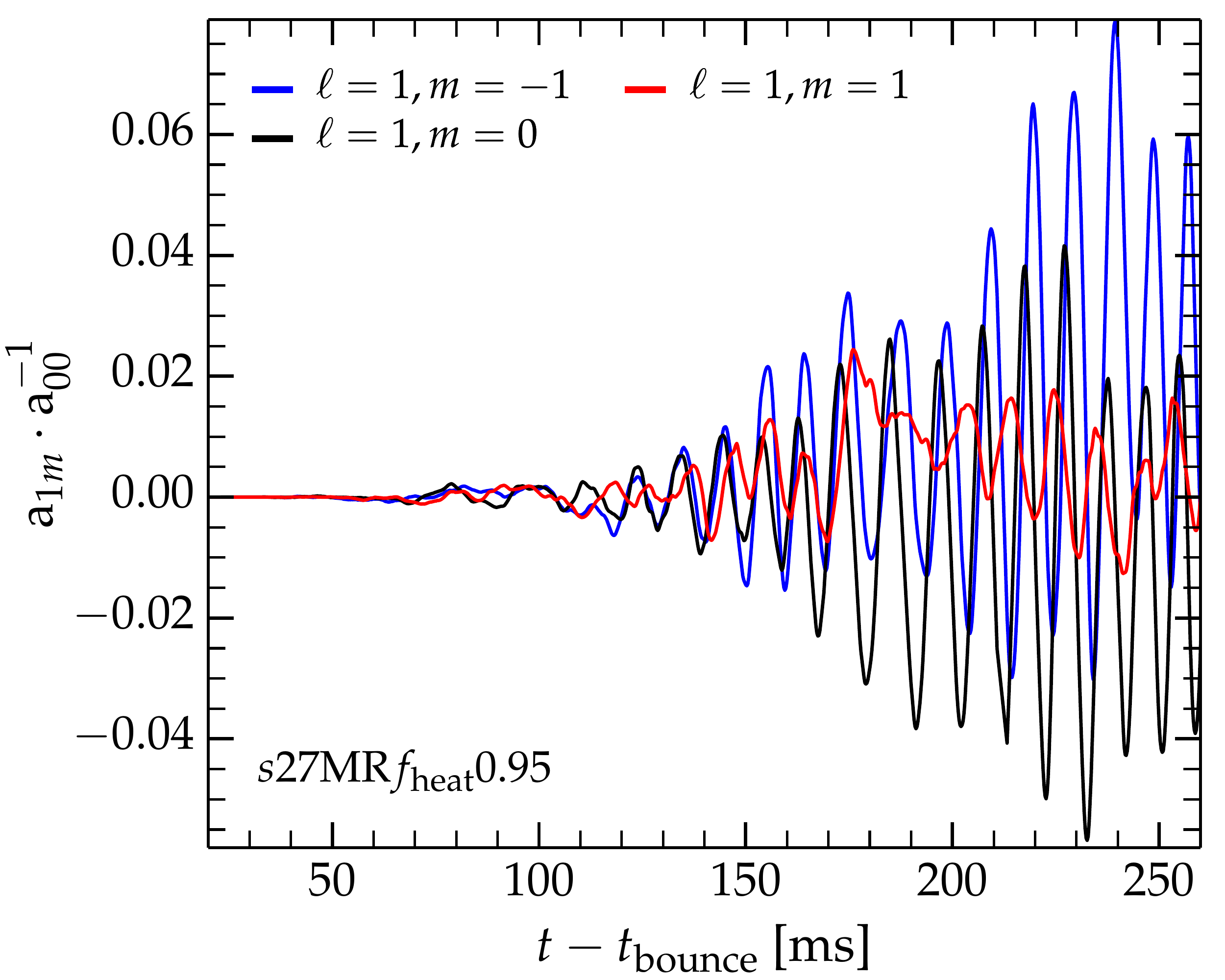}
\includegraphics[width=0.47\textwidth]{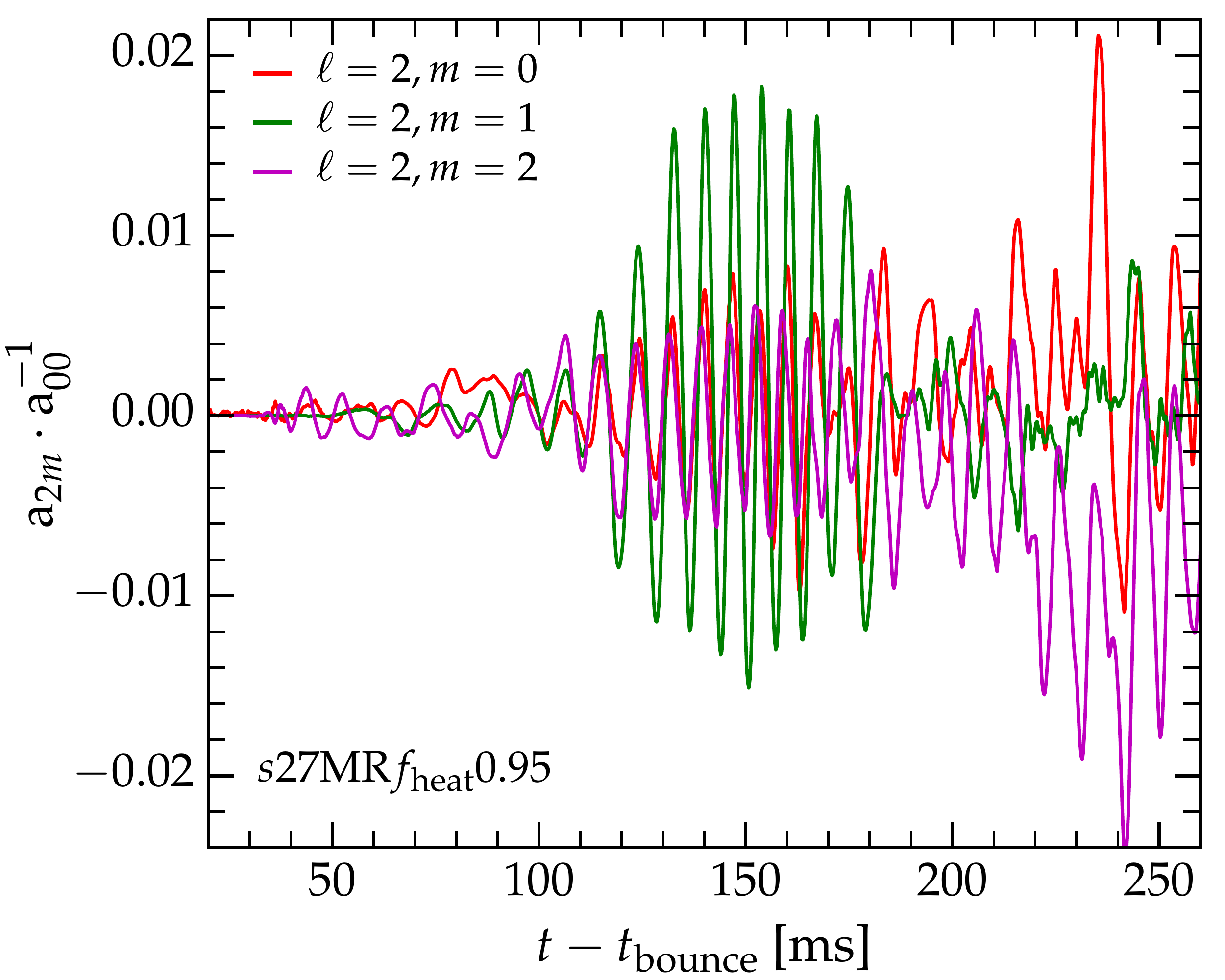}
\includegraphics[width=0.47\textwidth]{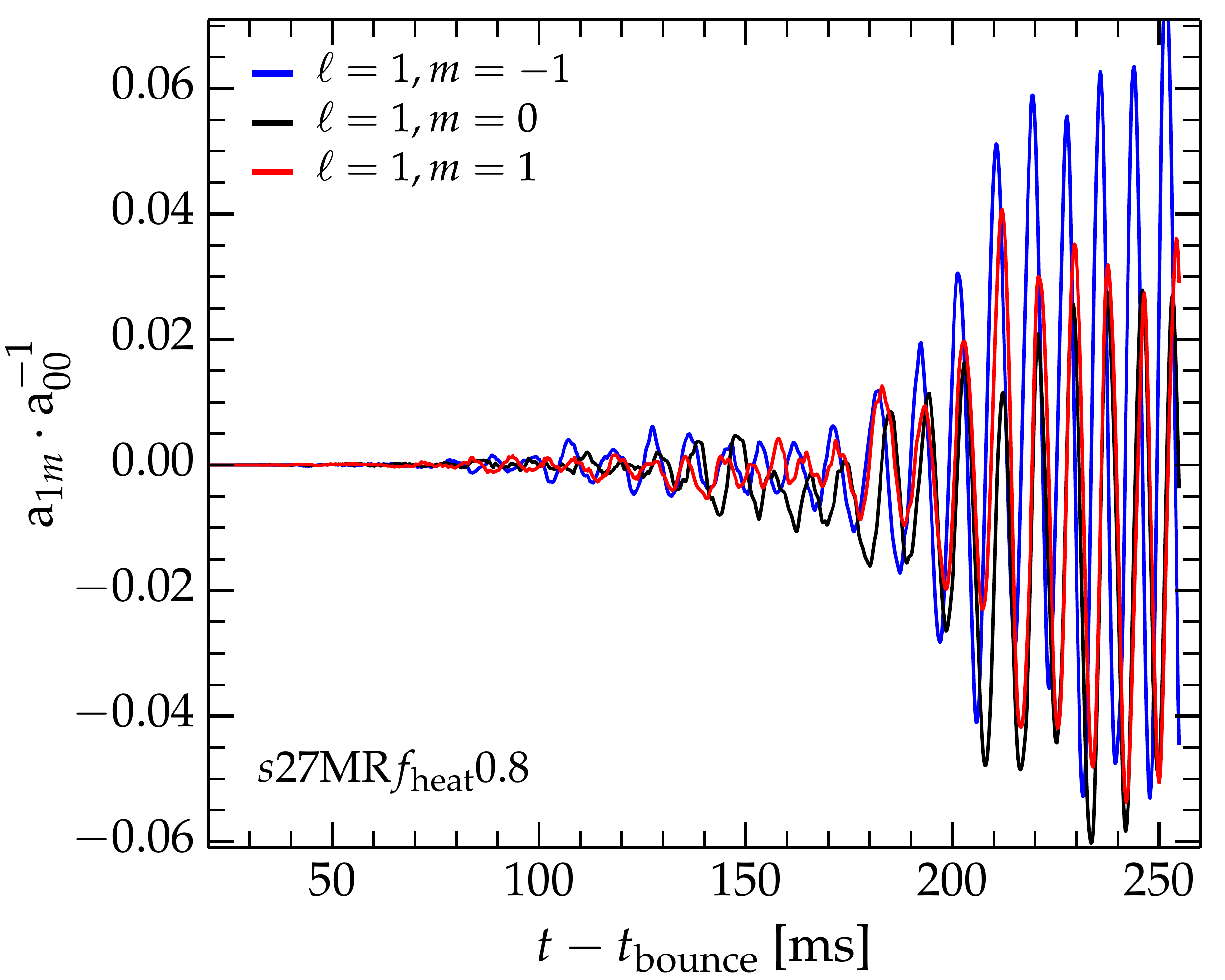}
\includegraphics[width=0.47\textwidth]{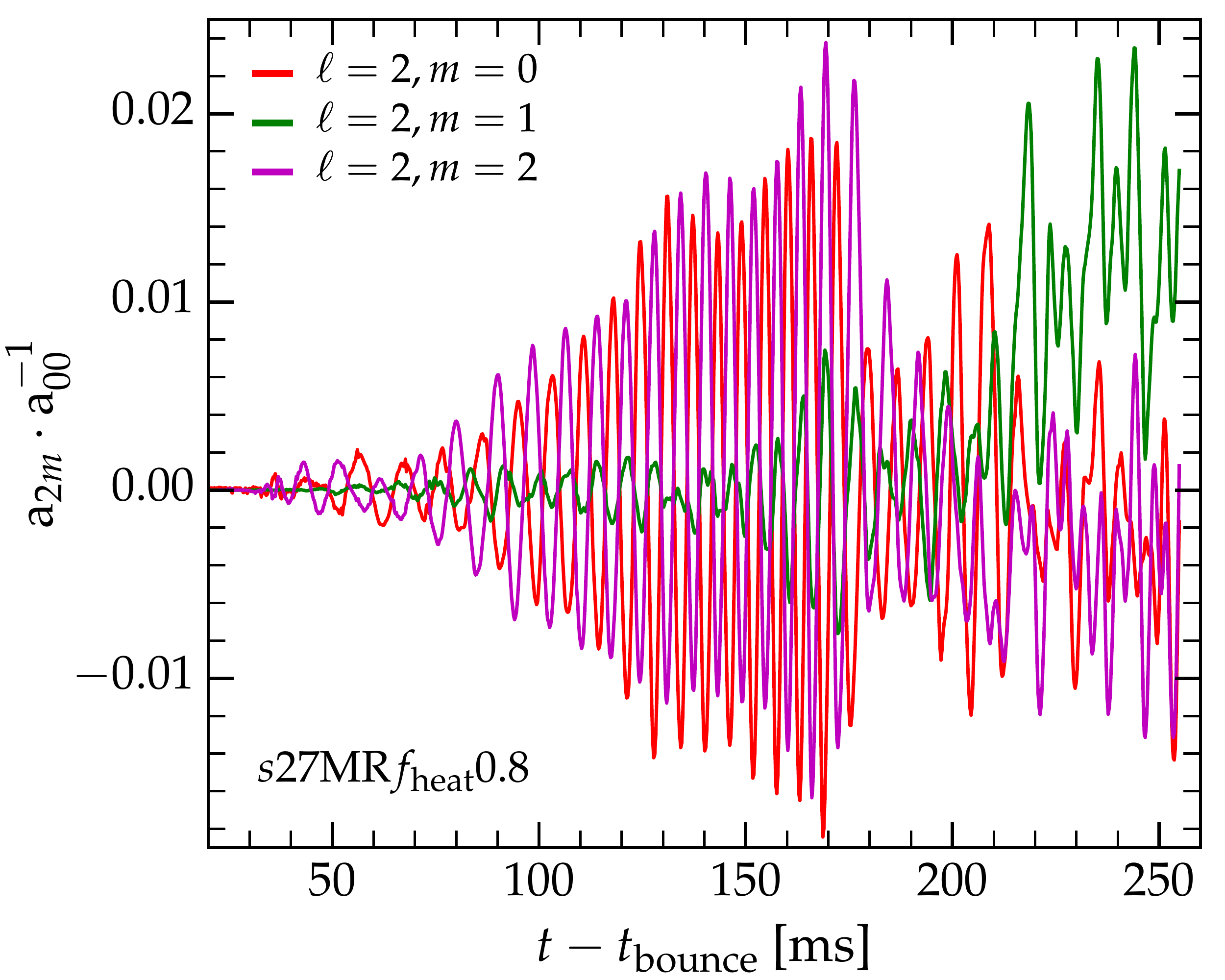}
\caption{Normalized mode amplitudes $a_{\ell m} \cdot a_{00}^{-1}$ of
  the shock front as a function of time for $\ell=1$ (left panels) and
  $\ell=2$ modes (right panels). { Only modes with $m\ge 0$
    are shown; modes with negative $m$ behave very similarly}. We
  show amplitudes for models $s27\mathrm{MR}f_\mathrm{heat}1.05$
  (strong neutrino heating, top row),
  $s27\mathrm{MR}f_\mathrm{heat}0.95$ (moderate neutrino heating,
  center row), and $s27\mathrm{MR}f_\mathrm{heat}0.8$ (weak neutrino
  heating, bottom row).  Note that the range in postbounce time shown
  in the top row for the exploding model
  $s27\mathrm{MR}f_\mathrm{heat}1.05$ is different from the postbounce
  time covered for the two non-exploding models that develop strong
  long-lasting SASI oscillations.}
\label{fig:a1s}
\end{figure*}

\begin{figure}[t]
\centering
\includegraphics[width=0.47\textwidth]{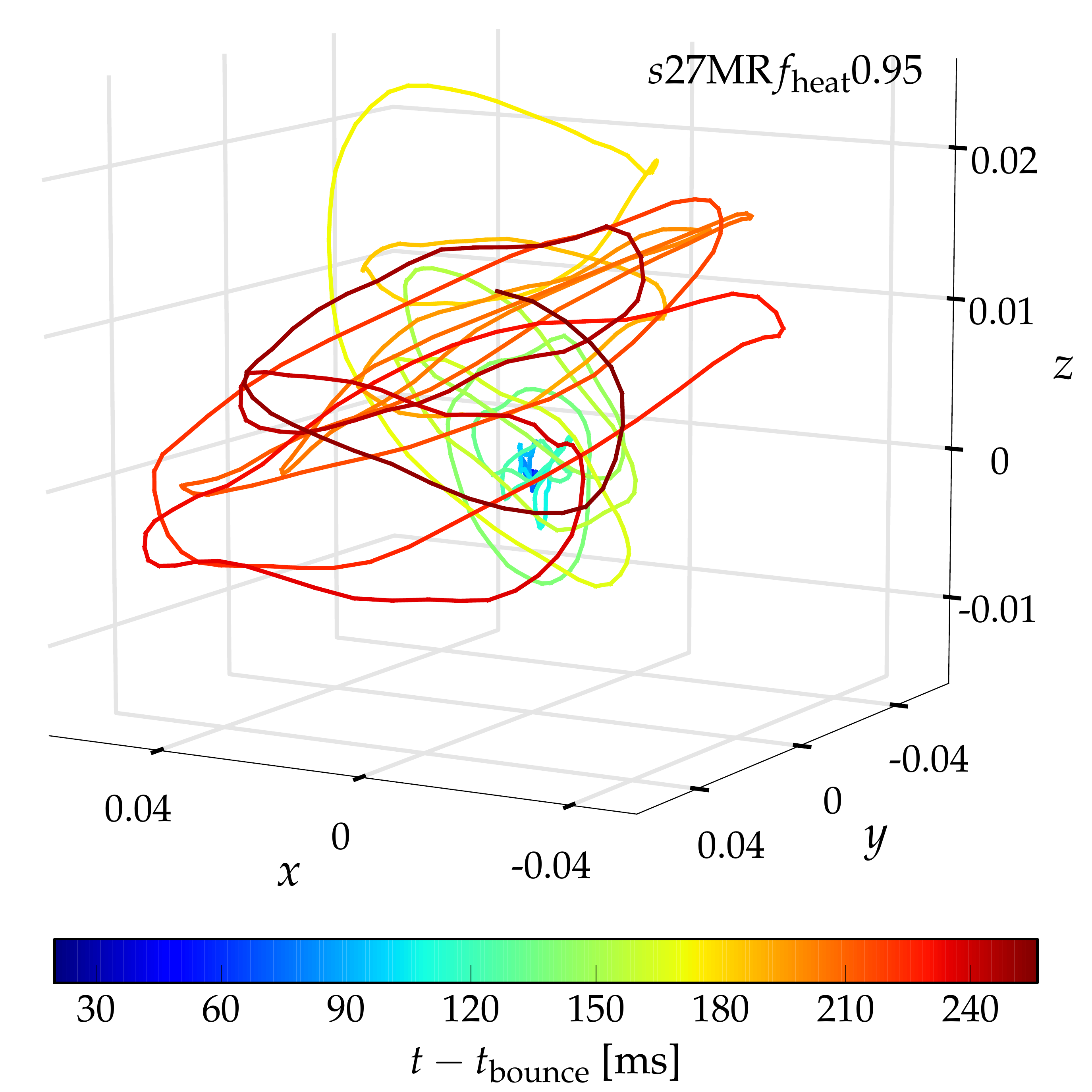}
\includegraphics[width=0.47\textwidth]{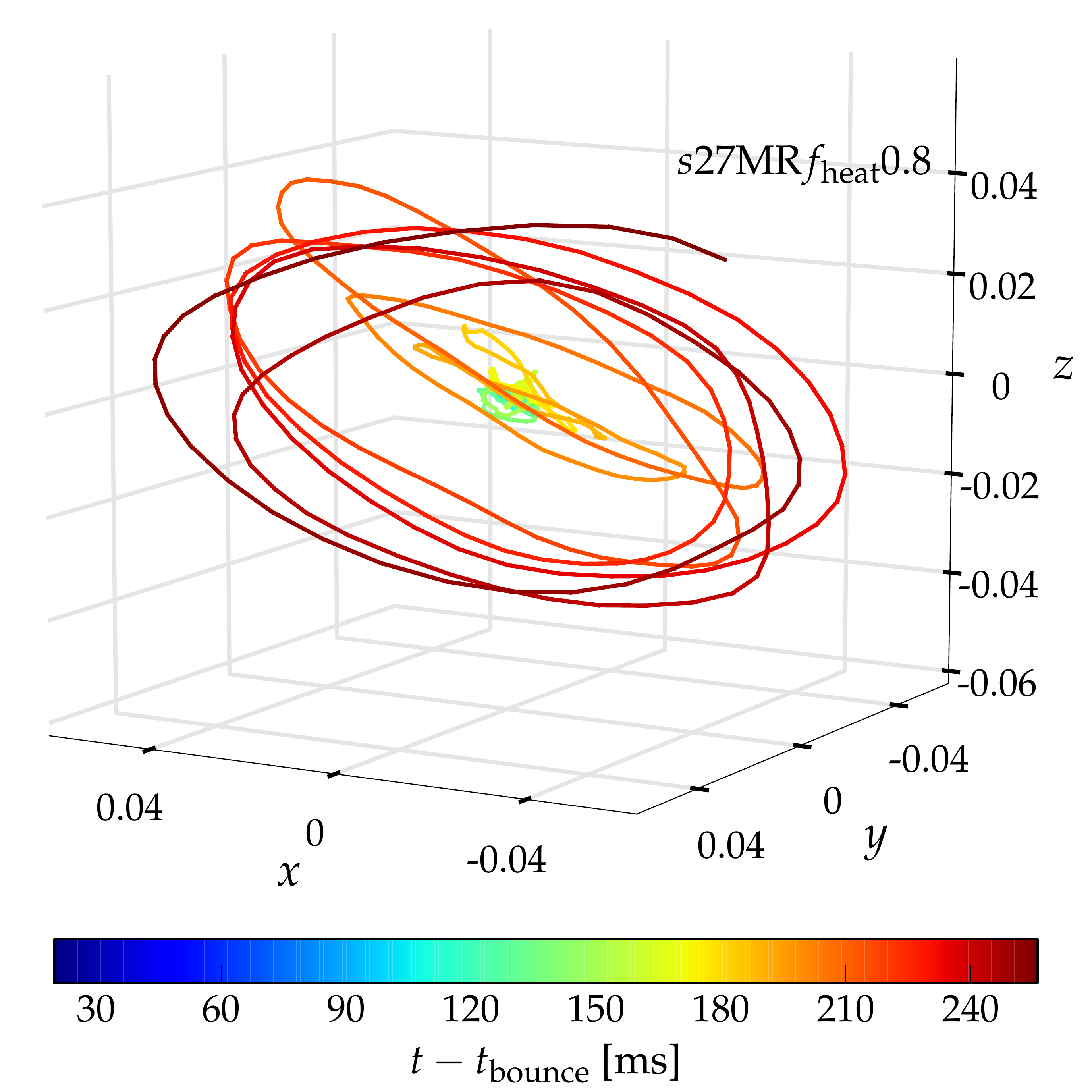}
\caption{Evolution of the normalized $\ell=1$ mode vector ${\bf
    a}_1/a_{00}$ for models $s27\mathrm{MR}f_\mathrm{heat}0.95$ (top
  panel) and $s27\mathrm{MR}f_\mathrm{heat}0.8$ (bottom panel). The
  viewing directions on each panel are chosen to be perpendicular to
  the plane of the spiral SASI motion when it reaches the largest
  amplitude. The color of the graphs demark time. Both models exhibit
  spiral SASI oscillations, but they are strongest in the model
  with weakest neutrino heating, $s27\mathrm{MR}f_\mathrm{heat}0.8$.}
\label{fig:spiral}
\end{figure}

Figure~\ref{fig:a1s} depicts the normalized mode amplitudes $a_{lm}
\cdot a^{-1}_{00}$ for $\ell = 1$ (left panels) and $\ell =2$ (right
panels) for the three previously introduced models with strong,
medium, and weak neutrino heating. The mode amplitudes grow gradually
with time in magnitude and in this reflect the evolution of the angular
deviation $\sigma$ of the shock radius in Figure~\ref{fig:shockradfheat}.
The relative asphericity of the shock is increasing with time in all
models.

In model $s27\mathrm{MR}f_\mathrm{heat}1.05$, the $\ell=1$ mode
amplitudes grow quickly at $50-80\,\mathrm{ms}$ after bounce, exhibit
$\sim$three periodic modulations with a period of
$\sim$$20\,\mathrm{ms}$ at nearly saturated magnitude, and then begin
to increase to larger values. The $\ell = 2$ modes start growing
earlier, but show less clear periodicity. The evolution of the
$\ell=1$ and $\ell=2$ modes suggest that some form of SASI is present
in model $s27\mathrm{MR}f_\mathrm{heat}1.05$, but a look at the top
row of specific entropy slices in Figure~\ref{fig:entropy_slices}
reveals that violent neutrino-driven convection is active, fully
developed, and { driving the local deviation from spherical
  symmetry at late times in this model.}

Models $s27\mathrm{MR}f_\mathrm{heat}0.95$ and
$s27\mathrm{MR}f_\mathrm{heat}0.8$ exhibit strong SASI oscillations in
their $\ell = 1$ and $\ell = 2$ modes that last for many cycles.  The
$\ell=2$ modes actually start growing first and the initial growth of
all modes exhibits exponential character until they reach saturation
on a timescale of $\sim$$50\,\mathrm{ms}$. The oscillation period is
$\sim$$10\,\mathrm{ms}$ and $\sim$$6\,\mathrm{ms}$ for $\ell = 1$ and
$\ell =2$, respectively.

From the $x-z$ specific entropy slices shown in
Figure~\ref{fig:entropy_slices} one notes that at
$\sim$$80\,\mathrm{ms}$ after bounce, there are signs of convection in
model $s27\mathrm{MR}f_\mathrm{heat}0.95$, but no convective plumes
are visible in model $s27\mathrm{MR}f_\mathrm{heat}0.8$ with the
weakest neutrino heating. The entropy slices at $150\,\mathrm{ms}$
after bounce show large shock deformations with $\ell = 2$ symmetry
and no clearly convective features in either model. Interestingly, in
both models, the $\ell = 2$ modes get damped and overtaken by $\ell =
1$ oscillations at $\sim$$170\,\mathrm{ms}$ after bounce
(cf.\ Figure~\ref{fig:a1s}), when the silicon interface reaches the shock,
leading to transient shock expansion. Accordingly, the late-time
entropy slices of these models in Figure~\ref{fig:entropy_slices}
exhibit predominantly $\ell=1$ asymmetry.

Although the accretion of the silicon interface damps the initially dominant
$\ell=2$ modes significantly (cf.~Figure~\ref{fig:a1s}), they again, but
only episodically, reach large amplitudes at later postbounce
times. This is uncharacteristic for linear growth of physical models
and may possibly be due to nonlinear interactions with the
then-dominant $\ell=1$ modes.

The $\ell=1$, $m=\{-1,0,1\}$ modes shown in Figure~\ref{fig:a1s} have
different phases with respect to each other. This is suggestive of
``spiral'' SASI oscillations as identified, e.g., by
\cite{blondin:07,fernandez:10,iwakami:14}. We analyze the vector
\begin{equation} 
  {\bf a}_1 = \left(a_{1-1}, a_{10}, a_{11}\right)\,\, , 
\end{equation} 
which gives the direction and magnitude of the $\ell = 1$ shock
deformation with respect to the center of the protoneutron star
\citep{hanke:13}. We visualize the time evolution of ${\bf
  a}_1/a_{00}$ with a line in 3D space in the top and bottom panels of
Figure~\ref{fig:spiral} for models $s27\mathrm{MR}f_\mathrm{heat}0.95$
and $s27\mathrm{MR}f_\mathrm{heat}0.8$, respectively. Each point on
the graph is color coded according to postbounce time
$t-t_\mathrm{bounce}$.  During the early postbounce evolution, $|{\bf
  a}_1|/a_{00}$ is small and does not exhibit any clear rotational
patterns in either of the models. After the silicon interface has
accreted through the shock, the $\ell = 1$ modes reach large
amplitudes. It is then (orange--red colors in Figure~\ref{fig:spiral})
that ${\bf a}_1/a_{00}$ clearly describes several complete spiral
cycles in both models. This confirms the spiral nature of the late
$\ell = 1$ SASI, which is qualitatively very similar to what
\cite{hanke:13} and \cite{couch:14a} found in their 3D simulations of
the same progenitor.

It is interesting to ask why we observe an early growth of the $\ell =
2$ SASI mode in our simulations while $\ell = 1$ is usually identified
to be the most unstable SASI mode.  { We speculate, fueled
  by Foglizzo (\emph{private communication}), that} one possible
explanation may be related to the trend found by \cite{foglizzo:07}
that higher values of $\ell$ are favored when the shock radius is
small.  Just before the accretion of the silicon shell interface, the
average shock radius $R_\mathrm{shock,avg}$ in models
$s27\mathrm{MR}f_\mathrm{heat}0.95$ and
$s27\mathrm{MR}f_\mathrm{heat}0.8$ is as small as $97\,\mathrm{km}$
and $75\,\mathrm{km}$, respectively.  The reduction in ram pressure at
the silicon interface lets the shock jump outward, possibly creating a
situation more favorable for $\ell=1$ oscillations than before. This
might be the reason for the sudden damping of the $\ell=2$ modes and
the development of the $\ell=1$ oscillations.

In the simulations of the same $27$-$M_\odot$ progenitor of
\cite{ott:13a}, \cite{couch:14a}, and \cite{hanke:13}, the $\ell=1$
modes reach large amplitudes before the accretion of the silicon
interface.  It dominates over $\ell = 2$ at least in the early
evolution in \cite{couch:14a} and \cite{ott:13a} (\citealt{hanke:13}
do not provide $\ell = 2$ amplitudes). In these simulations, the
average shock radius is nearly always above 
$100\,\mathrm{km}$ in the early postbounce phase. It drops below this
value early on in our present simulations with weak
($s27\mathrm{MR}f_\mathrm{heat}0.8$) and moderate
($s27\mathrm{MR}f_\mathrm{heat}0.95$) neutrino heating.  Following the
above argument, this may explain why only our simulations exhibit an
initially predominantly $\ell = 2$ SASI.

It is worth mentioning that, to the best of our knowledge, strong
excitation of predominantly $\ell=2$ modes in the 3D case was observed
only in the work of \cite{takiwaki:12}, who studied the 3D postbounce
hydrodynamics in a $11.2$-$M_\odot$ progenitor. However, in their
simulation, this mode undergoes only $2-3$ oscillations during the
simulated time, whereas in our case, we observe $\sim 30$ oscillation
cycles before $\ell = 2$-dominated dynamics ceases.  The $\ell = 2$
modes also reach large amplitudes in the 3D simulations of
\cite{iwakami:08}, \cite{ott:13a}, and \cite{couch:14a}, but their
amplitudes generally do not exceed those of the $\ell = 1$ modes.

\begin{figure*}[t]
\centering
\includegraphics[width=0.82\textwidth]{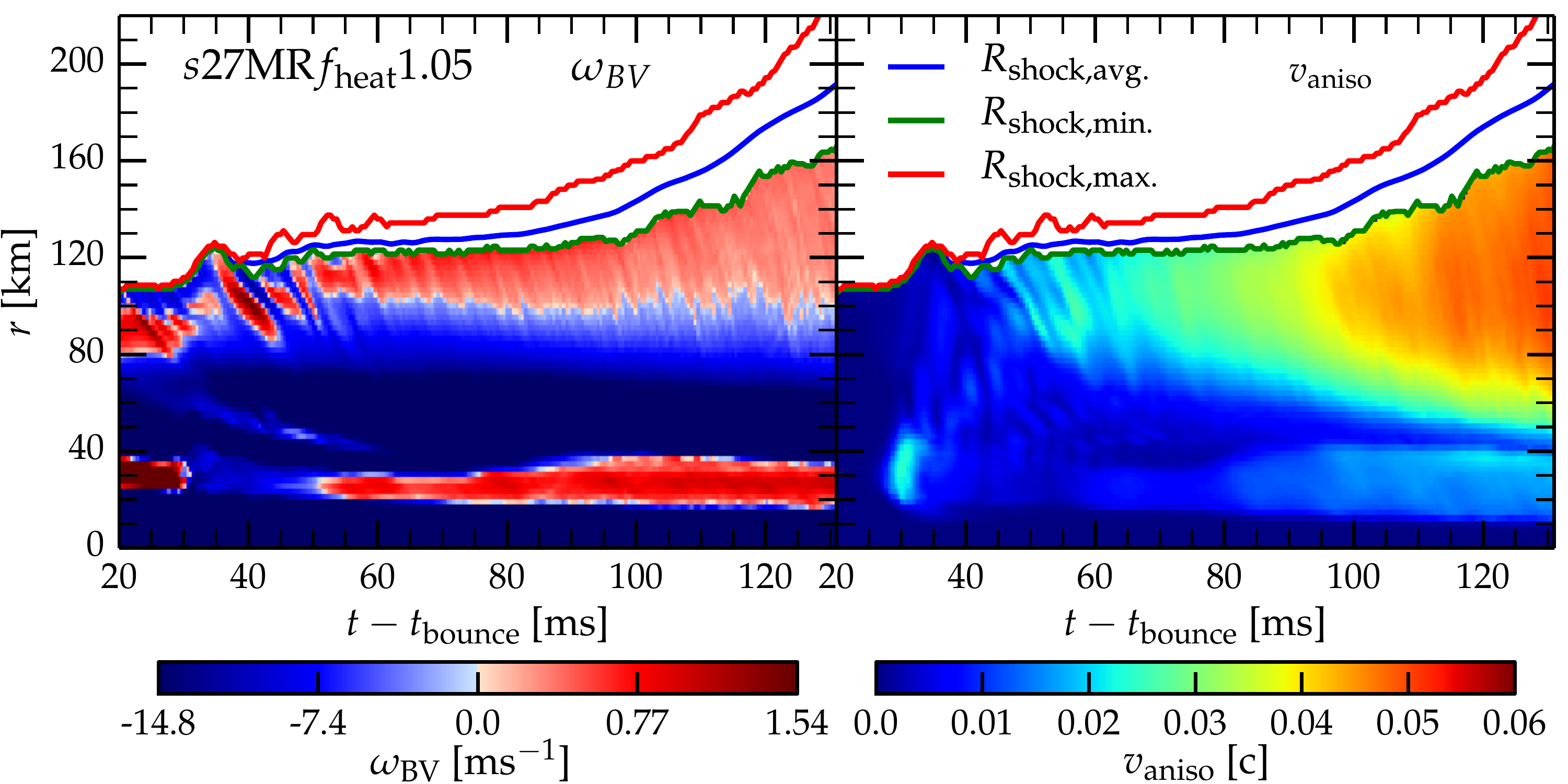}
\includegraphics[width=0.82\textwidth]{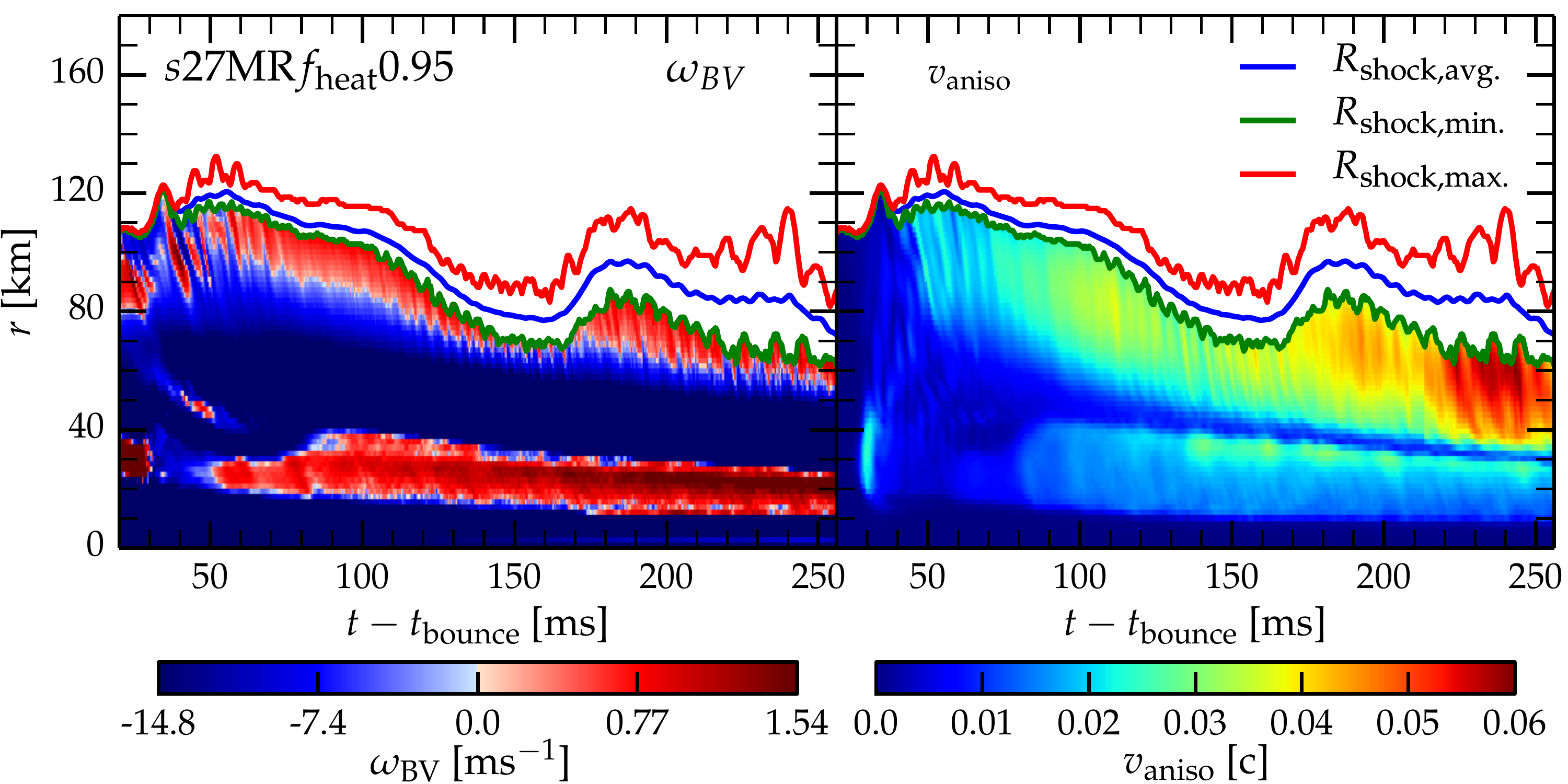}
\includegraphics[width=0.82\textwidth]{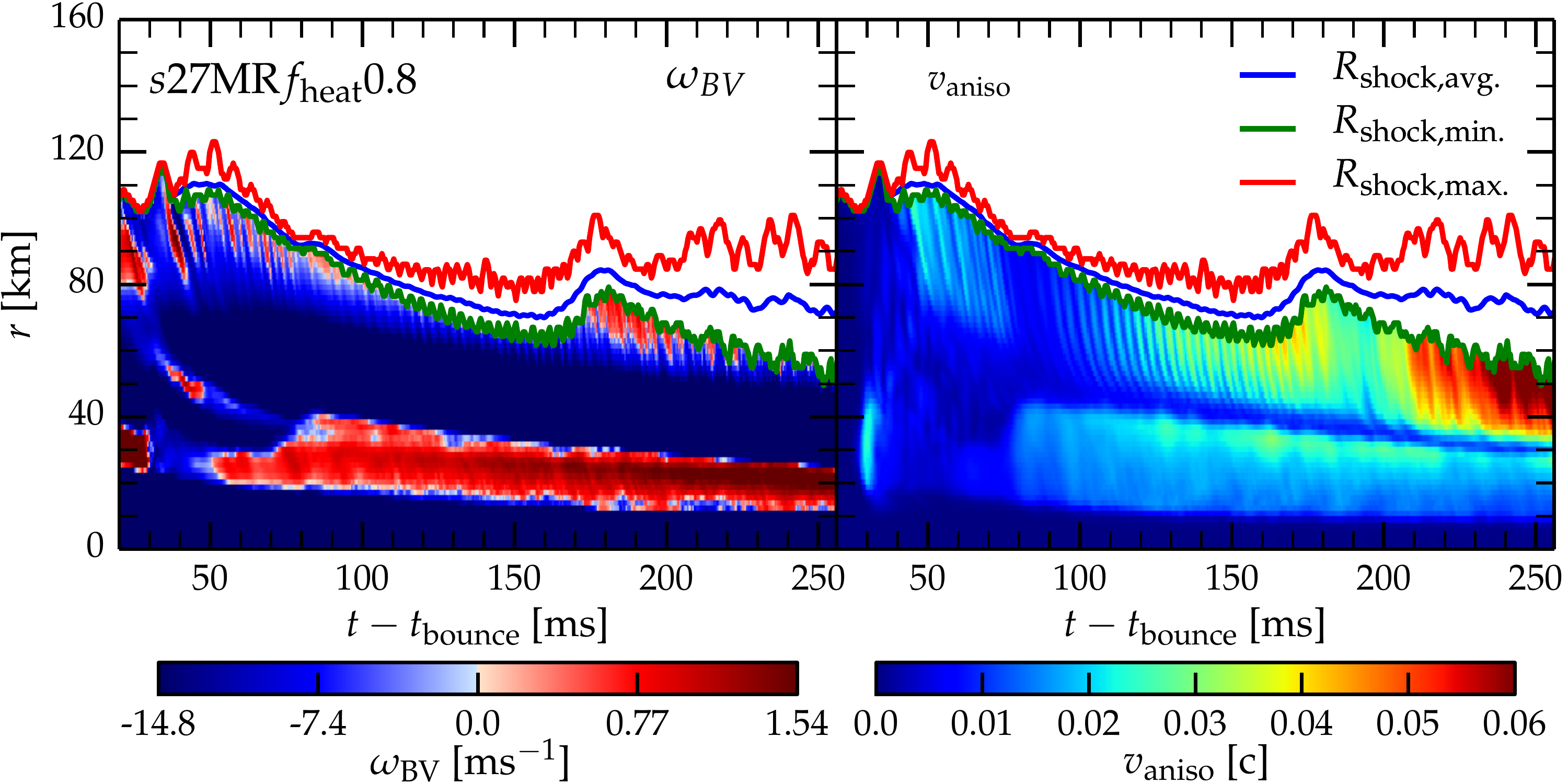}
\caption{Colormaps showing radial slices of the angle-averaged
  Brunt-V\"ais\"al\"a (BV) frequency $\omega_\mathrm{BV}$
  (Equation~\ref{eq:omegabv}; left panels) and anisotropic velocity
  $\upsilon_\mathrm{aniso}$ (Equation~\ref{eq:vaniso}; right panels) in
  models $s27\mathrm{MR}f_\mathrm{heat}1.05$ (top panels),
  $s27\mathrm{MR}f_\mathrm{heat}0.95$ (center panels), and
  $s27\mathrm{MR}f_\mathrm{heat}0.8$ (bottom panels). Also shown are
  the maximum (red curves), the average (blue curves), and the minimum
  (green curves) shock radii. We do not show
  $\omega_\mathrm{BV}$ and $\upsilon_\mathrm{aniso}$ outside the
  minimum shock radius. Note the differing temporal and radial scales 
  chosen for different models. 
  }
\label{fig:bv_vaniso}
\end{figure*}

\subsection{Neutrino-Driven Convection}
\label{sec:convection}

Neutrino heating in the gain region establishes a negative radial
entropy gradient (e.g., \citealt{herant:92}) and thus can drive
convection. In stable stars, convection occurs on a stationary
background. Not so in the postshock region of a core-collapse
supernova: material accreting through the stalled shock is advecting
towards the protoneutron star with velocities up to a few percent of
the speed of light. In order for convection to fully develop,
convective plumes must not only be buoyant with respect to the rest
frame of the background flow, but must be able to rise in the
laboratory (coordinate) frame against the background advection stream.

Depending on accretion rate (determined by progenitor structure;
e.g. \citealt{oconnor:11}), strength of neutrino heating
(i.e. steepness of the entropy gradient), and initial size of the
perturbations entering through the shock from which buoyant plumes can
grow, one can identify three different regimes of convection: (1)
dominance of advection, plumes do not even become buoyant in the rest
frame of the accretion flow; (2) plumes are buoyant in the rest frame
of the accretion flow, but are still advected out of the gain region
into the convectively stable cooling layer; (3) plumes are fully
buoyant and rise against the accretion flow. As we shall see, our
simulations cover all three of these regimes.

We analyze buoyant convection in our simulations with the Ledoux
criterion \citep{ledoux:47} and express its compositional
dependence in terms of the lepton fraction $Y_l$:
\begin{equation}
C_\mathrm{L}= - \left(\frac{\partial \rho}{\partial
  P}\right)_{s,Y_l}
\left[
\left(\frac{\partial P}{\partial s_\mathrm{T}}\right)_{\rho,Y_l}
\left(\frac{\mathrm{d} s_\mathrm{T}}{\mathrm{d} r}\right)
+
\left(\frac{\partial P}{\partial Y_l}\right)_{\rho,s_\mathrm{T}}
\left(\frac{\mathrm{d} Y_l}{\mathrm{d} r}\right)
\right],
\label{eq:ledoux}
\end{equation}
where $s_\mathrm{T} = s+s_\nu$ is the sum of entropies of the matter
$s$ and neutrino field $s_\nu$, while $Y_l = Y_e + Y_{\nu_e} -
Y_{\bar{\nu_e}}$ is the lepton fraction. Since our leakage/heating
scheme does not track the neutrino distribution function, we set $Y_l
= Y_e$ and $s_\nu = 0$ in Equation~(\ref{eq:ledoux}). This is a very good
approximation in the gain region, where neutrinos are almost free
streaming, but is less accurate in the protoneutron star where
neutrinos are trapped at densities above $\sim
10^{12}\,\mathrm{g\,cm^{-3}}$. A fluid parcel is convectively stable
if $C_{\rm L} \le 0$ and unstable otherwise. In the latter case, the
linear growth time of small perturbations to buoyant plumes is given,
approximately, by the inverse of the Brunt-V\"ais\"al\"a (BV)
frequency,
\begin{equation} 
\omega_\mathrm{BV} =
\mathrm{sgn}\left(C_{\mathrm{L}}\right)\sqrt{
\frac{\left|C_{\mathrm{L}}\right| g}{\rho}}\,\,,
\label{eq:omegabv}
\end{equation}
so $\omega_\mathrm{BV} > 0$ implies instability. Here $g$ is the local
free-fall acceleration, which we approximate as $-G M(r) r^{-2}$ in
our postprocessing analysis, where $M(r)$ is the mass enclosed within
radius $r$. A similar approach was used in, e.g.,
\cite{buras:06a,takiwaki:12,ott:13a}.

In addition, we compute the Foglizzo $\chi$ parameter \citep{foglizzo:06}, 
\begin{equation}
\chi = \int_{R_\mathrm{gain}}^{R_\mathrm{shock}}
\frac{\omega_\mathrm{BV}}{|\upsilon_r|}
dr \, ,
\label{eq:chi}
\end{equation}
where $\upsilon_r$ is the radial velocity in the gain region.  $\chi$
can be interpreted as the ratio of the advection timescale to an
average timescale of convective growth. Any small linear seed
perturbation \cite[coming, e.g., from turbulent convection in nuclear
  burning shells; e.g.,][]{arnett:11a,couch:13d,couch:15a} accreting
through the shock can at most grow by a factor of $\sim \exp{(\chi)}$
during its advection through the gain region.  For such linear-scale
perturbations, \cite{foglizzo:06} found that $\chi \gtrsim 3$ is
necessary for convection to develop in the gain region. The situation
is different for large seed perturbations for which the time integral
of buoyant acceleration is comparable to the advection
velocity~\citep{scheck:08}. In this case, a seed perturbation may
develop into a buoyant plume and stay in the gain region instead of
being advected out. The results of \cite{scheck:08} suggest that seed
perturbations of $\sim 1\%$ may be sufficient to allow fully developed
convection even when $\chi < 3$.  \cite{fernandez:14} pointed out that
$\chi$ is quite sensitive to the way it is calculated. We follow the
recent works of \cite{ott:13a,couch:14a,hanke:13}, who all used
instantaneous angle-averaged quantities to compute $\chi$ via
Equation~(\ref{eq:chi}).

If convection develops (either in regime 2 or 3, which we introduced
earlier in this section), its vigor can be measured using the
anisotropic velocity $\upsilon_\mathrm{aniso}$ defined
as~\citep{takiwaki:12}
\begin{equation}
\label{eq:vaniso}
\upsilon_\mathrm{aniso}(r) = \sqrt{\frac{\left\langle \rho \left[
      \left(\upsilon_r -\langle \upsilon_r \rangle_{4\pi} \right)^2 +
      \upsilon_\theta^2 + \upsilon_\varphi^2 \right] 
  \right\rangle_{4\pi}}{ \langle \rho \rangle_{4\pi} }}\,\,,
\end{equation}
where $\langle . \rangle_{4\pi}$ denotes an angular average at a 
fixed radius $r$. $\upsilon_\mathrm{aniso}$ measures the magnitude of the
velocity component that is not associated with a purely
spherically-symmetric radial background 
flow. $\upsilon_\mathrm{aniso}$ is high in regions of large 
angular variations in $\upsilon_r$ and large nonradial velocities
$\upsilon_\theta$ and $\upsilon_\varphi$.  

Convective activity in our simulations can be diagnosed via
Figure~\ref{fig:chi} (showing the amount of buoyant mass and Foglizzo
$\chi$), Figure~\ref{fig:entropy_slices} (showing colormaps of 2D
$x-z$ entropy slices at various postbounce times), and
Figure~\ref{fig:bv_vaniso} (showing the evolution of radial profiles
of the angle-averaged Brunt-V\"ais\"al\"a frequency
$\omega_\mathrm{BV}$ and $\upsilon_\mathrm{aniso}$).

In all models, within milliseconds of bounce, a convectively unstable
region with a steep negative entropy gradient develops inside the
radial shell ranging from $\sim 25\,\mathrm{km}$ to $\sim
40\,\mathrm{km}$ due to the propagation of the gradually weakening
shock. In our simulations, this phase occurs already during the 1D
evolution with \code{GR1D} (not shown here). This leads to the
development of strong \emph{prompt} convection within $\sim
20\,\mathrm{ms}$ after the start of the 3D simulations, as is evident
from the $\upsilon_\mathrm{aniso}$ profiles shown in
Figure~\ref{fig:bv_vaniso}. The $\chi$ parameter (Figure~\ref{fig:chi}) is
generally $< 3$ in all models, but prompt convection develops
nevertheless from numerical perturbations, which are $\gtrsim 1\%$ at
the time the profile is mapped from 1D to 3D and settles on the 3D
grid (cf.\ the discussion in \citealt{ott:13a} about perturbations
from the Cartesian computational grid). Prompt convection smoothes out
the negative entropy gradient on a timescale of $5-10\,\mathrm{ms}$,
leading to a rapid weakening and then to complete disappearance of
convection. The latter is most apparent from the dramatic decrease in
buoyant mass shown in the top panel of Figure~\ref{fig:chi}.

Deleptonization at the edge of the protoneutron star creates a
negative lepton gradient within $30-40\,\mathrm{km}$. It drives
convection in the protoneutron star, setting in at
$35-50\,\mathrm{ms}$ after bounce (Figure~\ref{fig:bv_vaniso}).
Protoneutron star convection {(albeit modeled only
  schematically, given the limitations of our neutrino treatment,
  cf. \S\ref{sec:methods})} is similar in all models, since it is
independent of neutrino heating in the gain region.

In model $s27\mathrm{MR}f_\mathrm{heat}1.05$, neutrino heating creates
a negative entropy gradient in the region between
$\sim$$80\,\mathrm{km}$ and the shock, leading to a convectively
unstable layer, as apparent from the upper left panel of
Figure~\ref{fig:bv_vaniso}. This triggers and sustains convection in
the postshock region starting at $t-t_\mathrm{b} \sim
50\,\mathrm{ms}$, at this early time aided by additional entropy
perturbations coming from variations in the shock radius. The amount
of buoyant mass (top panel of Figure~\ref{fig:chi}) has a local
maximum when convection first starts and exceeds this maximum only
once the explosion begins to develop in model
$s27\mathrm{MR}f_\mathrm{heat}1.05$. The Foglizzo $\chi$ parameter
shown (bottom panel of Figure~\ref{fig:chi}) suggests that much of the
convection, while clearly visible in the entropy slices of this model
shown in Figure~\ref{fig:entropy_slices}, is not fully bouyant in the
coordinate frame (regime 2). Only at $t \gtrsim 100\,\mathrm{ms}$
after bounce does $\chi$ grow beyond the linear-theory threshold value
of $3$ and the amount of buoyant mass increases, indicating that
convection is now fully buoyant and convective plumes begin to push
out the stalled shock, driving both its expansion and asymmetry
(regime 3). These general trends agree well with what was found by
\cite{burrows:12}, \cite{couch:13b}, \cite{dolence:13},
\cite{ott:13a}, and \cite{couch:14a} for 3D simulations with strong
neutrino heating that yielded explosions.

In model $s27\mathrm{MR}f_\mathrm{heat}0.95$ with moderate neutrino
heating, the buoyant mass peaks when neutrino-driven convection first
develops and then gradually declines with time. While we see clear
signs of convection in the entropy snapshot at $\sim$$80\,\mathrm{ms}$
after bounce in Figure~\ref{fig:entropy_slices}, convective plumes
never become fully buoyant in the coordinate frame in this model and
regime 3 of fully developed buoyant convection is never reached. At
$150\,\mathrm{ms}$ after bounce, convection has all but disappeared
and the buoyant mass has plummeted. At this point, SASI has taken
over from neutrino-driven convection as the dominant hydrodynamical
instability (cf.~\S\ref{sec:sasi}). It is the driving agent for the
large anisotropic motions visible at late times in
Figure~\ref{fig:bv_vaniso}.

Finally, in model $s27\mathrm{MR}f_\mathrm{heat}0.8$ with weak
neutrino heating, convective instability is weak and only
intermittent. As in the other models, the amount of buoyant mass peaks
at $\sim$$50\,\mathrm{ms}$ after bounce, but convection weakens
quickly and is almost gone at $80\,\mathrm{ms}$ after bounce, as is
obvious from the entropy snapshot of this model shown in
Figure~\ref{fig:entropy_slices}. SASI dominates the postbounce
hydrodynamics in this model and is responsible for the strong
anisotropic dynamics diagnosed via $\upsilon_\mathrm{aniso}$ in
Figure~\ref{fig:bv_vaniso} at later postbounce times.

\begin{figure}[t]
\centering
\includegraphics[width=1\columnwidth]{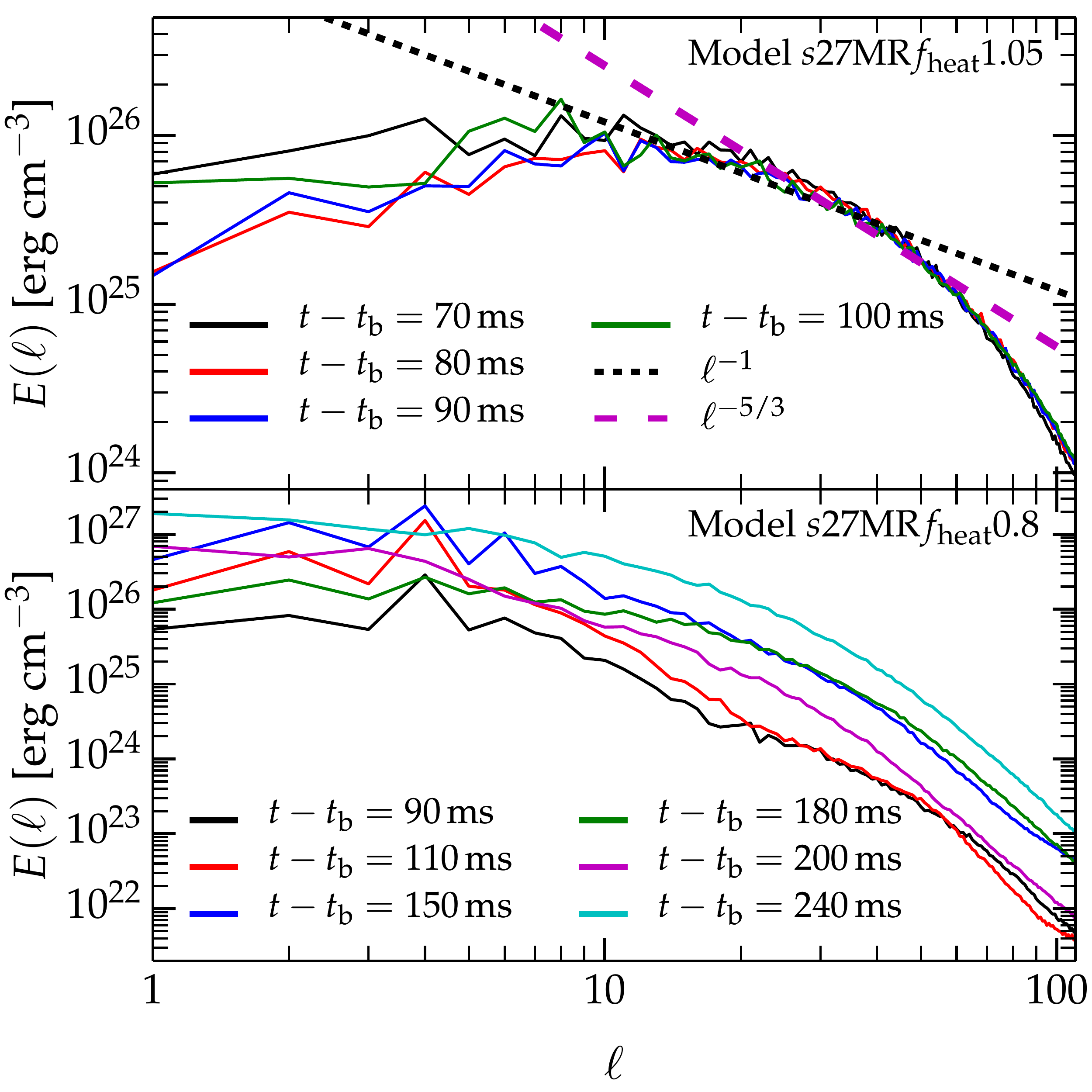}
\caption{{\bf Top panel:}
  Angular spectra $E(\ell)$ of the angular kinetic energy density of
  convective turbulent motion (Equation ~\ref{eq:el}) in model
  $s27\mathrm{MR}f_\mathrm{heat}1.05$ at a range of postbounce times
  before the onset of shock expansion. We overplot lines indicating
  $\ell^{-5/3}$ (Kolmogorov) and $\ell^{-1}$ scaling. The energy
  containing range is near $\ell = 5-10$ and should be linked by the
  inertial range to the dissipation scale at large $\ell$.  $E(\ell)$
  is most consistent with $\ell^{-1}$ scaling in the ``inertial
  range,'' which suggests that numerical viscosity affects the
  efficiency of kinetic energy from large to small scales.  {\bf
    Bottom panel:} Angular spectra $E(\ell)$ for model
  $s27\mathrm{MR}f_\mathrm{heat}0.8$ (weak neutrino heating) at
  various postbounce times. In this SASI-dominated model, turbulence
  is driven by shear and entropy gradients associated with secondary
  shocks. The $E(\ell)$ spectrum is highly nonstationary at all $\ell$.}
\label{fig:turb_spec}
\vspace{0.5ex}
\end{figure}

\subsection{Turbulence}
\label{sec:turbulence}

Turbulence has recently moved into the center of attention in
core-collapse supernova theory and simulation
\citep{murphy:11,murphy:13,couch:15a}. In the absence of very rapid
core rotation and strong magnetic fields (the most likely scenario for
the vast majority of massive stars; \citealt{heger:05,ott:06spin}),
there is no physical source of viscosity in the postshock gain layer
that could prevent neutrino-driven convection from developing into
high Reynolds number turbulence (see Appendix~\ref{sec:viscosity} for
a more detailed discussion of physical viscosity in the gain
layer). Similarly, shear flows and entropy gradients due to periodic
shock shape variations driven by SASI will seed turbulence behind
the stalled shock.

A growing number of core-collapse supernova studies analyzing
turbulence are showing that one of the key differences between 2D and
3D simulations is the well known (e.g., \citealt{kraichnan:67})
inverse and unphysical 2D turbulent cascade that drives kinetic energy
toward large scales in 2D instead of toward small scales in 3D (e.g.,
\citealt{hanke:12,dolence:13,takiwaki:14a,couch:13b,couch:14a,couch:15a}). 
Simulations
{ suggest} that kinetic energy at large scales is favorable for
explosion, which may explain why 2D simulations appear to explode more
easily than 3D simulations in many studies. Moreover, work by
\cite{murphy:13} and \cite{couch:15a} demonstrated that the effective
pressure generated by turbulent stress in the postshock region is an
important contribution to the overall pressure behind the shock and
likely pivotal in launching an explosion against the preshock ram
pressure of accretion.

Turbulence in the postshock region of core-collapse supernovae is
anisotropic in the radial direction and quasi-isotropic in nonradial
motions (\citealt{murphy:11,murphy:13,handy:14,couch:15a}). It is
mildly compressible (reaching preexplosion Mach numbers of $\sim$$0.3
- 0.5$; \citealt{couch:13d}) and only quasi-stationary. In the
following, we focus on the kinetic energy spectra of turbulence in our
simulations and compare neutrino-driven convection dominated and
SASI-dominated regimes of postbounce hydrodynamics.

We study the spectrum of turbulent motion in our simulations by
decomposing the kinetic energy density of the \emph{nonradial} motion
into spherical harmonics on a spherical shell in the gain
layer. Following previous work by \cite{hanke:12}, \cite{couch:13b},
\cite{dolence:13}, \cite{couch:14a}, and \cite{handy:14}, we define
coefficients
\begin{equation}
  \epsilon_{\ell m} (t)= \oint \sqrt{\rho(\theta,\phi)}\, \upsilon_t Y_{\ell}^{m} (\theta,\phi) d\Omega\,\,,
\end{equation}
where $\upsilon_t = \sqrt{\upsilon^2_\theta + \upsilon^2_\phi}$ and
where we average the $\sqrt{\rho} \upsilon_t$ part within the radial
shell $r\in(R_1, R_2)$. In our analysis, we use $R_1 =
0.7R_\mathrm{shock, min}$, $R_2 = 0.8 R_\mathrm{shock,min}$, where
$R_\mathrm{shock,min}$ is the minimum shock radius at the time we
carry out the spatial averaging { (we also tested
  variations of $R_1$ and $R_2$, i.e. $(0.7-0.9)R_\mathrm{shock,min}$
  and $(0.6-0.8)R_\mathrm{shock,min}$ and found no significant
  difference in the spectra)}. The total angular kinetic energy density
at a given $\ell$ is then
\begin{equation}
  \label{eq:el}
  E(\ell) = \sum_{m=-\ell}^\ell \epsilon^2_{\ell m} . 
\end{equation}
In order to calculate $E(\ell)$ at time $t$, we additionally average
$E(\ell)$ over the time interval $(t-\Delta t, t+\Delta t)$, where we
take $\Delta t = 5\,\mathrm{ms}$ in our analysis. We note that in the
turbulence literature, it is more common to express the turbulent
energy spectrum in terms of the wave number $k$ instead of
$\ell$. However, since we are decomposing the nonradial motion on a
spherical shell, spherical harmonics are the natural choice of basis.
We expect $E(\ell)$ to be a power law $\propto \ell^{-\alpha}$, with
$\alpha$ varying between different ranges in $\ell$. Any power-law
spectrum $E(k) \propto k^{-\alpha}$ corresponds to $E(\ell) \propto
\ell^{-\alpha}$ in the limit of large $\ell$ \citep[e.g., Chapter 21
  of][]{peebles:93} and, as pointed out by \cite{hanke:12}, the
power-law indices of $E(\ell)$ and $E(k)$ should correspond well to
each other already at $\ell \gtrsim 4$.

Studies of 3D turbulent flows in various scenarios have shown that the
spectrum of turbulent motion $E(\ell)$ consists of three different
regions \citep[e.g.,][]{pope:00}. The energy of turbulent flow is
supplied in the \emph{energy-containing range} at large spatial scales
comparable to the size of the turbulent region by creating large-scale
turbulent eddies with $\ell$ of $ \sim \,\mathrm{few}$.  In the energy
containing range, $E(\ell)$ is typically nearly constant or increases
mildly with $\ell$. The \emph{inertial range} is the range in $\ell$
in which energy cascades (i.e.\ is transferred) from large-scale
eddies down to small scales and $E(\ell)$ decreases with
$\ell^{-\alpha}, \alpha > 1$. In the \emph{dissipation range}, the
dependence of $E(\ell)$ on $\ell$ is significantly steeper than in the
inertial range, typically $E(\ell) \propto \exp(-\ell)$
\citep[e.g.,][]{pope:00}.  Our simulations do not contain any physical
viscosity (which would, in any case, be extremely small in the
postshock gain layer, cf.~Appendix~\ref{sec:viscosity}) and
dissipation is due to the numerical viscosity inherent to our
hydrodynamics scheme.

In Kolmogorov theory of isotropic, incompressible, stationary
turbulence (e.g., \citealt{landau:59}), $E(\ell) \propto \ell^{-5/3}$
in the inertial range. For the case of neutrino-driven convection in
the gain layer, we expect a similar or even steeper scaling, since (1)
turbulence is more or less isotropic in the nonradial directions
considered here (\citealt{murphy:13}), (2) turbulence has sufficient
time to fully develop, since the preexplosion, stalled-shock phase
lasts for many turnover cycles, and (3) higher Mach-number (more
compressible) flow generally leads to a more efficient turbulent cascade
to small scales, and thus a steeper power law (e.g., \citealt{garnier:09}).

The top panel of Figure~\ref{fig:turb_spec} shows $E(\ell)$ at various
postbounce times in model $s27\mathrm{MR}f_\mathrm{heat}1.05$, whose
gain-layer hydrodynamics is dominated by neutrino-driven convection
due to strong neutrino heating (cf.~\S\ref{sec:convection}). While
there are variations in $E(\ell)$ in the low-$\ell$ energy-containing
range, at $\ell \gtrsim 10$ the spectra are quite steady after $t -
t_b \sim 80\,\mathrm{ms}$, indicating that the flow is at least
quasi-stationary at intermediate and small scales in this model.
$E(\ell)$ should peak at $\ell$ corresponding to the size of the
convectively unstable gain region. At $90\,\mathrm{ms}$ after bounce
we infer from the top right panel of Figure~\ref{fig:bv_vaniso} a
radial extent of the turbulent region of $H \sim 70\,\mathrm{km}$ and
a typical radius of $R \sim 90\,\mathrm{km}$ (the center of the
convective region). The value of $\ell$ at which the spectrum
$E(\ell)$ peaks should correspond to the number of eddies with
diameter $H$ that fit into the turbulent region, $\ell_\mathrm{peak}
\sim (2\pi R) / H -1 \approx 7$. This is close to what is
realized by the spectrum at  $90\,\mathrm{ms}$ after
bounce shown in Figure~\ref{fig:turb_spec} for this model. At smaller
scales (larger $\ell$), the spectrum should first exhibit an extended
inertial range region with $E(\ell) \propto \ell^{-5/3}$ before
steepening in the dissipation range at very large $\ell$. This,
however, is not borne out by Figure~\ref{fig:turb_spec}. At
intermediate $\ell$ of $10 - 40$, the spectrum is much shallower than
$\ell^{-5/3}$ and most consistent with $\ell^{-1}$ and steepens only
at $\ell \gtrsim 40$ and quickly surpasses the $\ell^{-5/3}$
scaling. This kind of spectral behavior is qualitatively and
quantitatively consistent with what was found for neutrino-driven
turbulence in the simulations of \cite{dolence:13}, \cite{couch:14a},
and \cite{couch:15a}, who all used numerical methods and Cartesian
grid setups very similar to ours.

The bottom panel of Figure~\ref{fig:turb_spec} shows $E(\ell)$ at
various postbounce times in the SASI-dominated model
$s27\mathrm{MR}f_\mathrm{heat}0.8$ (weak neutrino
heating). Anisotropic motions in this model (and at postbounce times
$\gtrsim$$150\,\mathrm{ms}$ also in model
$s27\mathrm{MR}f_\mathrm{heat}0.95$ whose $E(\ell)$ is not shown) are
driven by entropy and vorticity perturbations caused by the SASI,
which is much more intermittent than neutrino heating. This is
reflected in the turbulent kinetic energy spectra that vary \emph{at
  all scales} with postbounce time and do not reach the
quasi-stationarity that we observe for neutrino-driven turbulent
convection in the top panel of Figure~\ref{fig:turb_spec}. The
variations in $E(\ell)$ in the SASI-dominated model can be directly
correlated with the strength of SASI. For example, the overall
magnitude of $E(\ell)$ grows from $90-150\,\mathrm{ms}$ after bounce,
which coincides with the increasing strength of SASI oscillations
seen in Figure~\ref{fig:a1s} for this model. At
$\sim$$180\,\mathrm{ms}$, $E(\ell)$ at large scales is decreased as a
result of damped SASI oscillations shortly after the silicon interface
advects through the shock (cf.~\S\ref{sec:sasi}). While there is much
variation in the overall magnitude of $E(\ell)$, the scaling of
$E(\ell)$ is significantly shallower than $\ell^{-5/3}$ and closer to
$\ell^{-1}$ at the scales one would naively be tempted to identify
with the inertial range. This is in agreement with the neutrino-driven
turbulent convection case.

Several authors (e.g., \citealt{dolence:13,couch:14a}) have argued
that the $\ell^{-1}$ scaling observed in contemporary 3D simulations
could be due to the physical nature of the postshock turbulent flow
that deviates significantly from the assumptions of Kolmogorov
turbulence. Our interpretation is different. An inertial range scaling
with $\ell^{-\alpha}$ with $\alpha \le 1$ is unphysical, since in the
limit of infinite resolution, the integral turbulent energy is
divergent. Neutrino-driven turbulence is essentially isotropic in the
nonradial directions, it is quasi-stationary, and only mildly
compressible. Local high-resolution studies of driven turbulence in
this regime generally find an inertial range with $\alpha \simeq 5/3$
for the incompressible transverse flow component and $\alpha > 5/3$
for the compressible part (e.g., \citealt{schmidt:06}). Those
simulations and simulations of turbulence in other regimes (e.g.,
\citealt{porter:98,sytine:00,kaneda:03,dobler:03,haugen:04,kritsuk:07,federrath:13}), however,
also find the appearance of a shallower region with $\alpha \sim 1$
near the end of the inertial range before the transition to the
dissipation range.  This corresponds to inefficient energy transport at
these scales and is referred to as the \emph{bottleneck effect}. This
is understood to be a physical feature of turbulence that is related
to a partial suppression of nonlinear interactions of turbulent eddies
of different scale near the regime of strongest dissipation
\citep{yakhot:93,she:93,falkovich:94,verma:07,frisch:08}.

\cite{sytine:00} carried out a resolution study with local
compressible (Mach 0.5) {freely-decaying} turbulence
simulations using the original PPM solver of \cite{colella:84}. Their
Figure~11 shows that their local simulations with $1024^3$ and $512^3$
cells resolve an inertial range with $\alpha = 5/3$. The bottleneck
with $\alpha < 5/3$ appears at the end of this range. However, with
decreasing resolution, the bottleneck shifts to progressively lower
wavenumbers, consuming more and more of the resolved inertial range.
Already at $256^3$, the inertial range is gone and energy injection
and dissipation scales are joined directly with $1 \lesssim \alpha <
5/3$. On the basis of their results and previous work by
\cite{porter:98}, \cite{sytine:00} argue that the numerical viscosity
of their PPM scheme provides dissipation that affects the flow
directly on spatial scales from 2 to $\sim$12 times the width of a
computational cell. This should be the equivalent of the dissipation
range. On somewhat larger scales, from $\sim$12 to $\sim$32 cell
widths, the flow is still affected by the viscosity of PPM indirectly,
creating the observed bottleneck effect. { We point out
  that these studies focused on a specific regime of turbulence,
  freely-decaying and isotropic, which is different from the one we
  observe in our simulations. In our simulations, turbulence is driven
  by buoyancy and is anisotropic. However, very recently,
  \cite{radice:15a} came to conclusions very similar to \cite{sytine:00}
  also for driven anisotropic turbulence.}

Our numerical hydrodynamics scheme is very similar to the PPM
implementation used by \cite{sytine:00}, but likely \emph{more
dissipative}, because we do not employ the original exact Riemann
solver of \cite{colella:84}, but the more dissipative HLLE solver
(cf.~\S\ref{sec:methods}). The numerical viscosity of our scheme is
thus larger than in the scheme of \cite{sytine:00} {(see
the comparison between HLLE and HLLC in \citealt{radice:15a}) } and 32
cell widths is only a lower bound on the scale that is affected by
numerical viscosity in our simulations. In our fiducial medium
resolution simulations for which we present $E(\ell)$ in
Figure~\ref{fig:turb_spec}, the cell width is $\sim$$1.4\,\mathrm{km}$
and the region that is turbulent has a radial extent of
$\sim$$70\,\mathrm{km}$ (cf.~Figure~\ref{fig:bv_vaniso}). Hence, we
have (in the best case) $70\,\mathrm{km} / 1.4\,\mathrm{km} \approx
50$ cells covering the turbulent region (of which $\sim$32 are
affected by numerical viscosity), which is much less than the 512
linear cell width needed by \cite{sytine:00} to resolve some inertial
range. We conclude that at the resolution employed here, \emph{we
cannot reasonably expect to resolve the inertial range in the
turbulent gain layer}. All that we are seeing here, and that the
simulations of \cite{dolence:13}, \cite{couch:14a}, and
\cite{couch:15a} show, is the \emph{contamination of the turbulent
energy spectrum by numerical viscous effects all the way up to the
energy containing range}. \emph{Turbulence is thus not resolved in
these and in the present 3D simulations}. This conclusion is further
supported by the low numerical Reynolds number of
$\mathrm{Re}$$\sim$$70$ that we find in
Appendix~\ref{sec:reynolds_number} for our simulations, suggesting
that our simulations are somewhere in between perturbed laminar flow
and turbulence. \cite{couch:15a} estimated $\mathrm{Re}\sim 350$ via a
simple comparison of the size of the convective region with the linear
grid spacing (e.g., \citealt{pope:00}). Using their approach, we find
$\mathrm{Re} \sim 180$. Authors carrying out simulations on spherical
grids have argued that they see $\alpha$ closer to $5/3$ and resolve
the inertial range \citep{hanke:12,handy:14}. However, the angular and
radial resolutions employed in these studies are significantly lower
than the effective resolutions provided by our 3D Cartesian grids and
it is not clear how turbulence could be resolved in their simulations
if not in ours.

\begin{figure}[t]
\centering
\includegraphics[width=1\columnwidth]{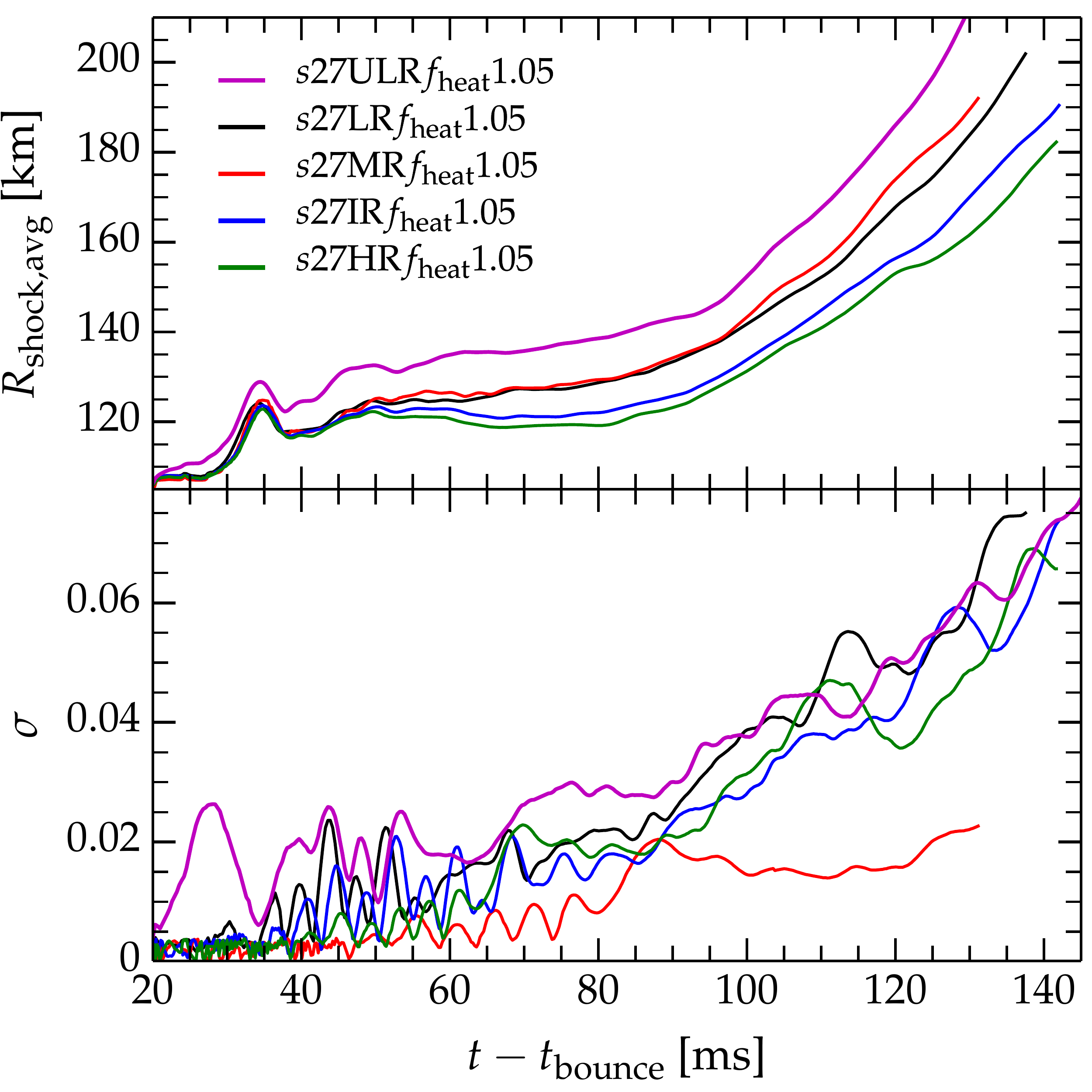}
\caption{{\bf Top panel}: Evolution of the average shock radii for
  five different resolutions in the strong neutrino heating regime
  ($f_\mathrm{heat}=1.05$; see Table~\ref{tab:results} for simulation
  details). Lower resolution leads to larger shock radii. {\bf Bottom
    panel}: Evolution of the normalized root mean square deviation
  $\sigma_\mathrm{shock}$ of the shock radius from its angle averaged
  value for the same five models.} 
\label{fig:shockradres_f1.05}
\vspace{0.5ex}
\end{figure}

\begin{figure}[t]
\centering
\includegraphics[width=1\columnwidth]{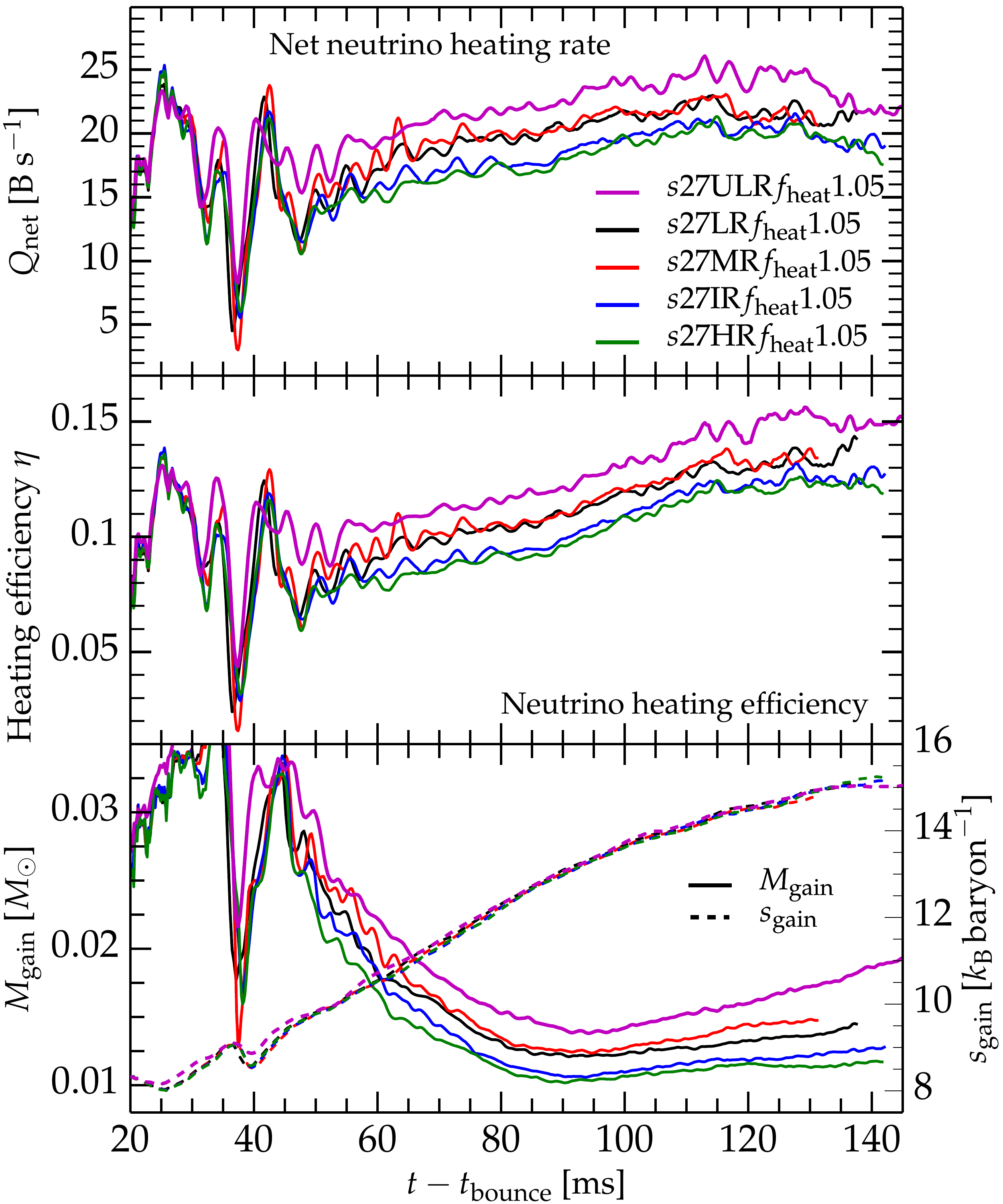}
\caption{Time evolution of the { integral} net neutrino
  heating $Q_\mathrm{net}$ (top panel), heating efficiency $\eta$
  (center panel), mass in the gain region $M_\mathrm{gain}$ (bottom
  panel, left ordinate), and the average entropy in the gain region
  $s_\mathrm{gain}$ (bottom panel, right ordinate) for the case of
  strong neutrino heating and five different resolutions. Low
  resolution results in artificially efficient neutrino heating and in
  an overestimate of the mass in the gain region.}
\label{fig:heat_f10}
\vspace{0.5ex} 
\end{figure}

\section{Results: Dependence on Numerical Resolution}
\label{sec:res}

\begin{figure}[t]
\centering
\includegraphics[width=1\columnwidth]{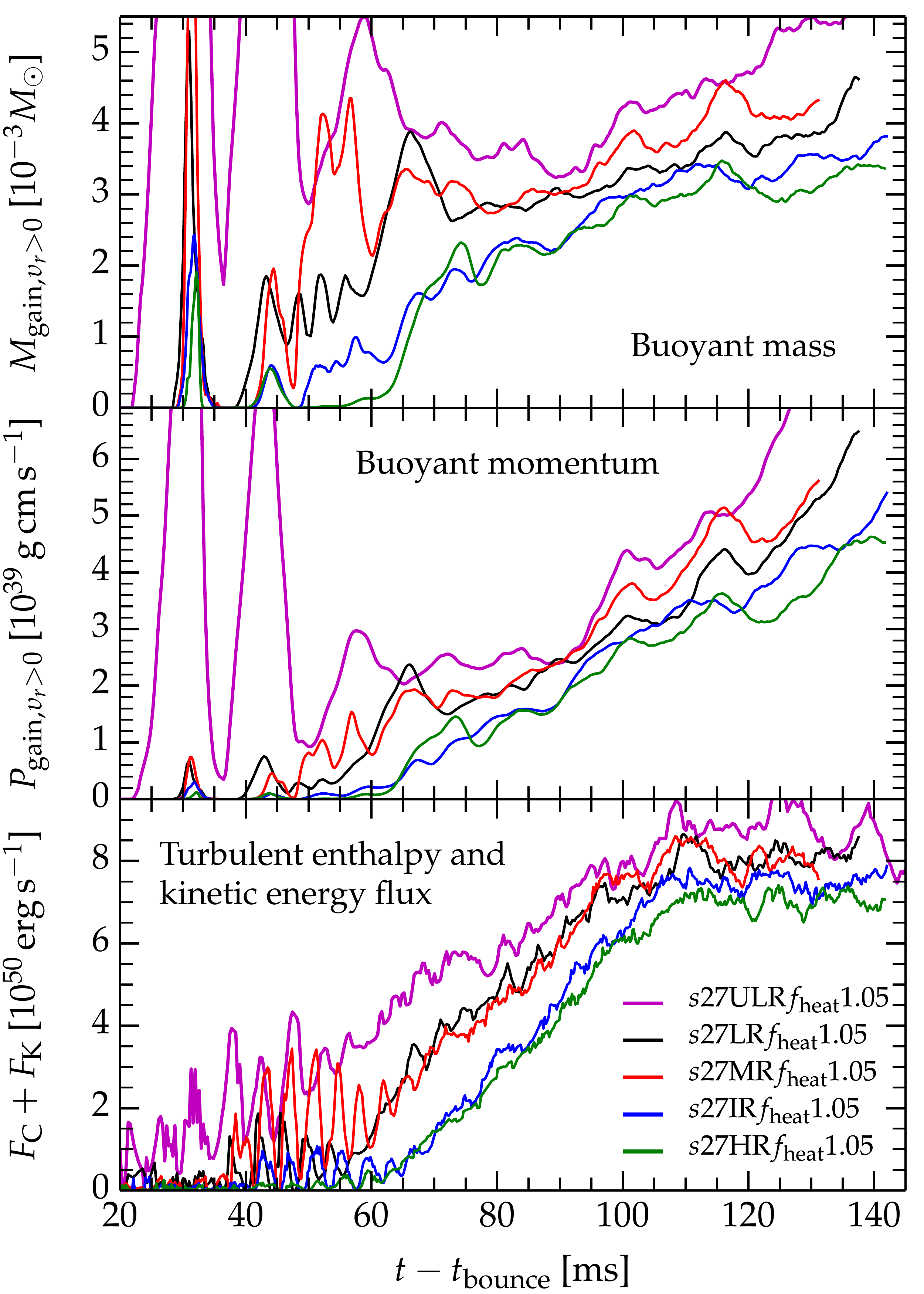}
\caption{Comparison of buoyant mass (top panel), buoyant momentum
  (center panel), and radial convective enthalpy and kinetic energy
  fluxes (bottom panel, cf.~Equation~\ref{eq:turbflux}) for
  simulations with five different resolutions of the strong neutrino
  heating case.  Higher-resolution simulations, in particular in the
  first $\sim$$100\,\mathrm{ms}$ after bounce (before shock expansion
  sets in), have smaller $M_{\mathrm{gain},\upsilon>0}$,
  $P_{\mathrm{gain},\upsilon>0}$, and $F_\mathrm{C}+F_\mathrm{K}$ than
  lower resolution simulations.}
\label{fig:buoyant_mass}
\end{figure}

\begin{figure*}[t]
\centering
\includegraphics[width=0.495\textwidth]{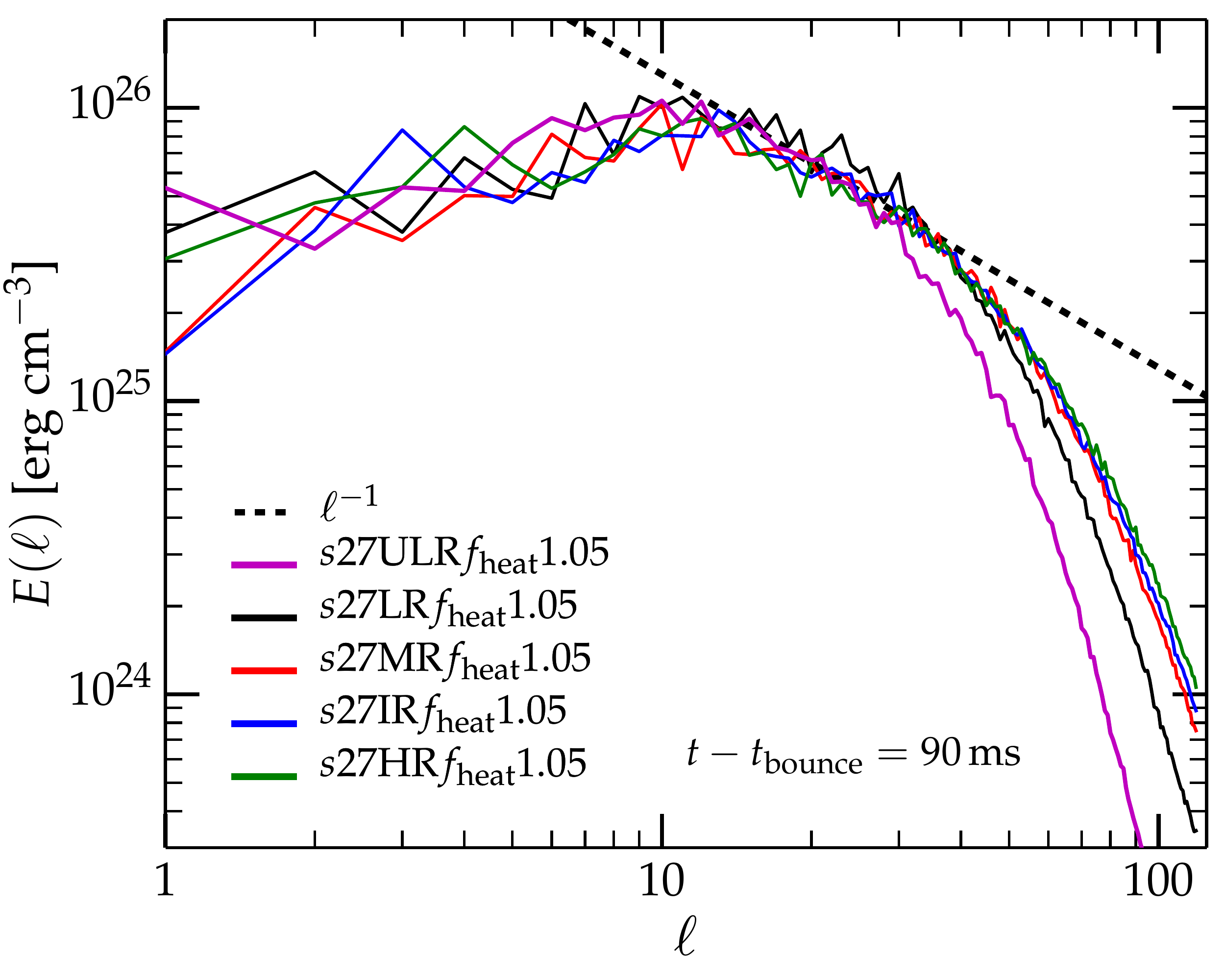}
\includegraphics[width=0.495\textwidth]{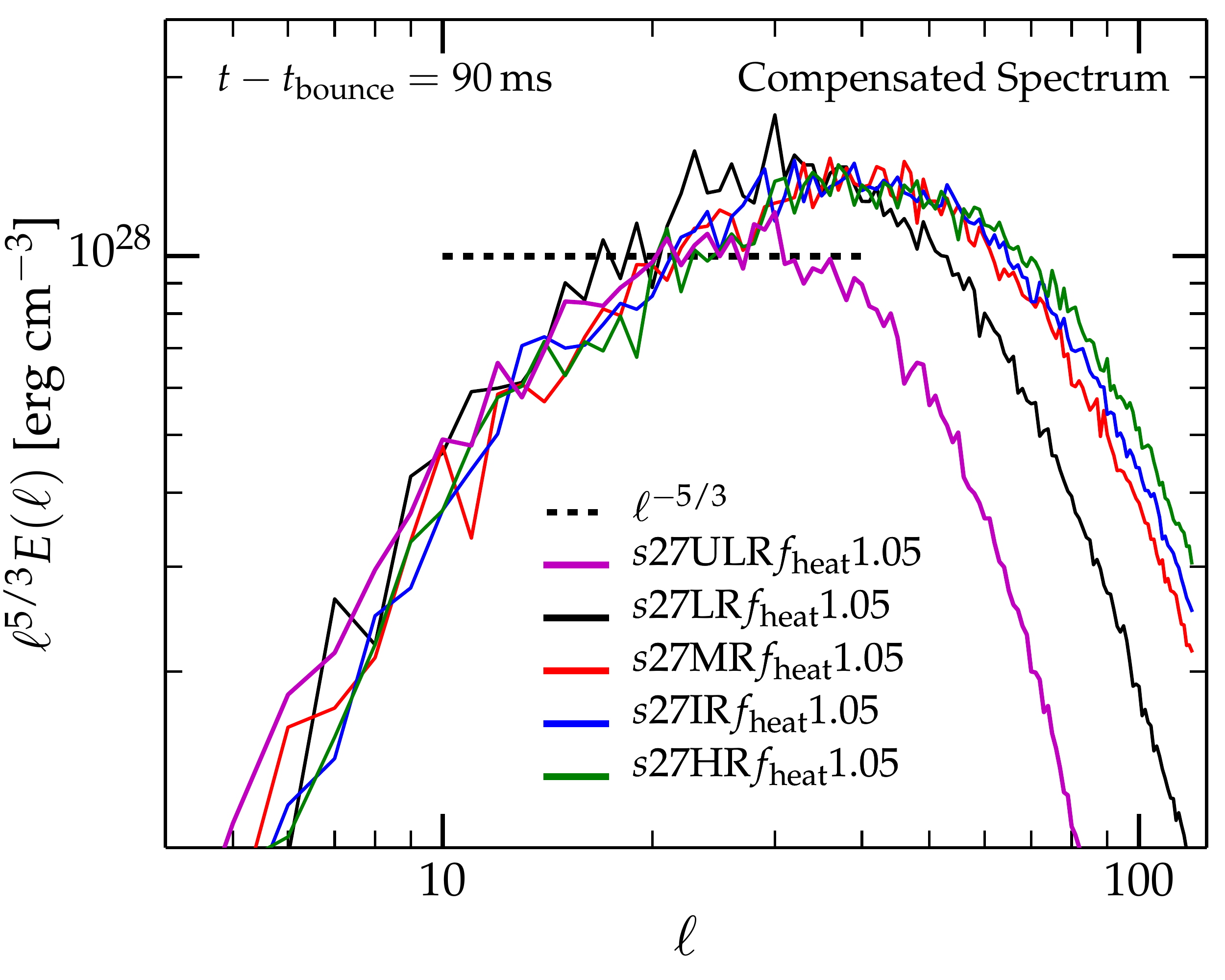}
\caption{{\bf Left panel:} Angular spectra of the turbulent kinetic energy density for
  five different resolutions at $90\,\mathrm{ms}$ after bounce in the
  strong neutrino heating case. The turbulent transport of energy to small
  scales becomes increasingly efficient with increasing resolution. Decreasing
  resolution leads to an onset of strong dissipation at smaller $\ell$.
{\bf Right panel:} Compensated ($\ell^{5/3}$ rescaled) turbulent spectra. The dashed line
  indicates the range in $\ell$ where we expect the inertial range and where the spectrum
  should be flat if $E(\ell) \propto \ell^{-5/3}$ were realized as predicted by theory.
}
\label{fig:turb_spec_res_dep}
\vspace{0.5ex}
\end{figure*}

\subsection{Strong Neutrino Heating,\\ Convection Dominated Regime}
\label{sec:res_comp_105}

We explore the impact of numerical resolution in the regime of strong
neutrino heating and neutrino-driven convection dominated 3D
hydrodynamics by running simulations of the $s27$ progenitor with a
total of five different resolutions with linear cell width in the
postshock gain layer varying by almost a factor of four.  Our baseline
$s27\mathrm{MR}f_\mathrm{heat}1.05$ model has a resolution on the AMR
level containing the postshock region and the shock with linear
resolution $dx_\mathrm{shock} = 1.416\,\mathrm{km}$. This correponds
to an effective angular resolution at a radius of $100\,\mathrm{km}$
of $d(\theta,\phi) = 0.81^\circ$.  In models
$s27\mathrm{ULR}f_\mathrm{heat}1.05$ (``ultra-low resolution''),
$s27\mathrm{LR}f_\mathrm{heat}1.05$ (``low resolution''),
$s27\mathrm{IR}f_\mathrm{heat}1.05$ (``intermediate resolution''),
$s27\mathrm{HR}f_\mathrm{heat}1.05$ (``high resolution''), this is
$3.784\,\mathrm{km}$, $1.892\,\mathrm{km}$, $1.240\,\mathrm{km}$, and
$1.064\,\mathrm{km}$, respectively.  These correspond to effective
angular resolutions at a radius of $100\,\mathrm{km}$ of $2.15^\circ$,
$1.08^\circ$, $0.81^\circ$, $0.71^\circ$, and $0.61^\circ$, for ULR,
LR, IR, HR, respectively (see also Table~\ref{tab:results}).

Figures~\ref{fig:shockradres_f1.05} and \ref{fig:heat_f10} give a
concise summary of the effects of resolution on the postbounce
hydrodynamics and on the development of a neutrino-driven
explosion. The overall trend is very clear: the lower the resolution,
the larger the average shock radius, the higher the neutrino heating
rate, the greater the heating efficiency, and the larger the mass in
the gain layer. 

While these overall trends are robust, there are some inconsistencies
in detail of note.  The mean specific entropy in the gain layer
(bottom panel of Figure~\ref{fig:heat_f10}) appears almost completely
independent of resolution. The asphericity of the shock front,
measured by the normalized root mean square deviation
$\sigma_\mathrm{shock}$ of the shock radius in the bottom panel of
Figure~\ref{fig:shockradres_f1.05} has no systematic resolution
dependence in its magnitude and variations. The fiducial MR simulation
is an outlier with the overall smallest $\sigma_\mathrm{shock}$. The
shock radius, neutrino heating, heating efficiency, and mass in the
gain region are very similar in the LR and MR models (differing in
$dx_\mathrm{shock}$ by $\sim$30\%) and at the end of its evolution,
the MR simulation actually has a slightly larger shock radius than its
LR counterpart. On the other hand, the IR and HR simulations, which
differ only by $\sim$$15\%$ in resolution, are consistent with each
other in all quantities except $\sigma_\mathrm{shock}$. The MR/IR
simulation pair differs in resolution by $\sim$$15\%$, yet their
results are much farther apart than those of the LR/MR pair that
differs by $\sim$$30\%$ in resolution. { These variations
  about the general trend are indicative of the possibility that many
  if not most (or all) of our simulations are not yet in the
  convergent regime.  Perhaps much higher resolutions in the
  convectively unstable layer may be needed to accurately and in a
  converged manner capture the hydrodynamics of core-collapse
  supernovae dominated by neutrino-driven turbulent
  convection.}

\cite{hanke:12}, \cite{couch:14a}, and \cite{takiwaki:14a}, who
carried out less extensive 3D parameter studies with similar or lower
resolutions, found the same trends with resolution observed in our
simulations. \cite{handy:14}, on the other hand, found improved
conditions for explosion with increasing resolution. However, they
studied angular grid spacings from $24^\circ$ down to only
$2^\circ$. Their highest resolution roughly corresponds to our ULR
case. At such coarse resolutions, which suppress nonradial convective
motions, it is not at all surprising that the conditions become more
favorable for explosion as increasing resolution begins to allow
nonradial motions. The \cite{handy:14} simulations thus probe the
resolution dependence of 3D postbounce hydrodynamics in a completely
different regime than our simulations.

Figure~\ref{fig:buoyant_mass} provides further evidence for why lower
resolution simulations are (artificially) favorable for
neutrino-driven explosions. The lower the resolution, the larger the
amount of buoyant mass (defined as the mass in the gain region with
positive radial velocity) and the greater the amount of positive
momentum in the gain region. The more mass is truly buoyant (and thus
in regime 3 of neutrino-driven convection discussed in
\S\ref{sec:convection}), the greater the neutrino heating rate and
efficiency (cf.\ Figure~\ref{fig:heat_f10}). Note, however, that by
comparing the total mass in the gain region given in the bottom panel
of Figure~\ref{fig:heat_f10} with the buoyant mass in the top panel of
Figure~\ref{fig:buoyant_mass}, one finds that that the truly buoyant
mass is at most $\sim$$20\%$ of the mass in the gain region. We expect
that this fraction will sensitively depend on progenitor structure and
will be higher in progenitors with lower postbounce accretion rates
than in the $27$-$M_\odot$\ progenitor that we study here.

The bottom panel of Figure~\ref{fig:buoyant_mass} shows the time
evolution of the sum of the angle-averaged ``turbulent'' radial fluxes
of enthalpy ($F_\mathrm{C}$, also known as ``convective flux'') and
kinetic energy ($F_\mathrm{K}$) near the shock. We follow
\cite{hurlburt:86} and \cite{handy:14} and define
\begin{eqnarray}
F_\mathrm{C} = \int_{4\pi} \rho \upsilon_r
\left(\epsilon+\frac{P}{\rho}\right)^\prime r^2 d\Omega, \nonumber \\
F_\mathrm{K} = \int_{4\pi} \rho \upsilon_r
\left(\frac{1}{2}\upsilon_i \upsilon_i \right)^\prime r^2 d\Omega,
\label{eq:turbflux}
\end{eqnarray}
where $\rho$ is the density, $\upsilon_r$ is the radial velocity, $\epsilon$ is
the internal energy, $P$ is the pressure, and $\upsilon_i$ is the $i$th
component of velocity. All primed quantities represent variations about the
angle-averaged mean {so that, for instance, $F_K$ measures the amount
of the turbulent part of the specific kinetic-energy (note that by construction
$\overline{\bar{\upsilon}_i \upsilon_i'} \equiv 0$, where $\bar{\cdot}$ denotes
the angular average)}. We evaluate the angular integrals in
Equation~\ref{eq:turbflux} at each time at a radius that corresponds to the
instantaneous minimum shock radius. A number of studies (e.g.,
\citealt{bhf:95,dolence:13,ott:13a,handy:14,couch:13b,couch:15a}) have argued
that buoyant convective/turbulent bubbles are locally important in driving shock
deformation and expansion. \cite{murphy:13} and \cite{couch:15a} have shown that
the additional effective ram pressure due to turbulence is crucial for the
relative ease of explosions in 2D and 3D compared with the 1D case. All these
effects are related to the convective/turbulent flux of kinetic energy and
enthalphy in the gain layer and near the shock (cf.~\citealt{yamasaki:06}).
Figure~\ref{fig:buoyant_mass} reveals that the sum $F_\mathrm{C} + F_\mathrm{K}$
near the shock decreases with increasing resolution, creating less favorable
conditions for explosion.

The radial convective/turbulent fluxes are dominated by flow at large
and intermediate scales. In Figure~\ref{fig:turb_spec_res_dep}, we
plot angular turbulent kinetic energy spectra $E(\ell)$
(cf.~equation~\ref{eq:el}) in the gain layer at $90\,\mathrm{ms}$
after bounce for all resolutions. The left panel shows the plain
$E(\ell)$ spectra, while the right panel shows ``compensated'' spectra
that are rescaled by $\ell^{-5/3}$ as is customary in studies of
Kolmogorov turbulence. A flat graph in the region where the inertial
range is expected would indicate consistency with Kolmogorov
turbulence. Given the spatial scale of the gain layer in our
simulations, we would expect the energy containing range to be around
$\ell \sim 7$ (cf.~\S\ref{sec:turbulence}) which should be followed by
an inertial range with $E(\ell) \propto \ell^{-5/3}$ before
dissipation sets in. None of our simulations, not even the HR case,
exhibits any inertial range. Where the inertial range should be,
$E(\ell)$ is most consistent with an $\ell^{-1}$ scaling, which is
indicative of a bottleneck due to viscous contamination because of
insufficient numerical resolution (cf.~\S\ref{sec:turbulence}).

Figure~\ref{fig:turb_spec_res_dep} does not clearly show large
differences of $E(\ell)$ in the energy-containing range with changing
resolution. However, note that at low $\ell$ the spectra are not fully
stationary (see Figure~\ref{fig:turb_spec}). One should also recall
that we here project out the radial part and that the important radial
kinetic energy and enthalpy fluxes decrease with increasing
resolution, which indicates less total energy/power at large scales
(Figure~\ref{fig:buoyant_mass}). The figure does, however, clearly
demonstrate that transport of turbulent energy to small scales becomes
increasingly efficient as the resolution is increased. The energy
contained at large $\ell$ increases systematically with resolution and
even appears to converge as the resolution gets close to the HR
case. However, the resolution decrements between the various
shown simulations are not constant and the three highest simulations
differ only by $\sim$$15\%$ in resolution, while MR and
LR differ by $\sim$$30\%$ and LR and ULR differ by a factor of
two. Since no inertial range is realized, we do not consider any of
our studied resolutions to be in the regime in which the flow is truly
turbulent. The HR simulation, at $\sim$$90\,\mathrm{ms}$ after bounce,
covers the entire postshock region with $\sim$$240^3$ computational
cells, but only the outer $\sim$$70\,\mathrm{km}$ are actually
convectively unstable and are effectively covered by $70.0
\,\mathrm{km} / 1.064\,\mathrm{km} \approx 66$
linear cell widths. According to \cite{sytine:00} this
resolution may still be a factor of $\gtrsim$$7-8$ too low for resolving
the inertial range.

\begin{figure}[t]
\centering
\includegraphics[width=1\columnwidth]{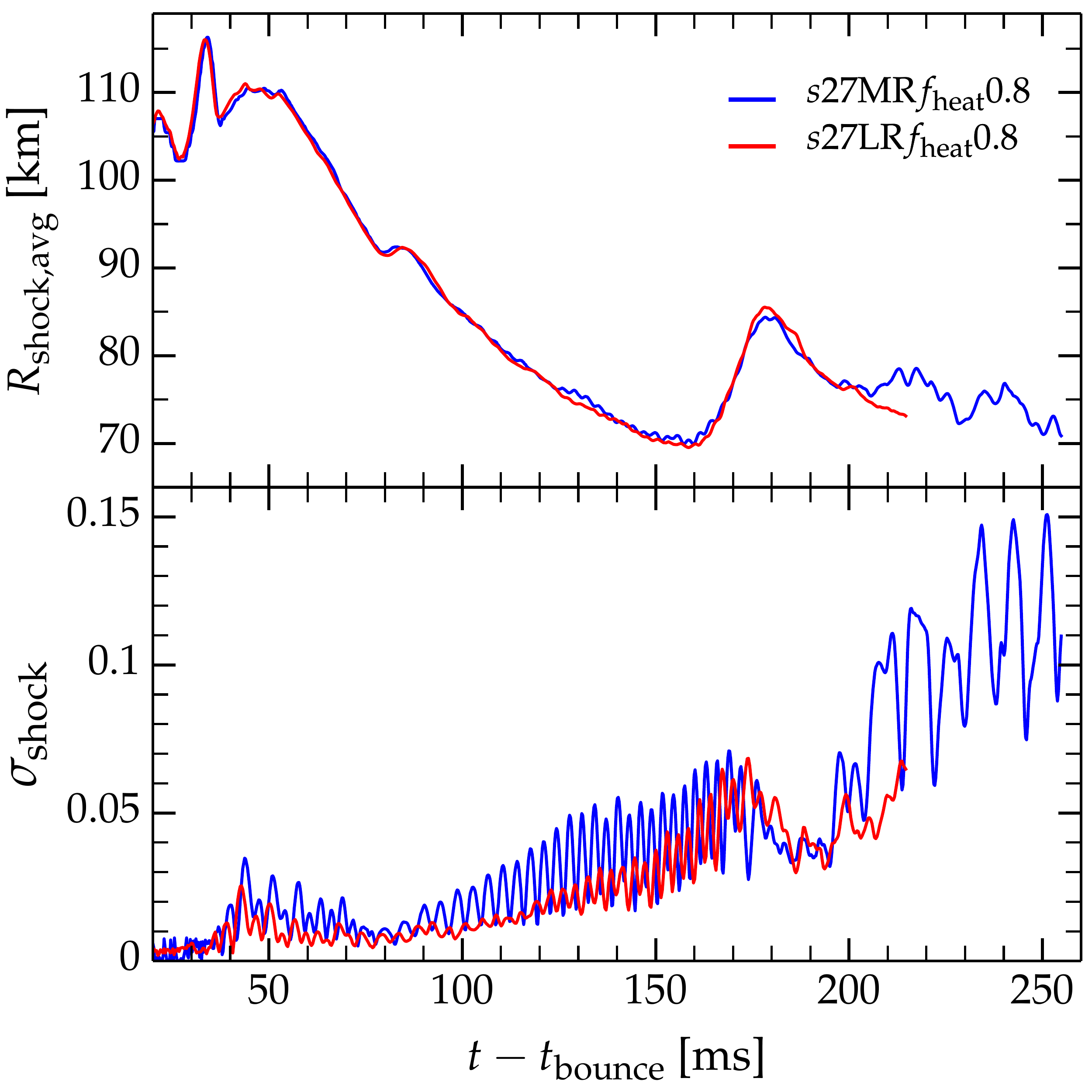}
\caption{{\bf Top panel}: Comparison of the average shock radius
  evolution in the MR and LR simulations of the SASI-dominated
  $f_\mathrm{heat}=0.8$ model with weak neutrino heating. The MR and
  LR resolutions differ by $\sim$$30\%$. The shock radius evolution is
  almost independent of resolution in this model and until
  $\sim$$200\,\mathrm{ms}$ after bounce. Then, the shock in the
  higher-resolution (MR) simulations expands somewhat, possibly
  related to the appearance of large-scale $\ell = 1$ SASI modes at
  this time (cf.~Figure~\ref{fig:alm_f08}).  {\bf Bottom panel}: The
  normalized root mean square deviation $\sigma_\mathrm{shock}$ of the
  shock radius from its angle averaged value in the MR and LR
  simulations. The oscillations in $\sigma_\mathrm{shock}$, which are
  due to SASI, are much stronger in the MR simulation, indicating
  that SASI is weaker in the LR simulation
  (cf.~Figure~\ref{fig:alm_f08}).}
\label{fig:shockradres_f0.8}
\vspace*{0.5ex}
\end{figure}

\begin{figure}[t]
\centering
\includegraphics[width=1\columnwidth]{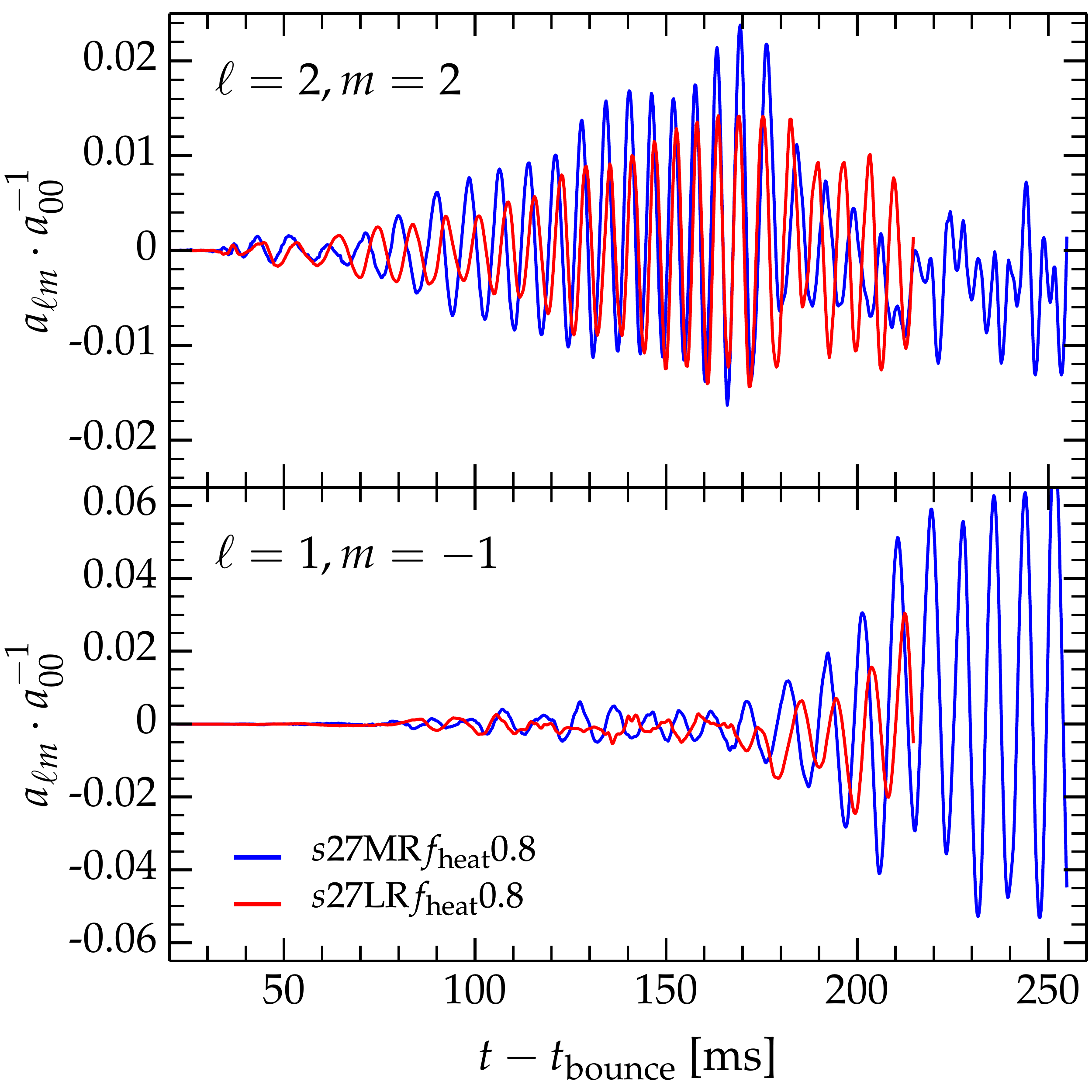}
\caption{Comparison of select normalized mode amplitudes $a_{\ell m}
  \cdot a^{-1}_{00}$ of the shock front between the LR and MR simulations of
  the SASI-dominated  $f_\mathrm{heat} = 0.8$ model. The top
  panel shows the $\ell = 2, m=2$ mode and the bottom panel shows the
  $\ell = 1, m = -1$ mode. The qualitative evolution of the modes are
  nearly independent of resolution and behave as discussed in
  \S\ref{sec:sasi} for this model. However, the magnitude of the mode
  amplitudes is generally lower in the lower-resolution
  simulation. The resolutions differ by $\sim$$30\%$.}
\label{fig:alm_f08}
\vspace*{0.5ex}
\end{figure}

\subsection{Weak Neutrino Heating, SASI Dominated Regime}

We investigate resolution dependence in the weak neutrino heating,
SASI-dominated case by comparing our baseline-resolution simulation
$s27\mathrm{MR}f_\mathrm{heat}0.8$ with a simulation carried out with
lower resolution, $s27\mathrm{LR}f_\mathrm{heat}0.8$, which uses the
same resolution of the LR simulation in the previous section.  MR and
LR resolutions differ by $\sim$$30\%$ (cf.\ Table~\ref{tab:results}).
Additional simulations with further decreased or increased resolution
would be advisable but were not possible for the SASI-dominated case
within the limitations of our computational resources.

The top panel of Figure~\ref{fig:shockradres_f0.8} compares the
evolution of the average shock radius in the MR and LR
simulations. They are qualitatively and quantitatively nearly
identical and significantly closer to each other than the LR and MR
simulations in the strong neutrino heating case discussed in the
previous Section~\ref{sec:res_comp_105}. We also find (and show in
Figure~\ref{fig:heat}) that { integral} net neutrino
heating, neutrino heating efficiency, and the mass in the gain region
are { very similar} in the MR and LR models throughout the
simulated postbounce time.

While the average shock radius evolves nearly identically in the MR
and LR cases, we find that deviations from the average due to SASI
oscillations are smaller in the LR case. This is apparent from the
bottom panel of Figure~\ref{fig:shockradres_f0.8}, which shows the
normalized root mean square deviation $\sigma_\mathrm{shock}$ of the
shock radius from its angle-averaged value. The oscillations in
$\sigma_\mathrm{shock}$ are due to SASI and their amplitudes are
much smaller in the LR simulation. Figure~\ref{fig:shockradres_f0.8}
depicts the evolution of the normalized $\ell = 2, m = 2$ and $\ell =
1, m = -1$ amplitudes (Equation~\ref{eq:alm}) of the shock front as
representative examples of the $\ell = \{1,2\}$ mode families in the
LR and MR simulations.  The evolution of these modes is qualitatively
similar in both LR and MR simulations, but the LR simulation shows
systematically lower mode amplitudes in both $\ell = 1$ and $\ell = 2$
until $\sim$$160\,\mathrm{ms}$ after bounce. At that time, the silicon
interface advects through the shock, leading to its transient
expansion, and to a profound change in the SASI mode structure
(cf.~\S\ref{sec:sasi}). In the LR simulation, the $\ell = 2$ mode
amplitudes decay less than in the MR case, but the $\ell = 1$ modes do
not grow as strongly as in the MR case. This deviation between MR
and LR SASI dynamics has an effect on the average shock radius, whose
MR and LR evolutions depart from each other towards the end of the LR
simulation at $\sim$$200\,\mathrm{ms}$ after bounce.

Our results show that the weak neutrino heating, SASI-dominated regime
of 3D postbounce hydrodynamics is sensitive to resolution and this
sensitivity is strongest in the development and non-linear dynamics of
SASI. \cite{sato:09} have shown that for SASI to reach convergence,
the numerical resolution must be sufficiently high to capture the full
advective-acoustic cycle of entropy/vorticity perturbations that
advect through the postshock region, are reflected at the protoneutron
star, and propagate back up to the shock. The LR simulation
($dx_\mathrm{shock} = 1.892\,\mathrm{km}$) has evidently too low
resolution, but since we only have two resolutions at hand, we cannot
with confidence say that the MR simulation ($dx_\mathrm{shock} =
1.416\,\mathrm{km}$) is in the convergent regime for SASI.

\section{Discussion and Conclusions}
\label{sec:conclusions}

Core-collapse supernovae are fundamentally three dimensional (3D). The
3D simulations presented in this paper add to the growing set of
modern 3D simulations that are beginning to elucidate the many facets
of postbounce hydrodynamics in neutrino-driven core-collapse
supernovae. Our results -- in agreement with \cite{hanke:13} and
\cite{couch:14a} -- show, beyond reasonable doubt, that 3D postbounce
hydrodynamics can be dominated by neutrino-driven convection or by the
standing accretion shock instability (SASI) or can involve both at the
same time or at different times. 

SASI is not an artifact of axisymmetry (2D), but is, at least in
current 3D results, generally associated with high postbounce
accretion rates, with moderate or weak neutrino heating, and with
failed 3D explosions in progenitors that explode in 2D
\citep{burrows:12,ott:13a,hanke:13,couch:14a}.  An interesting open
question is now if 3D SASI-dominated core-collapse supernovae can
still yield explosions or if their progenitors are part of the
possibly large fraction of massive stars that simply do not explode
and result in black holes
\citep{kochanek:14a,kochanek:15,clausen:15}. \cite{hanke:13} found
an explosion in at least one SASI-dominated case of a $25$-$M_\odot$
progenitor, but that simulation used an artificial contracting inner
boundary, dialed-in inner boundary neutrino luminosity, and a gray
neutrino transport scheme. Their more sophisticated energy-dependent
radiation-hydrodynamics 3D simulation of the same $27$-$M_\odot$
progenitor studied here shows SASI-dominated dynamics and does not
appear to yield an explosion.

There is broad consensus now that high (kinetic) energy at scales
comparable to the size of the postshock gain layer is favorable for
shock expansion and explosion. More (buoyant) nonradially moving mass
in the gain layer increases the efficiency of neutrino heating (e.g.,
\citealt{buras:06b,murphy:08}).  Large-scale convective radial fluxes
of buoyant material, associated with buoyant high-entropy bubbles (due
to neutrino-driven convection or SASI) can deliver heat and do
mechanical work on the shock
\citep{bhf:95,yamasaki:06,hanke:12,dolence:13,couch:13b,ott:13a,
  handy:14,couch:14a}. The effective pressure of turbulence at large
scales adds to the thermal pressure in the postshock region and
facilitates larger shock radii and thus helps explosion
\citep{murphy:13,couch:15a}.

If it is indeed energy/power/dynamics at large scales that is needed
to revive the stalled shock, then the results of our resolution and
turbulence study in this paper do not at all bode well for the
{ standard} neutrino mechanism in 3D.  We studied effective
angular resolutions in the postshock gain layer from $\sim$$2^\circ$
(which is the resolution used in \citealt{hanke:13} and the highest
resolution considered by \citealt{handy:14}) to $\sim$$0.6^\circ$.
Going from the lowest to the highest resolution, the neutrino heating
rate drops precipitously (by $\sim$$25\%$), and so do the total amount
of mass in the gain layer, the amount of buoyant mass, and the
convective fluxes of kinetic energy and enthalpy.  The result is a
smaller average shock radius and a slower transition to explosion with
increasing resolution. Our model with strong neutrino heating still
shows at least the onset of an explosion even in the highest
resolution, but in a more critical case, a low-resolution simulation
may incorrectly predict an explosion where a higher-resolution
simulation does not.

Our results, in agreement with the results of the simpler
``light-bulb'' simulations carried out by \cite{hanke:12}, show that
the higher the resolution in 3D, the more efficient becomes the
turbulent cascade of nonradial kinetic energy to small scales.
Moreover, comparing our results for the turbulent energy spectra with
what is expected from turbulence theory and local simulations of
mildly compressible turbulence, we find that even our
highest-resolution simulation does not resolve the inertial range of
turbulence. Instead, the realization of turbulence in our
  simulations is likely affected by numerical viscosity all the way up
  to the scale of energy injection. This reduces the efficiency of
the turbulent cascade to small scales and results in a shallow scaling
of the angular energy spectrum. The same is likely true also for the
simulations of \cite{dolence:13} and \cite{couch:14a}, who find
similarly shallow scalings. 

In our highest-resolution simulation, the turbulent gain layer is
covered by $\sim$66 linear computational cell widths. \cite{sytine:00}
argue that the numerical viscosity of the PPM scheme affects regions
of up to $\sim$32 cell widths and that $\gtrsim$512 linear cell widths
across a mildly compressible turbulent region are necessary to resolve
any inertial range with PPM. This would, in the best case, correspond
to $\sim$$7-8$ times our current resolution in the gain layer.
{ Should our conclusion be correct, then obtaining
  neutrino-driven explosions will just get harder when
  higher-resolution simulations become available that resolve the
  inertial range and efficiently transport energy to small scales.}
The standard neutrino mechanism may then need help to somehow corral
energy at large scales and/or a source of additional heating. For
example, large-scale perturbations from precollapse aspherical shell
burning were shown by \cite{couch:13d} to boost the vigor of
turbulence and thus could help.  Magnetic fields could help converge
flow to long-lived high-entropy bubbles \citep{obergaulinger:14} and
the dissipation of Alfv\'en waves propagating from a magnetized
protoneutron star into the gain layer may be an additional source of
heat \citep{suzuki:08}. Moderate rotation in combination with the
magnetorotational instability could also lead to additional heat input
into the gain layer \citep{thompson:05}.

Work in the immediate future will need to be directed towards better
understanding turbulence in the core-collapse supernova context. This
can be addressed first with local simulations that adopt flow
conditions characteristic of the gain layer and resolve a significant
inertial range. Such simulations should be able to test the
conclusions we have drawn on the basis of our global
simulations. Subsequently, high-resolution semi-global simulations
could be used to test the ramifications of not resolving the inertial
range.

\section*{Acknowledgments}
We thank Sean Couch, Peter Goldreich, and Mike Norman for helpful
discussions on turbulence and Thierry Foglizzo for help with
interpreting the behavior of SASI in our simulations. We furthermore
acknowledge helpful discussions with Adam Burrows, Joshua Dolence, Steve Drasco,
Rodrigo Fernandez, Sarah Gossan, Thomas Janka, Bernhard M\"uller,
Jeremiah Murphy, Evan O'Connor, Sherwood Richers, and other members of
our Simulating eXtreme Spacetimes (SXS) collaboration
(http://www.black-holes.org). This research is partially supported by
NSF grant nos.\ AST-1212170, PHY-1404569, PHY-1151197, PHY-1212460,
and OCI-0905046, by NSERC grant RGPIN 418680-2012,
by a grant from the Institute of Geophysics,
Planetary Physics, and Signatures at Los Alamos National Laboratory,
by the Sloan Research Foundation, and by the Sherman Fairchild
Foundation. CR and LR acknowledge support by NASA through Einstein
Postdoctoral Fellowship grant numbers PF2-130099 and PF3-140114,
respectively, awarded by the Chandra X-ray center, which is operated
by the Smithsonian Astrophysical Observatory for NASA under contract
NAS8-03060.  The simulations were performed on the Caltech compute
cluster ``Zwicky'' (NSF MRI award No.\ PHY-0960291), on supercomputers
of the NSF XSEDE network under computer time allocation TG-PHY100033,
on the NSF/NCSA Blue Waters system under NSF PRAC award ACI-1440083,
on machines of the Louisiana Optical Network Initiative, and at the
National Energy Research Scientific Computing Center (NERSC), which is
supported by the Office of Science of the US Department of Energy
under contract DE-AC02-05CH11231. The multi-dimensional visualizations
were generated with the open-source \code{VisIt} visualization package
(\url{https://wci.llnl.gov/codes/visit/}). All other figures were
generated with the \code{Python}-based \code{matplotlib} package
\citep[][\url{http://matplotlib.org/}]{hunter:07}.

\begin{appendix}

\section{A. Dissipation of turbulent motion}
\label{sec:viscosity}

The parameter that is used to indicate the onset of turbulence
is the physical Reynolds number
\begin{equation}
\mathcal{R}e = \frac{\ell u}{\nu}\,\,,
\end{equation}
where $\ell$ is a length scale of the flow, $u$ is a velocity scale of
the flow, and $\nu$ is the physical kinematic viscosity.  Laboratory 
experiments show that the transition from
laminar to turbulent float occurs at $\mathcal{R}e \sim 10^2 - 2\times
10^3$, depending upon the geometry of the experimental boundaries
\citep[cf.][]{arnett:14a}.

The kinematic viscosity is related to the efficiency of momentum 
transport by particles in the fluid.  Employing the Chapman-Enskog 
procedure to first order on the Boltzmann equation gives the kinematic 
viscosity 
\begin{equation}
\nu = \frac{5}{8} \frac{\sqrt{\pi m T}}{\sigma_{t} \rho}, 
\end{equation}
where $\sigma_t$ is the transport cross-section for particles in the 
fluid \citep[cf.][]{mekjian:13}.  Therefore, particles which 
have the smallest total cross section but large average momentum (i.e. 
electrons are unlikely to contribute) will be responsible for the 
viscosity in the medium.  Clearly, neutrons will have the smallest 
interaction cross section due to their neutrality.
Therefore, the kinematic viscosity in the postshock region is given by 
\cite{mekjian:13} (assuming the thermal DeBroglie wavelength is greater 
than the neutron s-wave scattering length $a_{sl}=-17.4 \, \mathrm{fm}$),
\begin{equation}
\nu_N \sim 0.2 
\left(\frac{\rho}{10^{10} \, \mathrm{g\,cm}^{-3}}\right)^{-1} 
\left(\frac{T}{10 \, \mathrm{MeV}}\right)^{1/2} 
\, \mathrm{cm}^{2}\, \mathrm{s}^{-1}.
\end{equation}

The convectively unstable gain layer has a typical length scale of
$\sim$$10^7\,\mathrm{cm}$ and typical velocities of
$\sim$$10^9\,\mathrm{cm}\,\mathrm{s}^{-1}$. Hence, for $\rho =
10^{10}\,\mathrm{g}\,\mathrm{cm}^{-3}$ and $T = 10\,\mathrm{MeV}$, we
obtain an estimate for the physical Reynolds number of
\begin{equation}
\mathcal{R}e \approx  10^{17}\,\,,
\end{equation}
which is larger than what would be predicted just using the 
Braginskii-Spitzer viscosity \citep{braginskii:58,spitzer:62} and 
clearly implies the system should be turbulent.

Momentum exchange due to neutrino emission, absorption, and scattering
has also been invoked as a source of viscosity that can damp turbulent
convection in core-collapse supernovae and protoneutron stars. The
neutrino viscosity in the opaque and semi-transparent regimes was
estimated by, e.g., \cite{burrows:88b,keil:96}, and
\cite{thompson:93}. Here we provide an estimate of the relevance of
neutrino viscosity in the gain region, where neutrinos stream
relatively freely.

The { specific momentum deposition rate due to neutrino
absorption in the gain region can be estimated as
\begin{equation}
\label{eq:dotP}
\dot P \sim 3 \frac{L_{\nu_e}}{4\pi r^2 c} \frac{\sigma_0}{m_\mathrm{b}} \left( 
\frac{\epsilon_\nu}{m_\mathrm{e} c^2}\right)^2, 
\end{equation}
where $L_{\nu_e}$ is the electron neutrino luminosity emerging from
the neutrinosphere, nuclei are assumed to be dissociated,
neutrino--nucleon interactions from \cite{burrows:06} are employed
(electron scattering is neglected), $\sigma_0$ is the characteristic 
neutrino cross section scale defined in \cite{burrows:06}, and the 
luminosity in all neutrino flavors is assumed equal }. Since most of these neutrinos
propagate in the radial direction, momentum will mostly be deposited
in that direction. This will not dampen stochastic turbulent flow, for
which momentum needs to be exchanged between turbulent
eddies. However, $\dot P$ can still be used as an upper limit for
momentum exchange between different turbulent eddies. Using
(\ref{eq:dotP}), one can estimate the timescale for momentum change in
the gain region due to $\dot P$:
\begin{eqnarray}
\label{eq:taumom}
\tau_P \sim \frac{P}{\dot P} \sim 106 \, \mathrm{ms} \left(
\frac{L_{\nu_e}}{10^{52}\, \mathrm{erg\,s^{-1}}} \right)^{-1} \left(
\frac{r}{100\, \mathrm{km}} \right)^2 \left( \frac{\epsilon_\nu}{10\,
  \mathrm{MeV}} \right)^2 \left( \frac{\upsilon_0}{0.01c}
\right), \nonumber
\end{eqnarray}
where $P$ is the characteristic momentum of the largest turbulent
eddies in the gain region and $\upsilon_0$ is their characteristic
velocity. The latter is roughly equal to $\upsilon_\mathrm{aniso}$
(Equation~\ref{eq:vaniso}). The timescale of convective motion of
eddies of size $\lambda$ in the gain region can be estimated as
\begin{equation}
  \tau(\lambda) \sim 3\,\mathrm{ms} \left( \frac{0.01c}{\upsilon_0}
  \right) \left( \frac{\lambda}{10\,\mathrm{km}} \right),  
\end{equation}
For $\lambda \sim 10\,\mathrm{km}$, which is a reasonable estimate for
the eddy scale, we get $\tau_P \gg \tau(\lambda)$, implying that
momentum exchange due to neutrinos is unimportant at large scales. At
smaller scales, the characteristic turbulent eddy velocity is given by
\cite[e.g.,][]{pope:00}
\begin{equation}
\label{eq:ulambda}
\upsilon(\lambda) = \upsilon_0 \left( \frac{\lambda}{\lambda_0}
\right)^{1/3},
\end{equation}
where $\lambda_0$ is the size of the largest eddies. 
Combining (\ref{eq:taumom}) and (\ref{eq:ulambda}), we get
$\tau_P \propto \lambda^{1/3}$. The characteristic timescale of
turbulent eddies scales with $\lambda$ as \cite[e.g.,][]{pope:00} 
\begin{equation}
\tau(\lambda) = \tau(\lambda_0) \left( \frac{\lambda}{\lambda_0}
\right)^{2/3},
\end{equation}
i.e., $\tau(\lambda)$ decreases with $\lambda$ faster than $\tau_P$
does, hence $\tau_P$ remains much larger than $\tau(\lambda)$ for any
$\lambda$. In other words, momentum exchange due to neutrinos cannot
damp turbulence in the gain region.  

\section{B. Effective Reynolds Number}
\label{sec:reynolds_number}

Our simulations do not include any explicit physical viscosity but rely
on the viscosity of the numerical scheme to model the unresolved scales
of the turbulent cascade, in accordance with the implicit large eddy
simulation (ILES) paradigm \citep{garnier:09}. The ILES procedure has
been shown to be robust and accurate for a number of turbulent flows as
long as the effective Reynolds number, defined as
\begin{equation}
  \mathrm{Re} = \frac{\upsilon_0 l_0}{\nu_N},
\end{equation}
$\nu_N$ being the ``numerical viscosity'', is sufficiently large,
e.g.,~\citep{zhou:14}. That is, as long as there is a sufficient separation
between the energy-containing scale $l_0$ and the dissipation scale $l_D$. How
large the scale separation should be in order for the ILES procedure to reach
convergence (in a statistical sense), is problem dependent. Nevertheless it is
useful to measure the range of scales covered by our
simulations in a quantitative way. This will also ease the comparison with
future simulations.

Unfortunately, estimating the effective Reynolds number in ILES
calculations is not trivial because the numerical viscosity does not
really behave like a physical viscosity, that is, it cannot easily be
associated with a given kinematic viscosity coefficient
$\nu_N$. Instead, it is a complex nonlinear function of the
hydrodynamic quantities. Nevertheless, in the framework of
Kolmogorov's theory of turbulence, it is possible to construct
measures of the Reynolds number that do not explicitly depend on
$\nu_N$.  In particular, our estimate of the Reynolds number is based
on the Taylor length~\citep[e.g.,][]{pope:00}:
\begin{equation}\label{eq:taylor.scale}
  \lambda^2 = \frac{5 E}{Z}\,\,,
\end{equation}
where $Z$ is the enstrophy
\begin{equation}\label{eq:enstrophy}
  Z = \int_0^\infty k^2\, E(k)\, \mathrm{d}k\,\,,
\end{equation}
and $E$ is the total energy
\begin{equation}\label{eq:total.energy}
  E = \int_0^\infty E(k) d k = \frac{1}{2} \rho_0 \upsilon_0^2\,\,.
\end{equation}
In the incompressible limit, the average kinetic energy dissipation 
rate is related to the enstrophy via the relation
\begin{equation}\label{eq:kolmogorov.hypothesis.2}
  \epsilon = 2 \nu Z\,\,,
\end{equation}
where $\nu$ is the kinematic viscosity. Furthermore, in Kolmogorov's
theory of turbulence the energy dissipation rate is assumed to be
\begin{equation}\label{eq:kolmogorov.hypothesis}
  \epsilon = C \rho_0 \frac{\upsilon_0^3}{l_0}\,\,,
\end{equation}
where $C$ is of order one (here assumed to be $C=1$) and $l_0$ is the
integral scale, i.e.,~the scale of energy containing
eddies. Substituting \eqref{eq:kolmogorov.hypothesis.2},
\eqref{eq:kolmogorov.hypothesis}, and \eqref{eq:total.energy} into
\eqref{eq:taylor.scale} and using the definition of the Reynolds
number, we obtain
\begin{equation}\label{eq:reynolds.number.1}
\mathrm{Re} = 5 \left(\frac{l_0}{\lambda}\right)^2\,\,.
\end{equation}

We compute the enstrophy in our numerical data as
\begin{equation}\label{eq:computing.enstrophy}
  Z = \sum_{\ell = 0}^\infty R_0^{-2}\, \ell\, (\ell + 1)\, E(\ell)
    \approx \sum_{\ell = 0}^{\ell = 120} R_0^{-2}\, \ell\, (\ell + 1)\, E(\ell)\,\,,
\end{equation}
where $R_0 = 100\ \mathrm{km}$ is the radius at which the spectra are computed
and we restrict our calculation to $\ell \leq 120$, because, for $\ell \gtrsim
120$, the floating point precision necessary to compute the associated Legendre
functions can exceed the limits of the double precision employed in our analysis
code. In computing \eqref{eq:computing.enstrophy}, we used the fact that the
$k^2$ factor in the Fourier expansion corresponds to (minus) the Laplacian in the
physical space and that, by definition,
\begin{equation}
  R_0^2\, \Delta\, Y_{\ell m} = - \ell (\ell + 1) Y_{\ell m}\,\,,
\end{equation}
so that a $k^2$ factor in the Fourier expansion corresponds to a $R_0^{-2} \ell
(\ell + 1)$ factor in the angular expansion.

$E$ is computed in a similar way to $Z$, summing the angular expansion
coefficients of the energy (Equation \eqref{eq:el}) up to $\ell =
120$.  From the values of $Z$ and $E$, we can infer $\lambda =
16.5\ \mathrm{km}$ for model $s27\mathrm{HR}f_\mathrm{heat}1.05$ at
$90\,\mathrm{ms}$ after bounce. The Taylor length is sometimes
interpreted as being the radius of the smallest coherent structures of
the turbulent flow, so it is not surprising that we find $\lambda$ to
be roughly $13$ cells, close to the scale at which we expect numerical
dissipation to be too strong for coherent structures to persist.

The integral scale is computed as
\begin{equation}\label{eq:integral.length}
  l_0 = \frac{\pi}{\ell_0 + 1} R_0\,\,,
\end{equation}
where we compute $\ell_0$ via
\begin{equation}
  \ell_0 \approx \frac{1}{E} \sum_{\ell = 0}^{120} \ell\, E(\ell)\,\,.
\end{equation}
We find, for model $s27\mathrm{HR}f_\mathrm{heat}1.05$, $\ell_0 = 4.1$
corresponding to $l_0 = 61.5\ \mathrm{km}$. The corresponding Reynolds number is
$\mathrm{Re} = 72$.

\begin{deluxetable*}{ccccc}
  \tablecolumns{5} \tablewidth{0pc}
  \tablecaption{Reynolds Number.}
  \tablehead{
    Model&
    $l_D/\mathrm{d} x_{\mathrm{shock}}$&
    $l_0\ [\mathrm{km}]$&
    $\lambda\ [\mathrm{km}]$&
    $\mathrm{Re}$}
    \startdata
    $s27\mathrm{\,U\,L\,R}f_\mathrm{heat}1.05$ & 0.92 & 68.80 & 21.08 & 53.25 \\
    $s27\mathrm{\,L\,R}f_\mathrm{heat}1.05$ & 1.56 & 65.37 & 18.55 & 62.06 \\
    $s27\mathrm{MR}f_\mathrm{heat}1.05$ & 1.82 & 60.97 & 16.52 & 68.14 \\
    $s27\mathrm{\,\,I\,\,R}f_\mathrm{heat}1.05$ & 2.05 & 62.57 & 16.45 & 70.03 \\
    $s27\mathrm{HR}f_\mathrm{heat}1.05$ & 2.33 & 61.55 & 16.20 & 72.21
    \enddata
  \tablecomments{$l_D/\mathrm{d} x_{\mathrm{shock}}$ is the ratio
    between the dissipation length, as measured from
    \eqref{eq:reynolds.number.2} from the $\mathrm{Re}$ and of $l_0$,
    and the grid resolution on the refinement level containing the
    shock, see Table~\ref{tab:results}. $l_0$ is the integral length
    \eqref{eq:integral.length}.  $\lambda$ is the Taylor length
    \eqref{eq:taylor.scale}. Finally $\mathrm{Re}$ is the effective
    numerical Reynolds number computed from
    \eqref{eq:reynolds.number.1}.}
  \label{tab:reynolds.number}
\end{deluxetable*}

As a sanity check, we can use another identity for $\mathrm{Re}$
\citep{pope:00}:
\begin{equation}\label{eq:reynolds.number.2}
  \mathrm{Re} = \left(\frac{l_0}{l_D}\right)^{4/3},
\end{equation}
from which we find the effective dissipation scale to be $l_D \approx 2.5\
\mathrm{km}$. This value is of the same order as the grid spacing, meaning that
the two estimates (\ref{eq:reynolds.number.1}) and (\ref{eq:reynolds.number.2})
for the Reynolds number are roughly consistent with each other, which lends
additional credence to our estimate of $\mathrm{Re}$.

Table \ref{tab:reynolds.number} collects $l_D$, $l_0$, $\lambda$ and
$\mathrm{Re}$ as computed from different resolutions. As expected, the
effective Reynolds number increases slowly with resolution: the
integral scale, $l_0$, stays roughly constant (with the exception of
the $s27\mathrm{\,U\,L\,R}f_\mathrm{heat}1.05$ model), while $\lambda$
decreases.  The dissipation scale, and hence the numerical viscosity
at the grid scale, seems to be increasing with the resolution. A
similar effect was also reported, at much higher resolutions and for
different problems, by \cite{donzis:08} and \cite{aspden:09}. Its
origins are unclear \citep{aspden:09}, but it is again a reminder that
numerical viscosity can behave very differently from the physical
viscosity. The Reynolds numbers reported in
Table~\ref{tab:reynolds.number} are disappointingly low, but this is
not unexpected given the very low resolution ($\sim$66 linear cell
widths across the turbulent region in even our highest-resolution
simulation) that our global simulations provide.

Note that, since we restricted our calculation of $Z$ to $\ell \leq
120$, we are systematically underestimating the enstrophy.  This means
that we might be underestimating the actual value of the effective
Reynolds number \citep{couch:15a}. However, we point out that our
measure is probably also affected by other uncertainties, such as in
the determination of $l_0$, and, more importantly, by possible
systematic errors coming from the fact that we rely on the validity of
Kolmogorov theory of turbulence, which has not yet been verified in
the context of neutrino-driven convection. Given all of these
uncertainties, our estimate of the Reynolds number should only be
taken as an order of magnitude indication. We remark that other
approaches for measuring the Reynolds number have been proposed
\citep[e.g.,][]{fureby:99, aspden:09, zhou:14}.  However, these rely
either on uncertain estimates of the numerical viscosity or on
explicit measures of the kinetic energy dissipation rate. The latter
are difficult to carry out in complex simulations where gravity,
radiation, and compressible effects are all present and must be
accounted for.

\end{appendix}

\end{document}